\documentclass[twocolumn]{aastex61}
\pdfoutput=1 
\usepackage{amsmath,amstext}
\usepackage{mathtools}

\usepackage[T1]{fontenc}
\usepackage{apjfonts} 
\usepackage{graphicx}
\usepackage{color}
\usepackage[figure,figure*]{hypcap}
\usepackage[hang]{subfigure}
\usepackage{array,multirow}

\usepackage{booktabs}
\usepackage{multirow}

\newcommand{\txn}[1]{\textnormal{#1}}
\newcommand{\overbar}[1]{\mkern 1.5mu\overline{\mkern-1.5mu#1\mkern-1.5mu}\mkern 1.5mu}
\newcommand{\bmath}[1]{\mbox{ \boldmath $\!#1\!$ \unboldmath}}

\definecolor{khpink}{rgb}{0.858, 0.188, 0.478}
\newcommand{\KH}[1]{\textcolor{khpink}{#1}}
\definecolor{dscolor}{rgb}{0.13,0.67,0.8}

\definecolor{mrcolor}{rgb}{0.0,0.5,0.0}

\definecolor{recolor}{rgb}{1.0,0.44,0.37}

\definecolor{pfcolor}{rgb}{0.54,0.17,0.89}

\newcommand{\HST}{\emph{HST}}
\newcommand{\JWST}{\emph{JWST}}

\newcommand{\Spitzer}{\emph{Spitzer}}

\newcommand{\beagle}{\textsc{beagle}}
\newcommand{\cloudy}{\textsc{cloudy}}

\newcommand{\multinest}{\textsc{MultiNest}}
\newcommand{\eazy}{\textsc{eazy}}
\newcommand{\fast}{\textsc{fast}}

\renewcommand{\t}{\hbox{$t$}}
\newcommand{\M}{\hbox{$\txn{M}$}}
\newcommand{\Mstar}{\hbox{$\M_{\star}$}}

\newcommand{\Mgas}{\hbox{$\M_\txn{gas}$}}
\newcommand{\Msun}{\hbox{$\M_{\sun}$}}
\newcommand{\scM}{\textsc{m}}
\newcommand{\scMstar}{\textsc{m$_{\star}$}}
\newcommand{\LMstarMsun}{\hbox{$\log(\M/\Msun)$}}
\newcommand{\MStarMuv}{\hbox{\Muv--\Mstar}}

\newcommand{\Logten}{\hbox{$\log_{10}$}}
\newcommand{\OverMuvfuncMstarZ}{\hbox{$\overMuv(\Mstar,z)$}}

\newcommand{\MpcCube}{\hbox{$\txn{Mpc}^{3}$}}

\newcommand{\yr}{\hbox{$\txn{yr}$}}
\newcommand{\yrInv}{\hbox{$\txn{yr}^{-1}$}}
\newcommand{\MsunyrInv}{\hbox{$\Msun\,\yrInv$}}

\newcommand{\Reffmaj}{\hbox{$\txn{R}_\txn{eff,maj}$}}
\newcommand{\Reffcirc}{\hbox{$\txn{R}_\txn{eff,circ}$}}
\newcommand{\Rmock}{\hbox{$\txn{R}_\txn{mock}$}}

\renewcommand{\d}{\hbox{$d$}}
\renewcommand{\t}{\hbox{$t$}}
\newcommand{\tprime}{\hbox{$\t^{\prime}$}}

\newcommand{\zformm}{\hbox{$z_{\scriptscriptstyle \txn{form}}^{\scriptscriptstyle \txn{max}}$}}
\newcommand{\tausfr}{\hbox{$\tau_\textsc{sfr}$}}

\newcommand{\SigmaGas}{\hbox{$\Sigma_\txn{gas}$}}
\newcommand{\SigmaSfr}{\hbox{$\Sigma_{\psi}$}}
\newcommand{\sfr}{\hbox{${\psi}$}}
\newcommand{\ssfr}{\hbox{${\psi_\textsc{s}}$}}

\newcommand{\logOH}{\hbox{$12 + \log (\txn{O}/\txn{H})$}}
\newcommand{\Z}{\hbox{$\txn{Z}$}}
\newcommand{\Zism}{\hbox{$\Z_\textsc{ism}$}}
\newcommand{\Zbarism}{\hbox{$\bar{\Z}_\textsc{ism}$}}
\newcommand{\Zsun}{\hbox{$\Z_\sun$}}

\newcommand{\overMuv}{\hbox{$\overbar{M}_\textsc{uv}$}}
\newcommand{\Muv}{\hbox{$M_\textsc{uv}$}}

\newcommand{\Betauv}{\hbox{$\beta$}}
\newcommand{\MuvBeta}{\hbox{\Betauv--\Muv}}

\newcommand{\thetab}{\hbox{$\bmath{\Theta}$}}
\newcommand{\Db}{\hbox{$\mathbf{D}$}}

\newcommand{\HII}{\mbox{H\,{\sc ii}}}
\newcommand{\nH}{\hbox{$n_\textsc{h}$}}

\newcommand{\logUs}{\hbox{$\log U_\textsc{s}$}}
\newcommand{\xid}{\hbox{$\xi_\txn{d}$}}

\newcommand{\tauVAA}{\hbox{$\langle \hat{\tau}_\textsc{v} \rangle_i$}}
\newcommand{\tauVperpP}{\hbox{$\hat{\tau}_{\textsc{v}\,\perp}^\prime$}}
\newcommand{\tauVperp}{\hbox{$\hat{\tau}_{\textsc{v}\,\perp}$}}
\newcommand{\tauV}{\hbox{$\hat{\tau}_\textsc{v}$}}
\newcommand{\mud}{\hbox{$\mu$}}

\newcommand{\MFMstarOne}{\hbox{$\M^*_{1,\scM}$}}
\newcommand{\MFMstarTwo}{\hbox{$\M^*_{2,\scM}$}}
\newcommand{\MFMstar}{\hbox{$\M^*_{\scM}$}}
\newcommand{\phiStarOne}{\hbox{$\phi^*_{1,\scM}$}}
\newcommand{\phiStarTwo}{\hbox{$\phi^*_{2,\scM}$}}

\newcommand{\alphaOne}{\hbox{$\alpha_{1,\scM}$}}
\newcommand{\alphaTwo}{\hbox{$\alpha_{2,\scM}$}}

\newcommand{\LFMstar}{\hbox{$\M^*_{\sc{UV}}$}}
\newcommand{\prob}{\hbox{$\txn{P}$}}
\newcommand{\conditional}[2]{\hbox{$\txn{P}(#1 \mid #2)$}}
\newcommand{\range}[3]{\hbox{$#1 \sim #2 \,\txn{--}\, #3$}}

\newcommand{\OIIIHb}{$[\txn{O}\textsc{iii}]\lambda 5007/\txn{H}\beta$}
\newcommand{\NIIHa}{$[\txn{N}\textsc{ii}]\lambda 6584/\txn{H}\alpha$}

\newcolumntype{L}[1]{>{\raggedright\let\newline\\\arraybackslash\hspace{0pt}}m{#1}}
\newcolumntype{C}[1]{>{\centering\let\newline\\\arraybackslash\hspace{0pt}}m{#1}}
\newcolumntype{R}[1]{>{\raggedleft\let\newline\\\arraybackslash\hspace{0pt}}m{#1}}

\shorttitle{JWST Mock Galaxy Catalog}
\shortauthors{Williams et al.}

\begin{document}

\title{The JWST 
Extragalactic Mock Catalog: Modeling galaxy populations from the UV through the near-IR over thirteen billion years of cosmic history}

\author[0000-0003-2919-7495]{Christina C. Williams}
\altaffiliation{NSF Fellow}\affiliation{Steward Observatory, University of Arizona, 933 North Cherry Avenue, Tucson, AZ 85721, USA}
\author{Emma Curtis-Lake} \affiliation{Sorbonne Universit\'es, UPMC-CNRS, UMR7095, Institut d'Astrophysique de Paris, F-75014, Paris, France}
\author[0000-0003-4565-8239]{Kevin N. Hainline} \affiliation{Steward Observatory, University of Arizona, 933 North Cherry Avenue, Tucson, AZ 85721, USA}
\author{Jacopo Chevallard} \affiliation{Sorbonne Universit\'es, UPMC-CNRS, UMR7095, Institut d'Astrophysique de Paris, F-75014, Paris, France}

\author[0000-0002-4271-0364]{Brant E. Robertson} \affiliation{Department of Astronomy and Astrophysics, University of California, Santa Cruz, 1156 High Street, Santa Cruz, CA 95064, USA}
\author{Stephane Charlot} \affiliation{Sorbonne Universit\'es, UPMC-CNRS, UMR7095, Institut d'Astrophysique de Paris, F-75014, Paris, France}
\author{Ryan Endsley} \affiliation{Steward Observatory, University of Arizona, 933 North Cherry Avenue, Tucson, AZ 85721, USA}
\author{Daniel P. Stark} \affiliation{Steward Observatory, University of Arizona, 933 North Cherry Avenue, Tucson, AZ 85721, USA}

\author[0000-0001-9262-9997]{Christopher N. A. Willmer}\affiliation{Steward Observatory, University of Arizona, 933 North Cherry Avenue, Tucson, AZ 85721, USA}
\author{Stacey Alberts}\affiliation{Steward Observatory, University of Arizona, 933 North Cherry Avenue, Tucson, AZ 85721, USA}
\author{Ricardo Amorin}\affiliation{Cavendish Laboratory, University of Cambridge, 19 J. J. Thomson Ave., Cambridge CB3 0HE, UK} \affiliation{Kavli Institute for Cosmology, University of Cambridge, Madingley Road, Cambridge CB3 0HA, UK }
\author{Santiago Arribas} \affiliation{Departamento de Astrofisica, Centro de Astrobiologia, CSIC-INTA, Cra. de Ajalvir, 28850-Madrid, Spain}
\author{Stefi Baum} \affiliation{University of Manitoba, Dept. of Physics and Astronomy, Winnipeg, MB R3T 2N2, Canada}
\author{Andrew Bunker} \affiliation{Department of Physics, University of Oxford, Oxford, UK}
\author{Stefano Carniani} \affiliation{Cavendish Laboratory, University of Cambridge, 19 J. J. Thomson Ave., Cambridge CB3 0HE, UK} \affiliation{Kavli Institute for Cosmology, University of Cambridge, Madingley Road, Cambridge CB3 0HA, UK }
\author{Sara Crandall} \affiliation{Department of Astronomy and Astrophysics, University of California, Santa Cruz, 1156 High Street, Santa Cruz, CA 95064, USA}
\author{Eiichi Egami} \affiliation{Steward Observatory, University of Arizona, 933 North Cherry Avenue, Tucson, AZ 85721, USA}
\author{Daniel J. Eisenstein} \affiliation{Harvard-Smithsonian Center for Astrophysics 60 Garden St., Cambridge, MA 02138}
\KH{\author{Pierre Ferruit} \affiliation{Scientific Support Office, Directorate of Science, ESA/ESTEC, Keplerlaan 1, 2201AZ Noordwijk, The Netherlands}}
\author{Bernd Husemann} \affiliation{Max Planck Institute for Astronomy, Konigstuhl 17, D-69117 Heidelberg, Germany}
\author{Michael V. Maseda} \affiliation{Leiden Observatory, Leiden University, PO Box 9513, 2300 RA, Leiden, The Netherlands}
\author{Roberto Maiolino} \affiliation{Cavendish Laboratory, University of Cambridge, 19 J. J. Thomson Ave., Cambridge CB3 0HE, UK} \affiliation{Kavli Institute for Cosmology, University of Cambridge, Madingley Road, Cambridge CB3 0HA, UK }
\author{Timothy D. Rawle} \affiliation{European Space Agency, c/o STScI, 3700 San Martin Drive, Baltimore, MD 21218, USA}
\author{Marcia Rieke}\affiliation{Steward Observatory, University of Arizona, 933 North Cherry Avenue, Tucson, AZ 85721, USA}
\author{Renske Smit} \affiliation{Cavendish Laboratory, University of Cambridge, 19 J. J. Thomson Ave., Cambridge CB3 0HE, UK} \affiliation{Kavli Institute for Cosmology, University of Cambridge, Madingley Road, Cambridge CB3 0HA, UK }
\author{Sandro Tacchella} \affiliation{Harvard-Smithsonian Center for Astrophysics 60 Garden St., Cambridge, MA 02138}
\author{Chris J. Willott} \affiliation{NRC Herzberg, 5071 West Saanich Rd, Victoria, BC V9E 2E7, Canada}

\email{ccwilliams@email.arizona.edu, curtis@iap.fr}

\begin{abstract}

We present an original phenomenological model to describe the evolution of galaxy number counts, morphologies, and spectral energy distributions 
across a wide range of redshifts ($0.2<z<15$) and stellar masses [$\LMstarMsun\ge6$]. Our model follows observed mass and luminosity functions of both star-forming and quiescent galaxies, and 
reproduces the redshift evolution of colors, sizes, star-formation and chemical properties of the observed galaxy population. 
Unlike other existing approaches, our model includes a self-consistent treatment of stellar and photoionized gas emission and dust attenuation based on the \beagle\ tool. The mock galaxy catalogs 
generated with our new model can be used to simulate and optimize extragalactic surveys with future facilities such as the James Webb Space Telescope (\JWST), and 
to enable critical assessments of analysis procedures, interpretation tools, and measurement systematics for both photometric and spectroscopic data. As a first application of this work, we make predictions for the upcoming JWST Advanced Deep Extragalactic Survey (JADES), a joint program of the \JWST/NIRCam and NIRSpec Guaranteed Time Observations 
teams. We show that JADES will detect, with NIRCam imaging, thousands of galaxies at $z\gtrsim6$, and tens at $z\gtrsim10$ at m$_{AB}\lesssim$30 (5$\sigma$) within the $236$ $\txn{arcmin}^2$ of the survey. The JADES data will enable 
accurate constraints on the evolution of the UV luminosity function at $z>8$, and resolve the current debate about the rate of evolution of galaxies at $z\gtrsim8$.  Ready to use mock catalogs and software to generate new realizations are publicly available as the JAdes extraGalactic Ultradeep Artificial Realizations (JAGUAR) package.

\end{abstract}


\section{Introduction}

Over the last two decades, deep extragalactic surveys with the {\it Hubble} ({\it HST}) and {\it Spitzer} Space Telescopes have revolutionized our understanding of galaxy evolution.  These surveys measured the buildup of galaxy populations from the local Universe to the current redshift frontier at $z\sim10$ \citep[for a review, see, e.g.][]{Stark2016}.   Meanwhile, 
ground-based 8m- and 10m-class telescopes have characterized the physical conditions of galaxies even  beyond $z\sim2-3$,  the peak in the cosmic star formation rate density 
(e.g. with Keck/MOSFIRE; \citealt{Kriek2015}, \citealt{Steidel2014}). Currently, further progress 
is hindered by the limited wavelength coverage of \HST, relatively low sensitivity of \Spitzer, and the atmospheric limitations that impede ground-based campaigns. 
However, the soon-to-launch  {\it James Webb Space Telescope} \citep[{\it JWST};][]{Gardner2006}  will detect galaxies well beyond the current redshift frontier, 
below the magnitude and stellar mass limits currently achievable with existing facilities, while its high spatial resolution will image early galaxies in exquisite detail. Furthermore, the unprecedented spectroscopic capabilities of {\it JWST} will enable spectroscopic observations of even the faintest galaxies detected with {\it HST} to date \citep[e.g.][]{Chevallard2017}

This innovative telescope, hosting the largest mirror ever to fly in space 
and a suite of state-of-the-art near-infrared instruments, will provide unique data to answer key open questions about the formation and evolution of galaxies. 
Specifically, the wavelength coverage provides the opportunity,
for the first time, to study the rest-frame optical properties of galaxies out to $z\sim 9$, and the rest-frame UV out to $z > 10$. 
Observations with {\it JWST} will enable precise constraints on the evolution of the stellar and chemical make up of galaxies, dust attenuation, and ionization sources across a broad range of redshift, stellar mass and luminosity \citep[e.g.][]{Mannucci2010, Reddy2015, Strom2017, Shapley2017}. These data are fundamental for understanding the formation of the Hubble sequence, the emergence of quiescent galaxies, and the variety of observed scaling relations between galaxy properties 
\citep[e.g.][]{Faber1976, Tully1977,Kauffmann2003,Tremonti2004,Franx2008,Maiolino2008,Speagle2014, vanderWel2014, Glazebrook2017}. 
In addition,  {\it JWST} will be used to target the exact epoch and sources of cosmic reionization at high redshift \citep[e.g.][]{Bunker2004, Finkelstein2012reion,Robertson2015, Stark2016}.  
Studies that address these topics will require large survey campaigns
using multiple instruments on board {\it JWST} including, the Near Infrared Camera \citep[NIRCam;][]{HornerRieke2004} 
and the Near Infrared Spectrograph, \citep[NIRSpec;][]{Bagnasco2007,Birkmann2016}. These sensitive instruments will provide new space-based observation modes including parallel imaging and spectroscopic observations, simultaneous imaging enabled by the dichroic on NIRCam, as well as the choice of fixed slit, high-multiplex or integral field spectroscopy on NIRSpec.

Maximizing the scientific return of the innovative and complex instruments on board {\it JWST} will require the development of original analysis tools and space-based observing strategies. As an example, the advent of space-based multi-object spectroscopy (with the NIRSpec Micro-Shutter Array; MSA) initiates an era where spectroscopic follow up of {\it JWST}-selected targets will demand the rapid analysis of imaging data to create slit-mask designs.  
Meeting these future challenges requires physically-motivated simulations of {\it JWST} data that should ideally match existing observations, while also extending to the unprecedented depths and redshifts that will be attained by \JWST.
 Such simulations enable critical tests of analysis procedures and processing tools, and aid the scientific interpretation by identifying potential observational biases  on measured galaxy properties \citep[e.g. galaxy sizes or UV continuum slope $\beta$;][]{Dunlop2012,Finkelstein2012beta,Rogers2013,CurtisLake2016,Bouwens2017size}.

Physically-motivated {\it JWST} simulations will require  
mock galaxy catalogs, which can be built using semi-analytic galaxy formation models \citep[e.g.][]{Blaizot2005,Cai2009,Bernyk2016,Mirocha2017,Furlanetto2017} or hydrodynamical simulations \citep[e.g.][]{Torrey2015, McAlpine2016}. However, such sophisticated approaches \citep[e.g.][]{Croton2006,Benson2012,Vogelsberger2014, Schaye2015} are intrinsically model-dependent.   As an example, semi-analytical models that match low-to-intermediate redshift stellar mass functions may provide widely different predictions for low-mass galaxies  [$\LMstarMsun \lesssim 8$] and at high redshifts \citep[$z\gtrsim4$, e.g.][]{Lu2014}, or underpredict the specific star-formation rates (sSFR) of sub-$L^\ast$ galaxies \citep[e.g.][]{SomervilleDave2015, Fontanot2009,Weinmann2012, Somerville2015}.
In an effort to reduce the model-dependency of mock  observation tools,  empirically driven approaches have been developed based on observed galaxy distributions and relations among physical quantities that replicate deep extragalactic surveys as observed from current facilities \citep[e.g.][]{Schreiber2017}.

As we look forward to future facilities that extend beyond current limitations, we must incorporate accurate descriptions of the spectral energy distributions (SEDs) of young, low-mass and high sSFR galaxies across cosmic time.
These populations are of particular 
importance both as low-redshift interlopers, as well as the high-redshift galaxies which are the prime science targets for {\it JWST}, 
and are now known to produce strong nebular emission lines that can contribute significant excesses to broad-band photometric fluxes 
\citep{SchaererDeBarros2009,Shim2011,Atek2011,Labbe2013,Stark2013, Schenker2013, Smit2014, Smit2015, RobertsBorsani2016, Rasappu2016}. Thus 
the treatment of nebular emission in mock catalogs tailored to reproducing high-redshift galaxies is especially important. 
Currently,  the treatment of nebular emission in mock catalogs based on galaxy formation models is often approximated in post-processing with subgrid prescriptions  
 \citep[e.g.][]{SomervilleDave2015, NaabOstriker2017}, although more advanced ones have been recently proposed based on simplified prescriptions for the dependence of line emission on metallicity, ISM conditions or ionization parameter \citep[e.g.][]{Kewley2013, Orsi2014, Shimizu2016}. A fully self-consistent treatment of stellar and nebular emission in hydrodynamical simulations is, however, still limited to small numbers of objects rather than full cosmological simulations \citep{Hirschmann2017}.

With this work, we present a new phenomenological model for the cosmic galaxy population designed to benefit future surveys
with {\it JWST} and other forthcoming facilities targeting the UV to near-infrared emission of galaxies.
Our model is designed to reproduce observations of galaxy properties from $0 < z < 10$, and enables extrapolations of galaxy distributions to $z\sim$15, allowing for the generation of mock catalogs that include physically-motivated counts, luminosities, stellar masses, morphologies, photometry and spectroscopic properties down to arbitrarily low stellar mass. 
Importantly, we incorporate a self-consistent modeling of stellar and nebular emission using the models of \cite{Gutkin2016} teamed with the \beagle\ tool \citep{Chevallard2016}, which enables the inclusion of strong nebular emission lines and nebular continuum emission in mock galaxy spectra and photometric SED.
These models cover the wide parameter space required to model the range of physical conditions  expected in local and extremely high redshift galaxies ($z>10$) without resorting to simple prescriptions of emission line ratios.

Simulations using our model have already proven invaluable to optimize the design of a large ($\sim720$ hours) observational program, 
the \JWST\ Advanced Deep Extragalactic Survey (JADES), 
a joint program of the NIRCam and NIRSpec Guaranteed Time Observations (GTO) teams.
 In particular, mock catalogs produced using our model have been used to optimize the selection of photometric filters and spectral dispersers, the depth of the observations and area covered. This mock catalog tool, called JAdes extraGalactic Ultradeep Artificial Realizations (JAGUAR), and related \JWST\ simulations will also provide a fundamental aid for the scientific interpretation of future \JWST\ data,  and has enabled us to make realistic science predictions for the future GTO survey.

The outline of this paper is as follows. In Section \ref{methods}, we provide a conceptual overview of our procedure for producing mock galaxies and assigning their properties. In the subsequent sections, we describe the phenomenological model that underlies JAGUAR quantitatively. In Sections \ref{SFG_MF} and \ref{QG_MF}, we describe the procedure for producing star-forming and quiescent galaxies (respectively) across cosmic time, including their masses, redshifts, luminosities and SED  properties. In Section \ref{morphs} we describe the procedure for assigning morphological parameters to both star-forming and quiescent galaxies. In Section \ref{characterization}, we characterize a realization of our model (a JAGUAR mock catalog) by presenting comparisons to measurements made from current surveys between $0<z<10$.
In Section \ref{results}, we present our predictions for the science results of JADES that are enabled by this tool. Finally, in Section \ref{summary} we summarize this work. We release 
ready-to-use realizations
\footnote{http://fenrir.as.arizona.edu/jaguar} as described below, as well as a \textsc{Python} package for JAGUAR that can be used to generate catalogs to any area or depth.  
Throughout this work we assume a $\Lambda$CDM cosmology with H$_0$=70 km s$^{-1}$ Mpc$^{-1}$, $\Omega_M$ = 0.3, $\Omega_\Lambda$ = 0.7. When necessary, we assume
a \citet{Chabrier2003} stellar initial mass function (IMF).

\section{Methods Overview}\label{methods}

\begin{figure*}
\begin{center}
\includegraphics[width=0.7\textwidth]{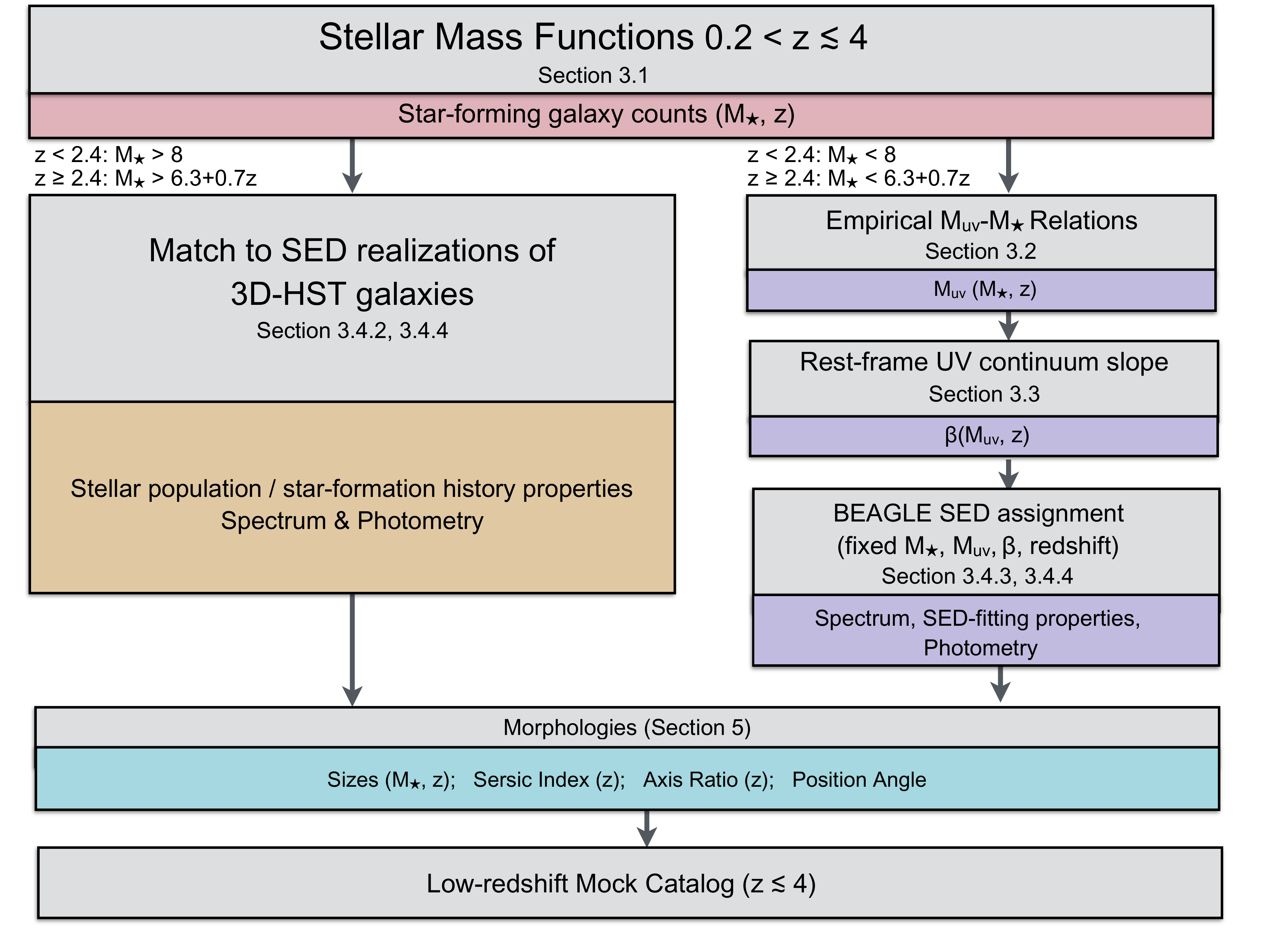}
\caption{Diagram summarizing the procedures for generating star-forming galaxies at $z \lesssim 4$. \Mstar\ is defined as \LMstarMsun. High-mass galaxies (defined as $\Mstar > 8$ for $z<2.4$ and $\Mstar > 6.3+0.7z$ for $z\geq2.4$; illustrated by the left pathway) and low-mass galaxies (right pathway) are generated differently as indicated. These criteria are defined in Section \ref{section:matching_sf_mock_to_parent_cat} as the approximate mass completeness limits in the 3D-HST catalog, which we use to assign real galaxy SEDs to high-mass mock galaxies. 
Gray boxes indicate the empirical relationships, distributions, or data on which mock galaxy properties are based, and colored boxes indicate the mock galaxy property which is generated in that step. Quiescent galaxies are generated at $z<4$ following a different procedure which is described in Section~\ref{QG_MF} and illustrated in Figure~\ref{flowchart_qg}.   }
\end{center}
\label{flowchart_zlt4}
\end{figure*}

\begin{figure*}
\begin{center}
\includegraphics[width=0.7\textwidth]{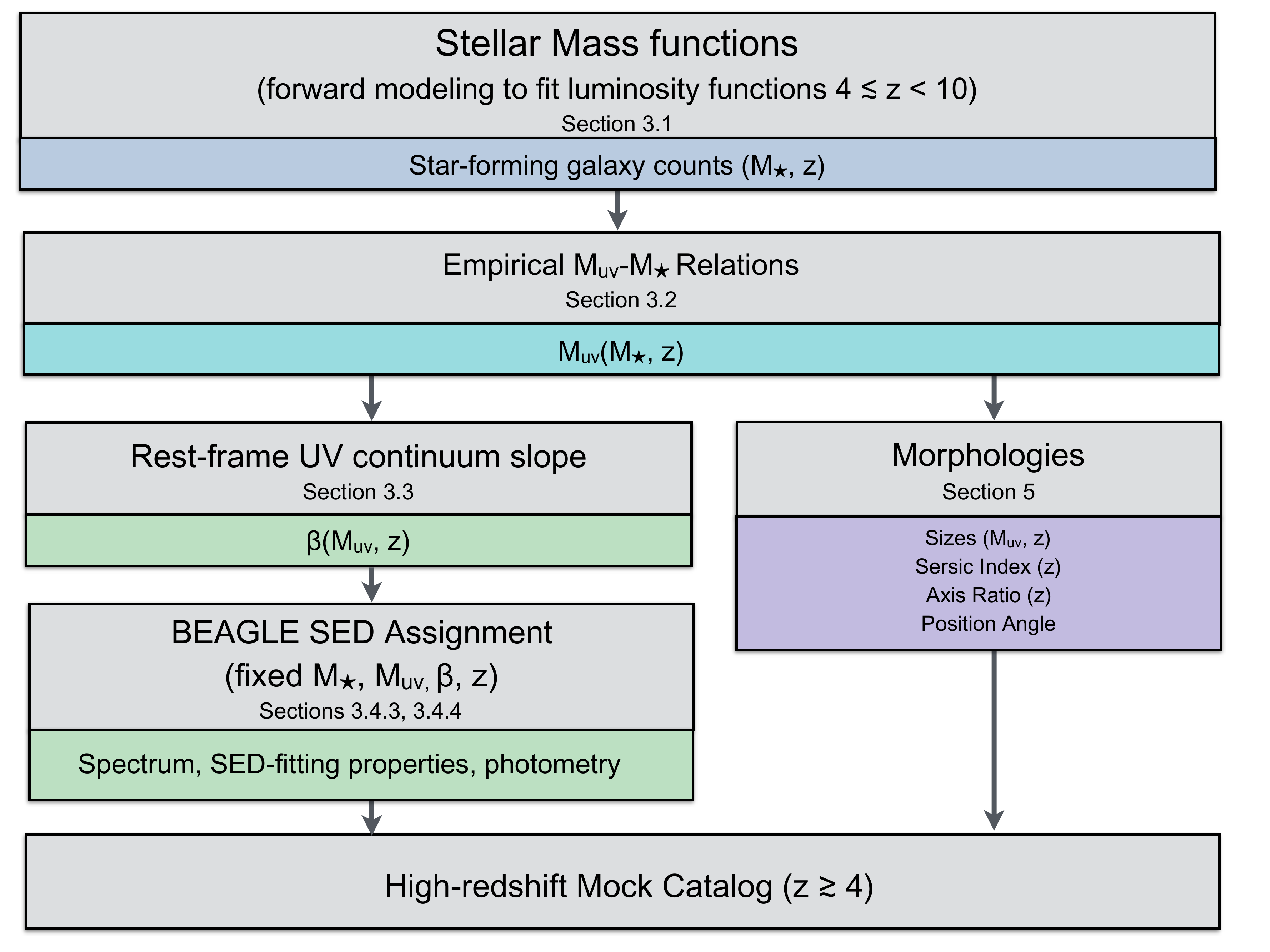}
\caption{Diagram summarizing the procedures for generating the star-forming galaxies at $z\gtrsim4$. \Mstar\ is defined as \LMstarMsun. Gray boxes indicate the empirical relationships, distributions, or data on which mock galaxy properties are based, and colored boxes indicate the mock galaxy property which is generated in that step. All star-forming mock galaxies at $z>4$ are generated following these procedures. Quiescent galaxies are generated at $z>4$ following a different procedure which is described in Section~\ref{QG_MF} and illustrated in Figure~\ref{flowchart_qg}. }
\label{flowchart_zgt4}
\end{center}
\end{figure*}

The foundation of our model consists of observed stellar mass and UV luminosity functions that have been measured from $0<z<10$. 
We use these observations to model 
the evolution of stellar mass functions for both star-forming and quiescent galaxies, which are then used to generate each mock population at all redshifts.\footnote{We note that we do not attempt to include galaxies composed of metal-free, `Pop\textsc{iii}' stars, since no empirical constraints exist on such objects.} 
We assign integrated properties such as the UV absolute magnitude \Muv\ and UV continuum slope $\beta$ (where $f_{\lambda}\propto \lambda^{\beta}$; for star-forming galaxies only), and structural properties based entirely on empirical relations or distributions. 
Finally, the model assigns spectra that are consistent with these integrated properties to each mock galaxy, which we use to produce the broadband photometry. 
Summaries of our overall procedure for star-forming galaxies are shown in Figures~\ref{flowchart_zlt4} and \ref{flowchart_zgt4}, which indicate the sections that describe the relevant quantitative procedures for assigning various properties to mock galaxies.

\subsection{Generating galaxy counts}
\label{galaxy_counts_general}

Here we describe the procedure we follow to generate galaxy number counts (i.e. the expected number of galaxies of a given mass, at fixed redshift and on-sky area). 
We first model the evolution of stellar mass functions across cosmic time using 
 continuously-evolving Schechter functions for both star-forming and quiescent galaxies. 
We then generate the expected number of star-forming or quiescent galaxies for a given redshift bin over a given survey area by integrating their respective model mass function, multiplying by the co-moving volume and drawing from a Poisson distribution with this mean.  By computing the cumulative distribution function (CDF) of the stellar mass function we can then effectively draw this number of galaxies from the mass function using inverse transform sampling.

For star-forming galaxies 
at $z\gtrsim 4$, stellar mass function measurements become increasingly difficult and uncertain. 
With current facilities, this epoch represents a transition to rest-frame UV selections tracing young stars (with {\it HST}) instead of rest-frame optical selections that trace stellar mass (which would require {\it Spitzer}/IRAC whose sensitivity is lower). 
At $z\gtrsim 4$ the UV luminosity function becomes much more easily measurable than the stellar mass function with current facilities. 
Therefore, at redshifts $z\gtrsim4$ 
 we rely on observed UV luminosity functions to constrain the number densities of star forming galaxies.
(Quiescent galaxies at $z>4$ are instead based on an informed extrapolation, which is discussed in Section~\ref{QG_MF}).

While generating galaxy counts from a stellar mass function is straightforward, using a UV luminosity function requires 
modeling a theoretical or an empirical relation linking a galaxy's UV luminosity, or \Muv,  to its stellar mass, hereafter $\Mstar = \LMstarMsun$.  
The observed connection between UV luminosity and stellar mass is not a simple monotonic relationship; 
galaxies exhibit a range of UV luminosities at fixed stellar mass that likely depends on other galaxy properties including stellar population age and metallicity, dust, and gas content.
We will describe this distribution in terms of a Gaussian scatter about an average \MStarMuv\ relation, where the standard deviation of \Muv\ at fixed \Mstar\ is given by $\sigma_{uv}$ (assumed independent of \Mstar).
We can then express the probability of a galaxy of stellar mass \Mstar\ to have a given \Muv\ as
\begin{equation}
\frac{d \prob}{d\Muv}(\Mstar,z) = \mathcal{N}[\Muv,\OverMuvfuncMstarZ,\sigma_{uv}]
\end{equation}
\noindent
where the mean relationship between UV luminosity at a given stellar mass and redshift is
\begin{equation}
\bar{M}_{UV}(\Mstar,z)\equiv \int \Muv \frac{d\prob}{d\Muv}(\Mstar,z)\,d\Muv\, .
\end{equation}

Once such a relation for \MStarMuv\ and its scatter has been adopted 
(see Section~\ref{MstarMuv}), the observed UV luminosity function $\Phi(\Muv, z)$  can be modeled as the convolution of the stellar mass function $\Phi(\Mstar,z)$ 
with the distribution of \Muv\:
\begin{eqnarray}\label{eq:MFconvMstarMUV}
\Phi(\Muv, z) = \nonumber\\
\int_{0}^{\infty}\Phi(\Mstar,z)&\,\mathcal{N}&[\Muv,\overMuv(\Mstar,z),\sigma_{uv}]\,d\Mstar \, ,
\end{eqnarray}
\noindent
where $\Phi(\Muv,z)$ represents the number of galaxies per co-moving volume with UV absolute magnitude \Muv\  
as a function of redshift. Therefore, to calculate star-forming galaxy counts at $z\gtrsim4$ where we have the best constraints from the UV luminosity function, we forward model the continuously evolving stellar mass function, convolved with an empirical characterization of $\mathcal{N}[\Muv,\OverMuvfuncMstarZ,\sigma_{uv}]$ 
in order to fit with observed UV luminosity functions over the range $4\lesssim z\lesssim 10$. \footnote{Uncertainties in  stellar masses complicate measurements of the stellar mass function, because the intrinsic stellar mass function must be convolved with the uncertainties in the stellar mass estimates.  
However, for this work we model the case where the intrinsic stellar masses are known perfectly.}

This procedure enables us to produce one continuously evolving stellar mass function that, when sampled randomly as outlined above, produces star-forming galaxy counts that follow observed stellar mass functions at $z\lesssim4$, the \MStarMuv\ distribution given by $\mathcal{N}[\Muv,\OverMuvfuncMstarZ,\sigma_{uv}]$, and the observed UV luminosity functions at $z\gtrsim4$. 
We will describe the characterization of the empirical \MStarMuv\ distribution,  $\mathcal{N}[\Muv,\OverMuvfuncMstarZ,\sigma_{uv}]$, 
that we use to forward model the stellar mass function in Section~\ref{MstarMuv}. In Section~\ref{evolvemassfn} we will describe our procedure to fit the observed evolving stellar mass function over $0.2<z\lesssim4$, 
and forward model by convolving the mass function with $\mathcal{N}[\Muv,\OverMuvfuncMstarZ,\sigma_{uv}]$ 
at $z\gtrsim4$, to produce galaxy counts.

\subsection{Generating integrated galaxy properties}

For each object generated in the mock galaxy population,
we use redshift and stellar mass to assign other integrated galaxy properties 
including UV absolute magnitude \Muv\ and continuum slope \Betauv\  for star-forming galaxies, as well as type-dependent structural parameters.  
To assign the integrated properties we use empirical relations, plus appropriate scatter, to generate smoothly redshift-evolving distributions of 
\MStarMuv\ (Section~\ref{MstarMuv}), \MuvBeta\ (Section~\ref{betamuv}), size-mass ($z<4$) and size-UV luminosity (at $z>4$; see Section~\ref{morphs}).
Figures~\ref{flowchart_zlt4} and \ref{flowchart_zgt4} provide more
details on this procedure.
The integrated properties inform the assignment of a fully consistent SED to the mock galaxies, from which we derive {\it JWST}, {\it HST} and {\it Spitzer} filter photometry.  These SEDs are created using \beagle\ and span a range of physical properties as described in the following section.

\subsection{Modeling galaxy SEDs with the \beagle\ tool}
\label{BEAGLE_general}

\beagle\ (\citealt{Chevallard2016}; C16) is a new-generation tool for the modeling and interpretation of spectro-photometric galaxy SEDs based on a
self-consistent approach to describe stellar emission and its transfer through the interstellar (ISM) and intergalactic (IGM) media. 
In this Section, we first describe the general characteristics of \beagle\ and the models integrated therein, and then summarize our two methods
for assigning SEDs to mock galaxies according to whether or not the realized properties overlap with those of observed galaxies from current surveys.

In \beagle, the emission from simple stellar populations of different ages, \tprime\, and metallicities, \Z\ (the mass fraction of all elements heavier than Helium), is described by the latest version of the
\citet{Bruzual2003} population synthesis code. 
Stellar emission is computed using the MILES stellar library \citep{Sanchez2006}
and includes new prescriptions for the evolution of massive stars \citep{Bressan2012, Chen2015} and their spectra \citep{Hamann2004, Leitherer2010}. 
We account for the (continuum+line) emission of gas photoionized by young stars by considering the large grid of photoionization models of \citet{Gutkin2016}.
These are based on the standard photoionization code \cloudy\ \citep[version 13.3;][]{Ferland2013} and assume `ionization bounded' nebulae, i.e. a zero escape fraction of H-ionizing photons. The models are described in terms of `effective', i.e. galaxy-wide parameters following the prescription of \citet{Charlot2001}. 
Adjustable model parameters include the ionization parameter \logUs, which sets the ratio of H-ionizing photons to H atoms at the edge of the Str\"omgren sphere, the interstellar metallicity \Zism, and the dust-to-metal (mass) ratio \xid, which traces metal depletion onto dust grains.
Since the gas density \nH\ and depletion factor \xid\ do not significantly affect emission line ratios at sub-solar metallicities \citep[see figure~3 and 5 of][]{Gutkin2016}, and most of our galaxies exhibit $\log(\Z/\Zsun)\lesssim -0.5$ (see Fig~\ref{fig:physicalParametersMock}), we fix $\nH = 10^2 \, \txn{cm}^{-3}$, the typical value measured in \range{z}{2}{3} galaxies \citep[e.g.][]{Sanders2016, Strom2017}, and $\xid=0.3$, a value similar to what measured in the Solar neighborhood (although, see Section~\ref{BPT_comparison}).  
We account for attenuation by dust of the emission from stars and photoionized gas
using the two-component model of \citet{Charlot2000},
parameterized in terms of the total attenuation optical depth \tauV, and the fraction of this arising in the diffuse ISM $\mu$.  
The mean effects of intergalactic medium absorption are
included following the model of \citet{Inoue2014}.

For mock galaxies with properties that are observable using current facilities  
we use \beagle\ to generate a distribution of model SEDs consistent
with the observations and assign these SEDs to the mock objects. 
To achieve this, we 
fit SED models from \beagle\ to the multi-band photometry of galaxies
in two CANDELS fields using the 3D-HST catalog \cite{Skelton2014}.
When performing parameter estimation, 
\beagle\ employs the nested sampling algorithm \citep{Skilling2006} as implemented in \multinest\ \citep{Feroz2009}. 
This procedure creates a range of statistically acceptable SED fits 
for each observed galaxy in a subset of the 3D-HST sources (see Sections~\ref{highmassgal} and \ref{Q_SEDs}, while for more detail of the \beagle\ output see C16, Section~3.3) which are then used to produce a parent catalog. 
This parent catalog is used to assign SEDs to mock objects
with high stellar mass  (i.e. those with mass above $\LMstarMsun > 8$, or, above the mass-completeness of the 3D-HST catalog if larger in that redshift bin) and low redshift ($z<4$), where the $\lambda\lesssim4.5\mu \textrm{m}$ 
photometry provides firm constraints on stellar mass. 
The SEDs are assigned by finding the closest match in stellar mass and redshift for each mock galaxy within the parent catalog, allowing us 
to encapsulate the observed diversity of galaxy SEDs at $z<4$ with
relatively few assumptions.

For mock galaxies with realized properties that extend beyond 
current measurements of real sources, we can leverage the
capabilities of \beagle\ to produce theoretical SEDs and generate
model spectra for the mock objects.
In this second method, we generate a parent catalog built of theoretical SEDs
covering a range of model parameters that can be matched to mock galaxy 
stellar mass, redshift, and, for star-forming galaxies,
 \Muv, and $\beta$ (see Sections~\ref{BEAGLE_SF_grid} and \ref{Q_SEDs}).
 We use this method at low stellar masses [$\LMstarMsun < 8$] where current galaxy survey sampling of the population is less complete, and at $z\ge4$ where SED coverage in the rest-frame optical is only available from imaging taken with IRAC, the $3.6-8\mu m$ camera on  {\it Spitzer} \citep{Fazio2004}.

\section{Generating Star-forming galaxies across cosmic time}\label{SFG_MF}

Here we describe the phenomenological model and quantitative procedure for generating counts, redshifts, stellar masses, luminosities,
and photometric and spectroscopic properties
for mock star-forming galaxies. Galaxies 
are assigned masses and redshifts according to evolving stellar mass functions, as described in Section~\ref{evolvemassfn}. 
In Sections~\ref{MstarMuv} and \ref{betamuv} we describe the procedure for assigning integrated star-forming galaxy properties (\Muv\ and $\beta$) based on empirical distributions. Finally, in Section~\ref{section:SF_SED_parent_catalog}, we describe the procedure for assigning SEDs to star-forming galaxies.

\subsection{Generating star-forming galaxy counts}\label{evolvemassfn}

In generating a mock galaxy catalog, we aim to reproduce measurements of the star-forming galaxy stellar mass functions at low redshift ($z\lesssim4$) and the UV luminosity function at high redshift ($z\gtrsim4$).  Our primary mass function constraints come from \citet[][hereafter T14]{Tomczak2014}, while our UV luminosity function constraints are adopted from \citet{Bouwens2015LF} at $4\lesssim z \lesssim 8$ and the newest $z\sim10$ estimate presented in \cite{Oesch2017}. 

T14 provide measurements of the stellar mass function of star-forming and quiescent galaxies in eight redshift bins in the range $0.2<z<3$.  They employed imaging data from the FourStar galaxy evolution (ZFOURGE) survey \citep{Straatman2016} covering the CDFS, COSMOS and UDS fields with 5 near-IR medium-bandwidth filters spanning the $J$ and $H$ bands, as well as broad-band $K_S$ imaging. Specifically they used the regions that also overlap with CANDELS $J_{125}$ and $H_{160}$ imaging (to $\sim26.5$ depth to 5$\sigma$), covering a total area of $\sim$316 arcmin$^2$.  Additionally, imaging from  NEWFIRM Medium-band Survey \citep{Whitaker2011} was used in the AEGIS and COSMOS fields, employing the same filter sets as the ZFOURGE survey to shallower depths but wider area to leverage better constraints of the high-mass end of the mass function.  Each of the fields also benefit from further imaging that allows comprehensive sampling of galaxy SEDs over the wavelength range $0.3-8\mu$m, with the field-specific filter-sets and imaging programs summarized in Section~2.4 of  \cite{Straatman2016}.

T14 inferred photometric redshifts and rest-frame colors (used to separate galaxies into star-forming or quiescent based on the $UVJ$ diagram of \citealt{Whitaker2011}) 
using the template-based \eazy\ code \citep{Brammer2008}, while stellar masses were estimated using \fast\ \citep{Kriek2009}. Within \fast, they used the original \citet{Bruzual2003} population synthesis code at fixed solar metallicity, employing a \citet{Chabrier2003} IMF, and a declining exponential star-formation history. The 80\% mass completeness limits of their sample increase from $\LMstarMsun\sim7.75$ at $z\sim0.5$ to $\LMstarMsun\sim9.25$ at $z\sim3$. T14 fit their resulting stellar mass functions with a sum of two \citet{Schechter1976} functions: 
\begin{equation}\label{eq:DoubleSchechter}
\begin{aligned}
\Phi(\Mstar)\d\Mstar = &\,\Phi_1(\Mstar)\d\Mstar + \Phi_2(\Mstar) \, \d\Mstar\\
=&\,\ln10\,\phiStarOne\,10^{(\scalebox{.7}{\Mstar-\MFMstarOne})(1+\scalebox{.7}{\alphaOne})}\textrm{exp}(-10^{\scalebox{.7}{\Mstar-\MFMstarOne}})\d\Mstar\\
&+\ln10\,\phiStarTwo\,10^{(\scalebox{.7}{\Mstar-\MFMstarTwo})(1+\scalebox{.7}{\alphaTwo})}\textrm{exp}(-10^{\scalebox{.7}{\Mstar-\MFMstarTwo}})\d\Mstar \, ,
\end{aligned}
\end{equation}
where $\Mstar=\LMstarMsun$, as defined in Section~\ref{galaxy_counts_general},  $\Phi(\Mstar)$ indicates the number of galaxies per \MpcCube\ with stellar masses between $\Mstar$ and $\Mstar+\d\Mstar$,  and \MFMstarOne, \MFMstarTwo, \phiStarOne, \phiStarTwo, \alphaOne and \alphaTwo\ are the six free parameters of the function.\footnote{Schechter function parameters used to describe a mass function are suffixed by an `\scM' to distinguish them from those used to describe a luminosity function.}  In a single Schechter function, $\M^*_{\scM}$ is the mass at the turnover, or ``knee'' of the mass function, $\phi^*_{\scM}$ is the characteristic number density of galaxies at the turnover, and $\alpha_{\scM}$ is the low-mass slope.  In the double-Schechter function used in T14, they explicitly set $\MFMstarOne=\MFMstarTwo = \MFMstar$  meaning that they fit with a single ``knee'' but the different normalizations and faint-end slopes of each  function enable them to fit the observed steepening of the mass function to low masses (see Figure~\ref{fig:zlt4MFevolution}).

At $z>4$ stellar masses become progressively less well constrained from measurements, in part because the rest-frame optical SED (a key region containing the Balmer break at $\sim 3600$ \AA, and the 4000 \AA\ break), shifts into the infrared where current facilities have low sensitivity. 
Additionally, high equivalent width (EW) emission lines can add to the flux in the reddest photometric bands, leading to an over-prediction of galaxy stellar masses (\citealt{Schaerer2010}, \citealt{Stark2013}, \citealt{CurtisLake2013}, \citealt{DeBarros2014}).
As a result, relative uncertainties on stellar mass measurements are high \citep[e.g. 0.4 dex at $10^{10} \Msun$ at $z = 4$, increasing with redshift and decreasing mass;][see also \citealt{Mobasher2015}]{Grazian2015} and may
contribute to the large scatter of mass function measurements in the literature \citep[nearly $\sim$1 dex in counts; see Figure~9 in][Figure~11 in \citealt{Davidzon2017}]{Song2016}. Therefore, 
to generate galaxy counts at $z > 4$ we leverage
the constraints provided by the observed UV luminosity function between $4 \lesssim z \lesssim 8$ from \citet[][]{Bouwens2015LF} with luminosity function measurements  
with mean redshifts at $<z>=[3.8, 4.9, 5.9, 6.8, 7.9]$  
 using data from the HST Legacy Fields, 
as well as the $z\sim10$ luminosity function of \cite{Oesch2017}. The binned UV luminosity function measurements we use for this work are overall consistent with many other results in the literature at \Muv$<-17$ \citep[e.g.][]{Mclure2013LF, Finkelstein2015, Atek2015, Laporte2015, Castellano2016, Yue2017, Livermore2017, Ono2017, Bouwens2017LF}.

We choose to model the redshift evolution of the six mass function parameters across the entire redshift range of the mock, i.e. $0.2<z<15$.  This ensures a smooth evolution in number counts across the transition from mass to luminosity function-based constraints. At $z<3.8$ (the mean redshift of the B-dropout sample used to produce the \cite{Bouwens2015LF} $z\sim4$ luminosity function) we use the measured mass functions of T14 to directly constrain their redshift evolution, while at $z\geq3.8$ we use our model of the redshift-evolving  \MStarMuv\ relation (see Section~\ref{MstarMuv}) to fit to the observed luminosity functions with mass function parameters.  However, it is important to note that this is not a direct prediction of the shape or evolution of the $z\gtrsim4$ mass functions that we expect to measure with {\it JWST}. Our $z\gtrsim4$ mass functions are dependent on our model of the \MStarMuv\ relation and additionally, we do not yet know how incomplete the current \Muv-selected samples at $z\gtrsim4$ may be.  

\begin{figure}
\begin{center}
\includegraphics[trim=0cm 2cm 0cm 2cm, clip, width=0.45\textwidth]{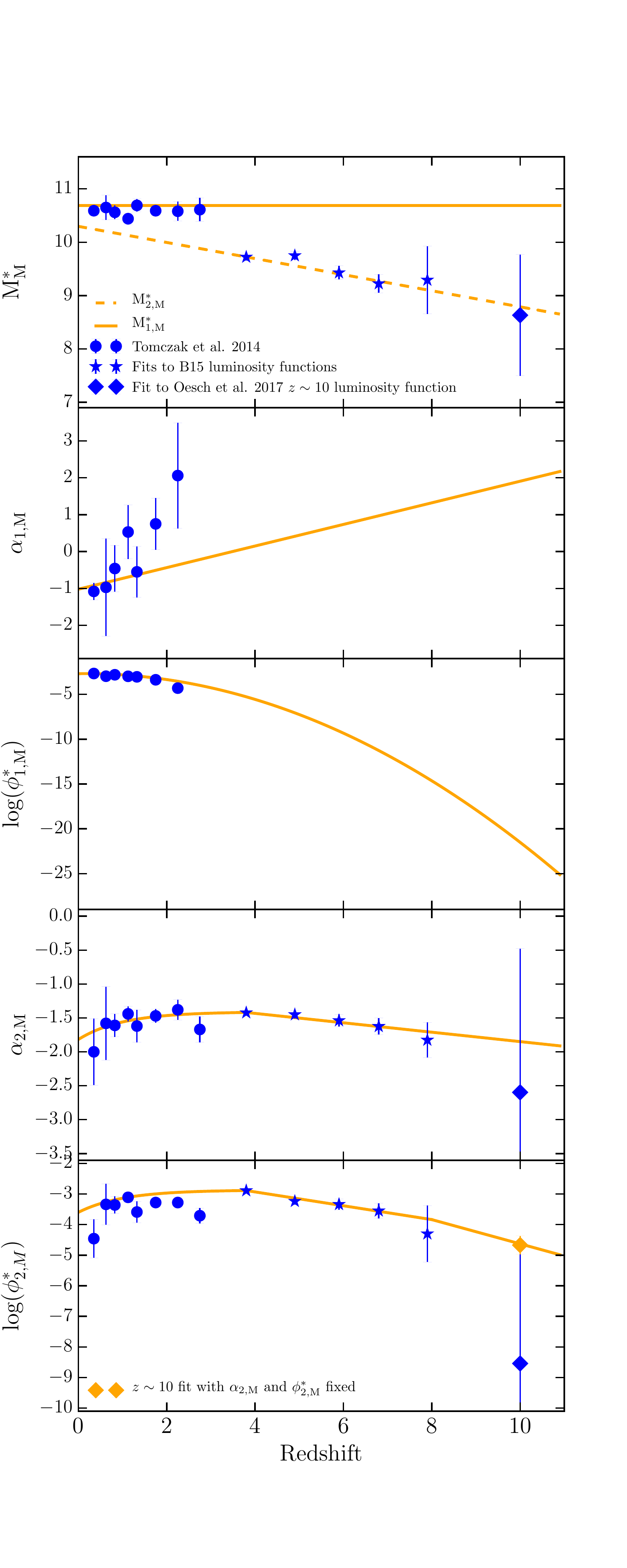}
\caption{Redshift evolution of each parameter of the double Schechter function adopted in our model. The orange lines show the adopted evolution, while circles represent the original maximum likelihood estimates from T14, with $\MFMstarOne=\MFMstarTwo$ set explicitly in the fitting. Stars (diamonds) show the median 68\% confidence intervals for the parameter estimates from our MCMC fitting (described in Appendix~\ref{app:MF_fit_to_LFs}) to the \cite{Bouwens2015LF} luminosity functions at $z\sim4,5,6,7$ and 8 (\cite{Oesch2017} $z\sim10$, see text for details).  The orange diamond in the lowest panel shows the $z\sim10$ luminosity function fit when fixing the values of \alphaTwo\ and \MFMstar.  The errors on the $z\sim10$ estimates are symmetric, so we choose to reduce the y-axis range of the panels displaying the redshift evolution of \alphaTwo\ and $\log(\phiStarTwo)$ for clarity.}
\end{center}
\label{fig:MF_param_evolution}
\end{figure}

To determine a suitable form for the redshift-evolution of the Schechter function parameters, we first need to know what \textit{mass} function parameters can reproduce the observed UV luminosity functions at $z\gtrsim4$.  The details of this fitting are given in Appendix \ref{app:MF_fit_to_LFs} and
we plot the Schechter (mass) function parameters that best fit the $z\gtrsim4$ luminosity function observations in Figure~\ref{fig:MF_param_evolution}, as well as the individual maximum-likelihood estimates of T14 at $z<4$. 
The estimates of \MFMstar\ derived from the measured luminosity functions at $z\gtrsim4$ are significantly lower than the T14 measurements.  However, if we fit the evolution of a double Schechter function with different ``knees'', as in Equation \ref{eq:DoubleSchechter}, 
we can use \MFMstarOne\ to fit to the high-mass end of the $z<3$ mass functions while \MFMstarTwo\ (plus the fast evolution of \phiStarOne) can be used to account for the rapid evolution in the bright end required to fit to the $z\gtrsim4$ luminosity functions.  We therefore choose to set the $z\ge3.8$ evolution of \MFMstarTwo, \alphaTwo\ and \phiStarTwo\ using a weighted least-squares linear regression to the luminosity function fits.  We extrapolate the linear fit of \MFMstarTwo\ to $z<3.8$ but re-fit the T14 measured mass functions allowing the other five Schechter function parameters to vary.  In fact, this choice of \MFMstarTwo\ evolution somewhat under-estimates the high-mass end of the $2<z<3$ mass functions (see Figure~\ref{fig:zlt4MFevolution}).   It is entirely possible that the reason for the strong evolution in $\M^*_{\scM}$ seen between the $z<3.8$ and $z\geq3.8$ samples is due to the \Muv-selected samples missing a population of dusty, high mass star-forming galaxies.  If they exist, these objects will be revealed by \textit{JWST}, but currently we lack firm constraints on their number density evolution.  We are basing this mock catalog on current observational constraints, and so choose to favor the fit to the $z\sim4$ luminosity function over the $2\lesssim z\lesssim3$ mass function at the high-mass end as it allows us to produce a model with number counts that vary relatively smoothly with redshift.  As such, a caveat of our model is that we are not modeling the dusty star-forming galaxies currently missed in UV-selected samples, and mildly under-represent the high mass end of the $2\lesssim z\lesssim3$ mass functions.  A model that simultaneously fits the $z\sim2.75$ T14 data and the $z\sim4$ \cite{Bouwens2015LF} luminosity function would require a strong gradient discontinuity in \MFMstarOne\ that would lead to a step discontinuity in the number counts of galaxies at high stellar masses.

When fixing the evolution of the Schechter function parameters at $z\gtrsim4$ we use a weighted least-squares linear regression excluding the point at $z\sim10$, which has noticeably lower number densities than can be accounted for by a simple linear relation in all three parameters. 
In fact, the exact form of the redshift evolution of the UV luminosity function, and associated cosmic star formation rate density (CSFRD) above $z\sim8$ has been an area of active debate in the literature, with e.g. \cite{McLeod2016} presenting measurements of the $z\sim9-10$ luminosity function that is consistent with a smooth decline in the CSFRD.  For our fiducial mock catalog we choose to base the model on the \cite{Oesch2017} results in order to provide a conservative limit on $z\gtrsim8$ galaxy number counts likely to be detected with \textit{JWST}.
We defer further discussion of this issue to Section~\ref{results}.
The constraints at $z\sim10$ are not strong enough to constrain the likely evolution in \MFMstarTwo, \phiStarTwo\ \textit{and} \alphaTwo.  
We thus choose to re-fit the $z\sim10$ luminosity function with \alphaTwo\ and \MFMstarTwo\ fixed to the values defined by the extrapolated linear fits at $z=10$, giving $\log(\phiStarTwo / \textrm{Mpc}^3 \textrm{dex}^{-1})=-4.67\pm0.3$.
We then require the gradient of the evolution in $\log(\phiStarTwo)$ to decrease further at $z>8$ so that this value is reached by the relation at $z=10$.

At $z<3.8$ we re-fit to the T14 mass functions using a Bayesian multi-level modeling approach (see Appendix \ref{app:MLmodelling}), which allows us to derive the best-fit redshift evolution of the Schechter function parameters by fitting to the mass function measurements in each redshift bin simultaneously.  This approach is more powerful than fitting a functional form to the published Schechter parameter estimates as it accounts for parameter covariance self-consistently.  At $z<3.8$ we choose a functional form for the redshift evolution for \alphaTwo\ and \phiStarTwo\ that asymptotically approaches the value of the best-fit linear relation at $z=3.8$, but decreases rapidly at the lowest redshifts.  Without this dip to low redshifts, the mass function is too shallow with too-high a normalization at low masses.  We accept a mildly discontinuous evolution at $z=3.8$ 
because allowing the functional form in either \alphaTwo\ or \phiStarTwo\ to increase and turn over by $z=3.8$ (to give smooth evolution at $z=3.8$)  produces mass functions that cross over at low masses, a situation that we are trying to avoid by requiring that our model is monotonically increasing at given mass with decreasing redshift.

\begin{figure}
\begin{center}
\includegraphics[width=0.5\textwidth]{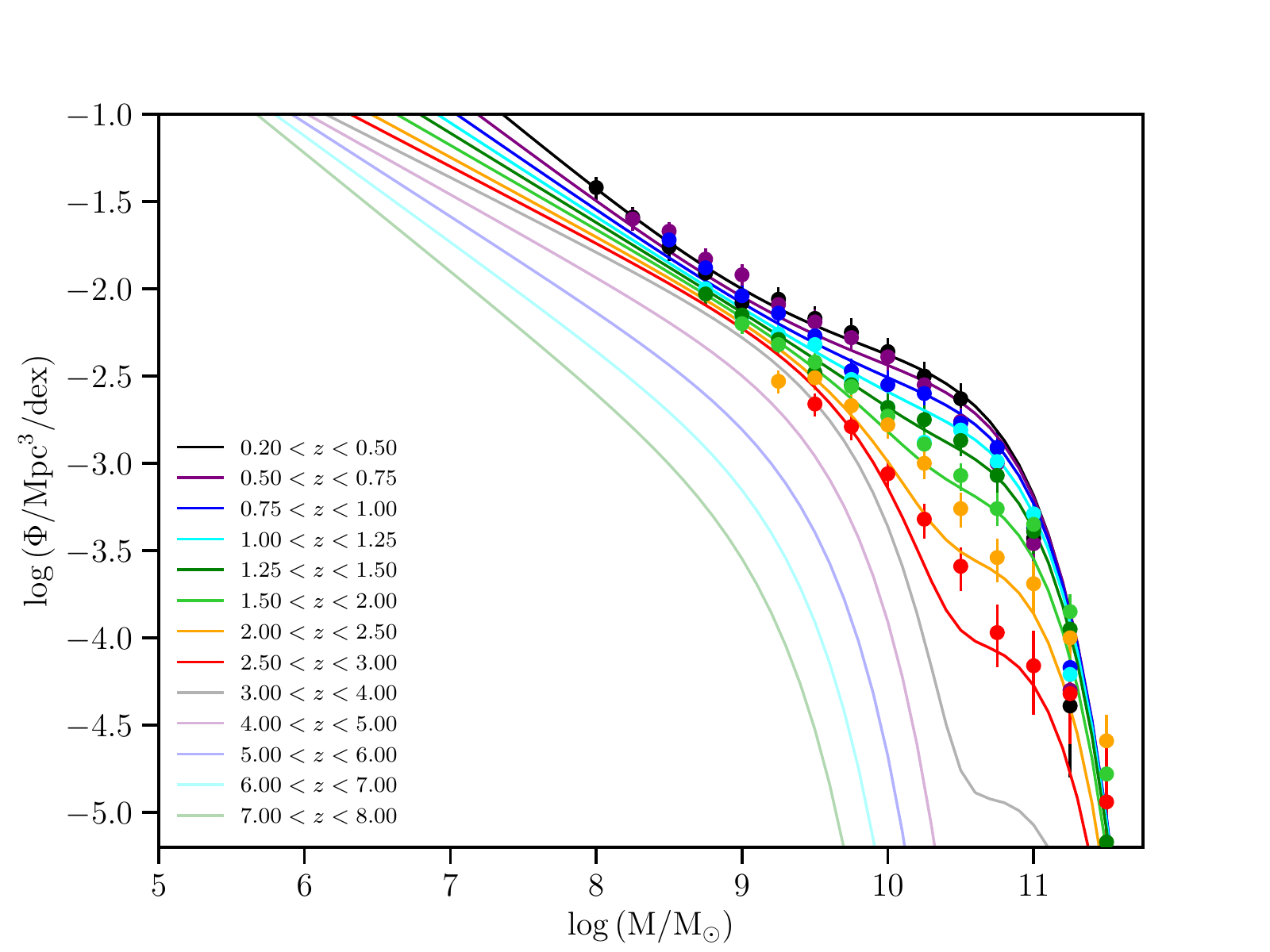}
\caption{Evolution of the star-forming galaxy mass function from $0.2 < z < 8.0$ (lines), plotted with the observations from T14 (circles). 
The parameters of this fit to the MF evolution are given in Equations \ref{eq:MFredshiftEvolutionStart}-\ref{eq:MFredshiftEvolutionEnd} and Table~\ref{table:MFhyperParams}. } 
\label{fig:zlt4MFevolution}
\end{center}
\end{figure}

The redshift evolution of each Schechter function parameter is summarized below:

\begin{align}\label{eq:MFredshiftEvolutionStart}
\MFMstarOne(z) &= a_1\\
\log[\phiStarOne(z)] &= b_1 + b_2\,z + b_3\,z^2\\
\alphaOne(z) &= c_1 + c_2\,z\\
\MFMstarTwo(z) &= D_1 + D_2\,z\\
\log[\phiStarTwo(z)] &= e_1 \, [1-\exp(-z)] + e_2 & z<3.8\\
                                                 &=E_1 + E_2\,z                        & 3.8\leq z<8\nonumber\\
                                                 &=E'_1 + E'_2\,z                         & z\geq8\nonumber\\
\alphaTwo(z) &= f_1 \, [1-\exp(-z)]+f_2 & z<3.8\label{eq:MFredshiftEvolutionEnd}\\
                                     &= F_1 +F_2\,z & z\geq3.8\nonumber
\end{align}
where the parameters $D_1$, $D_2$, $E_1$, $E_2$, $F_1$ and $F_2$ are all determined from the linear regression to the forward-modeled luminosity function fitting, and $E'_1$ and $E'_2$ are chosen to fit to the $z\sim10$ luminosity function while maintaining continuous evolution in log[$\phiStarTwo(z)]$ at $z=8$.  The parameters $e_2$ and $f_2$ are fixed to the values required to produce continuous evolution at $z=3.8$ with $e_2 = E_1+3.8\,E_2 - e_1[1-\exp(-3.8)]$ and $f_2 = F_1+3.8\,F_2 - f_1[1-\exp(-3.8)]$.  The remaining free parameters $a_1$, $b_1$, $b_2$, $b_3$, $c_1$, $c_2$, $e_1$ and $f_1$, are then constrained using the Multi-level modeling (see Appendix\ref{app:MLmodelling}) to the published T14 star-forming mass functions (their Table~1).

\begin{table}
\begin{center}
\label{table:MFhyperParams}
\caption{The values of the parameters used in our model of the mass function evolution, as described in Equations \ref{eq:MFredshiftEvolutionStart}-\ref{eq:MFredshiftEvolutionEnd}. For those parameters determined using the multi-level model fitting to $z<4$ mass functions, we report the median of the posterior distribution function, its 1$\sigma$ confidence interval, as well as the prior used in the fitting.}
\begin{tabular}{ C{0.2\columnwidth-2\tabcolsep} C{0.2\columnwidth-2\tabcolsep} C{0.2\columnwidth-2\tabcolsep} C{0.4\columnwidth-2\tabcolsep}} 
 \hline
 parameter & median & 1$\sigma$ uncertainty & prior/source of fits\\
 \hline
 $a_1$ & 10.69  & 0.04  &  $\mathcal{N}(0,50)$\\ 
 $b_1$ & -2.68 & 0.16  & $\mathcal{N}(0,50)$ \\ 
 $b_2$ & { }0.06 &  0.24 & $\mathcal{N}(0,50)$ \\ 
 $b_3$ & -0.19 & 0.08 & $\mathcal{N}(0,50)$, $\in[-\infty,0]$ \\ 
 $c_1$ & -1.02 & 0.16 & $\mathcal{N}(0,50)$ \\
 $c_2$ & { }0.29 & 0.13 & $\mathcal{N}(0,50)$\\
 $D_1$ & 10.30 & 0.10 & Linear fitting $4\lesssim z<8$\\
 $D_2$ & -0.15 &  0.02 & Linear fitting $4\lesssim z<8$\\
 $e_1$ & { }0.73 &  0.26 & $\mathcal{N}(0,50)$, $\in[0,\infty]$\\
 $e_2$ & -3.60 &  - & $=E_1+3.8E_2-e_1[1-\exp(-3.8)]$ \\
 $E_1$ & -2.03 & 0.41 &   Linear fitting $4\lesssim z<8$ \\
 $E_2$ & -0.23 & 0.09 &   Linear fitting $4\lesssim z<8$\\
 $E'_1$ & -0.67 & - &  $\phi^*_{2,\scM}$ fit to $z\sim10$ LF \\
 $E'_2$ & -0.40 & - & $\phi^*_{2,\scM}$ fit to $z\sim10$ LF\\
 $f_1$ & { }0.41 &  0.17 &  $\mathcal{N}(0,50)$, $\in[0,\infty]$\\
 $f_2$ & -1.82 &  - &  $=F_1+3.8F_2-f_1[1-\exp(-3.8)]$\\
 $F_1$ & -1.16 & 0.10 &  Linear fitting $4\lesssim z<8$\\
 $F_2$ & -0.07 & 0.02 &  Linear fitting $4\lesssim z<8$\\
 \hline
\end{tabular}
\end{center}
\end{table}

\begin{figure*}
\begin{center}
\includegraphics[width=0.8\textwidth]{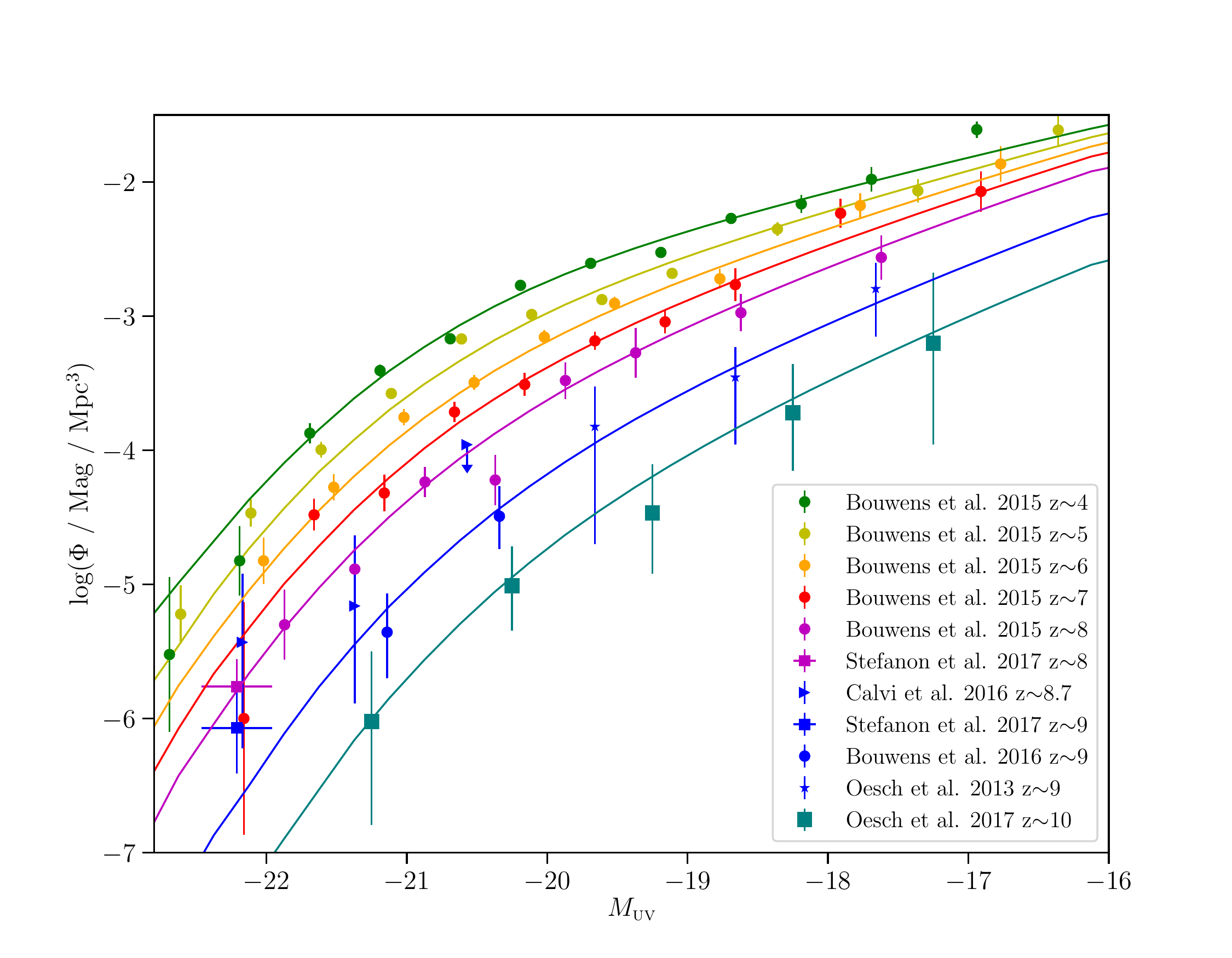}
\caption{ The UV luminosity function at $z\gtrsim4$ of our continuously evolving phenomenological model (solid lines; described in Section~\ref{SFG_MF}), evaluated at 
the mean redshift of the dropout samples used in the fitting. Points are observations at the same mean redshifts as indicated by the colors \citep{Bouwens2015LF,Stefanon2017,Calvi2016,Bouwens2016LF,Oesch2013_z9LF,Oesch2017}. 
Our forward modeling approach explicitly fits to the binned UV luminosity functions of \citet{Bouwens2015LF,Oesch2017}.
 }
\end{center}
\label{fig:LF}
\end{figure*}

We report the median values and associated uncertainties along with the values for the model parameters defined by linear fits to the $z\gtrsim4$ individual mass function estimates in Table~\ref{table:MFhyperParams}.  The chosen redshift evolution of each parameter is plotted as the orange lines in Figure~\ref{fig:MF_param_evolution}.  The resulting mass function comparisons to the T14 measurements at $z<4$ are plotted in Figure~\ref{fig:zlt4MFevolution}, and the luminosity function comparisons are shown in Figure~\ref{fig:LF}.

\subsection{The evolution of the \MStarMuv\ relation}
\label{MstarMuv}

In this Section, we describe our method to characterize the relation (slope, intercept, scatter) between  \Muv\   
and  \Mstar\ of galaxies at redshifts $0.2 \le z \le 15$. 
Hereafter, we use the definition of \Muv\ adopted, e.g., in \citet{Robertson2013}, as the average magnitude at rest-frame wavelength within a flat filter centered at 1500 \AA\ and with a width of 100 \AA, which is the definition adopted by \beagle\ \citep[][]{Chevallard2016}. 
This definition of \Muv\ differs slightly from that used to measure the UV luminosity functions in \cite{Bouwens2015LF}, which define \Muv\ to be at rest-frame wavelength of 1600\AA. We calculate the typical color correction based on the mean $\beta$ as a function of \Muv\ and redshift presented in \citet{Bouwens2014beta} and find that the typical difference in magnitudes between 1500 and 1600\AA\ is negligible ($|\delta\Muv|\lesssim0.05$). This correction is significantly smaller than the k-correction applied to estimate \Muv\ at 1600\AA\ from broad-band photometry in the first place ($|\delta\Muv|\lesssim 0.1$) and so we apply no conversion between rest-frame 1500\AA\ and 1600\AA\ \Muv\ values.

The \MStarMuv \ distribution and its evolution are critical components of our underlying phenomenological model, and are required to statistically assign UV luminosities to mock galaxies generated from our continuously evolving stellar mass function model. However, we note the following uncertainties 
to this procedure.
At all redshifts, galaxies exhibit a diversity of mass-to-light ratios, which depend on the stellar population properties (age, metallicity), star-formation history, and dust content of a galaxy. As a result, the exact form of the relation between \Muv\ and \Mstar\ and its dependency on galaxy properties are largely unknown. In general, brighter galaxies at UV wavelengths correspond to more massive objects \citep[e.g.][]{Stark2009, Lee2011, Gonzalez2011}, and this holds out to $z\sim7$ \citep{Duncan2014,Salmon2015,Grazian2015,Song2016}. Although the relation between \Muv\ and \Mstar\ follows a general trend of decreasing \Muv\ with \Mstar\ out to $\LMstarMsun \sim 10$, at higher masses the average \Muv\ becomes fainter due to the appearance of a population of fainter objects. This trend could be attributable to several effects, such as increased dust content and older average stellar ages among massive galaxies \citep[e.g.][]{Spitler2014}.
Characterizing the relationship is further complicated by the difficulty of measuring stellar mass owing to emission line contamination at high-redshift \citep{Labbe2013, Stark2013} and at low stellar masses \citep{Whitaker2014}, and the lack of a direct photometric probe of \Muv\ at intermediate redshifts ($0.6<z<1.5$). 
The procedure we outline here  has a direct impact on the resulting UV luminosity functions (see Sections~\ref{galaxy_counts_general}, \ref{evolvemassfn} and \ref{sect:LFevol}). We have therefore developed a straightforward description of the \MStarMuv\ distribution and its evolution that is designed to encapsulate the diversity of real galaxies.

\subsubsection{Characterizing the evolution of \MStarMuv\ from observations}

\begin{figure*}
\begin{center}
\includegraphics[width=.9\textwidth,trim=1cm 2cm 0cm 3cm,clip]{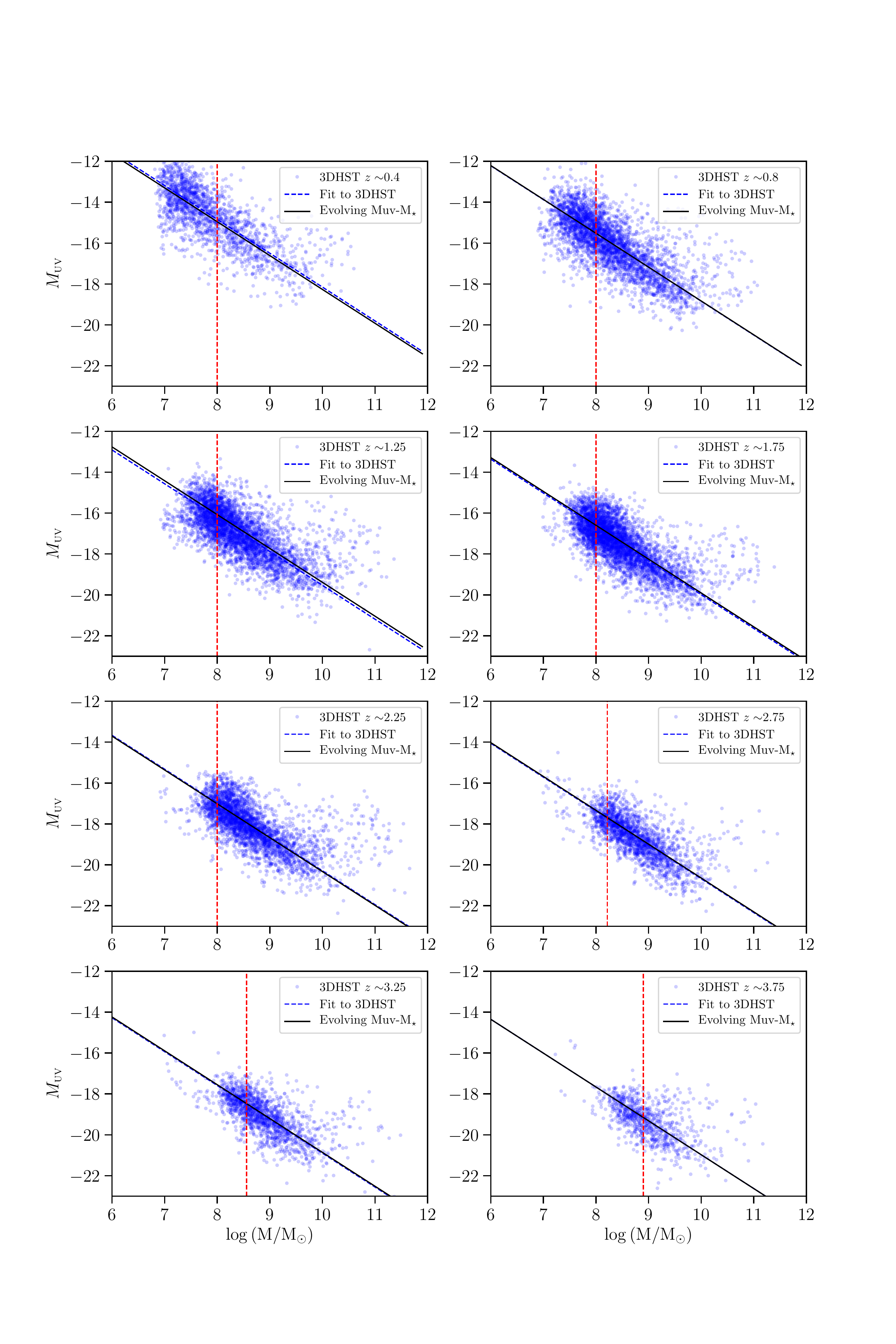}
\caption{The \MStarMuv\ relation for realizations of the SED-fitting of observed star-forming galaxies ($\log{(\mathrm{sSFR})} < -10$ Gyr$^{-1}$) in the 3D-HST survey (blue points) with high-confidence measurements of \Muv, \M, and redshift as described in Sections~\ref{MstarMuv} and \ref{section:3DHST_SEDfitting}. Blue dashed lines indicate the best-fitting linear relationship under the assumption of a fixed slope as described in the text. Solid black lines indicate the smoothly evolving redshift-evolution of the \MStarMuv\ relation in our model as characterized in Figure~\ref{MstarMuvevol} and Equation \ref{eq:MstarMuveqn}. The red dashed line indicates the stellar mass above which mock galaxies are matched to 3D-HST realizations, which is the larger between $\LMstarMsun>8$ and $\LMstarMsun>6.3+0.7z$ (the evolving mass limit exceeds $\LMstarMsun>8$ at $z>2.4$). }
\label{MstarMuvdist}
\end{center}
\end{figure*}

We characterize the \MStarMuv\ relationship at $z\lesssim4$ using measurements from the 3D-HST catalog (using SED-fitting with \beagle; see description in Section~\ref{section:3DHST_SEDfitting}). As discussed extensively in \citet{Stefanon2017_900nm}, selection effects can heavily influence the observed shape of the \MStarMuv\ distribution. Therefore, we avoid including observed galaxies whose \Muv\ or \Mstar\ measurements are poorly constrained by the \beagle\ fits. 
Specifically we only use galaxies with $\delta\LMstarMsun<1$, $\delta z < 1$ and $\delta\Muv<1$ (where e.g. $\delta z$ is the 68\% credibility interval on redshift.) 
The limits imposed were chosen to avoid biasing the 
characterization of the \MStarMuv\ distribution
with overly strict \Muv\ or \Mstar\ cuts, which we discuss further below.

 In Figure~\ref{MstarMuvdist}, we plot the \MStarMuv\ distributions for the 3D-HST galaxies with well-constrained \Muv, \Mstar\, and redshift measurements. 
As discussed above, these distributions show a trend of 
increasing stellar mass with decreasing \Muv\ at low stellar mass.
At high stellar mass the \Muv\ values tend to be fainter than the linear relation, 
as observed in  \citet{Spitler2014}.
Rather than attempting to fully model this mass-dependent behavior, especially given the Malmquist biases that begin to affect the higher-redshift bins, we adopt the following two-step procedure to describe the \MStarMuv\ distributions at $z\lesssim4$.
We fit the observed \MStarMuv\ distribution under the simplest assumption of a linear relationship to extrapolate to low masses, while at higher masses (\LMstarMsun$\gtrsim$8-8.5,  depending on the mass limit at a given redshift), we assign \Muv\ values by sampling from real galaxies of the same mass. The matching procedure  allows us to maintain the observed flattening of the distribution at high masses,  and is fully described in Section~\ref{highmassgal} below.

Figure~\ref{MstarMuvdist} illustrates the 
substantial scatter in the observed \MStarMuv\ distributions at $z\lesssim4$.
Owing to the large scatter, the best fitting slope will depend strongly on the uncertainties on the data points, 
and the size of the uncertainties may depend on \Muv, \Mstar, and also plausibly on redshift. Indeed, we find that when fitting with both slope and normalization as free parameters neither parameter is well constrained,
and the best fitting slope is highly variable between redshift bins. 
Therefore, we adopt a fixed slope for the \MStarMuv\ relation at all redshifts and fit only the intercept at each redshift. This procedure essentially fits the average redshift-dependent mass to light ratio,
which has lower uncertainty and is less dependent on the error on individual galaxy measurements and stellar mass-dependent systematics. 
Several studies have reported the measurement of constant slope for UV-selected galaxies, with normalization evolving in redshift \citep[][]{Duncan2014, Salmon2015, Grazian2015, Song2016, Stefanon2017_900nm}, and find a
reasonable description of the data. The blue dashed line in Figure~\ref{MstarMuvdist} shows our best fit relation to the
\MStarMuv\ distribution in each redshift bin, where the slope is fixed to a value of -1.66. We find excellent agreement with the observed distribution at all redshifts. For reference we also indicate the stellar mass limits in each redshift bin above which we assign \Muv\ values by sampling from real galaxies (red dashed lines). Fitting the \MStarMuv\ distribution only above these mass limits instead has a negligible effect on the result at $z<3$. At $z\sim3.75$ where there are fewer well-constrained measurements, fitting above this mass limit would increase the \MStarMuv\ intercept by $\sim0.1$ mag, an indication that fitting only at the high-mass end biases the characterization of the \MStarMuv\ due to the high-mass end flattening. Therefore, we choose to proceed using all galaxies with well-characterized stellar mass, redshift and \Muv.

\begin{figure}
\includegraphics[width=.5\textwidth]{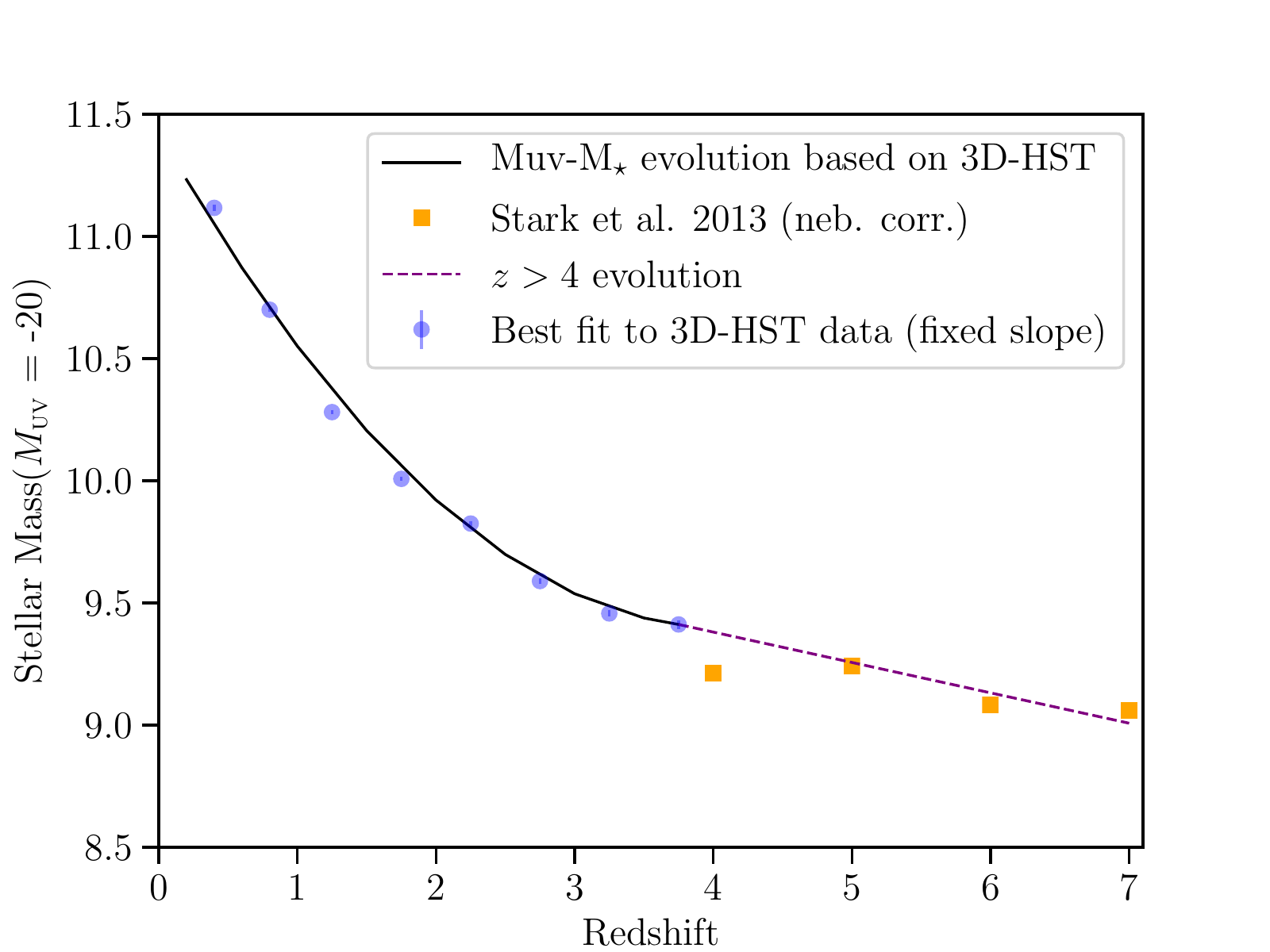}
\caption{The redshift evolution of the intercept of the \MStarMuv\ distribution, $M_\textsc{\MStarMuv}^0(z)$, defined at $\Muv=-20$. Blue points indicate the best fit intercept to the 3D-HST data presented in Figure~\ref{MstarMuvdist} assuming a fixed slope. Error bars are smaller than the size of the symbol. Orange points are based on relation presented in \citet{Stark2013} that includes the correction for nebular emission lines. 
}
\label{MstarMuvevol}
\end{figure}

To set the full redshift evolution of the \MStarMuv\ relation, we
combine the intercept values for the best fit relations with fixed slope at each redshift $z\lesssim4$ with measurements of the \MStarMuv\ intercept at $z>4$.
We use the average observed value of stellar mass for bright (\Muv = -20) galaxies
at $4<z\le7$ to set the overall normalization in each $z>4$ redshift bin, while assuming the same constant \MStarMuv\ slope. We utilize the high-redshift stellar mass measurements shown in figure~7 of \citet{Stark2013}, where the measured stellar masses were fit while including the contribution to the SED from nebular emission lines. The normalization value at $\Muv = -20$ shows an overall decline between $4\lesssim z<7$, indicating a decrease in the average mass to light ratios of galaxies with increasing redshift. The measured values for the \MStarMuv\ intercepts at all redshift bins, evaluated at $\Muv=-20$, are shown as points in Figure~\ref{MstarMuvevol}.

\subsubsection{Continuous redshift evolution of \MStarMuv }

Our model of the redshift evolution of the \MStarMuv\ relation is defined by fitting to the evolving intercept values shown in Figure~\ref{MstarMuvevol}.  These intercepts show a rapid decline from $z\sim 0-3$, with a shallower decline at $z>4$. We find that at $z\le4$ the intercept measurements are adequately described by a quadratic function, and a linear function at $z>4$.
To ensure these two functions remain continuous at $z\sim 4$, we include the boundary condition that the derivatives of the two functions are equal at the center of the highest redshift bin where we use the 3D-HST data ($z=3.75$). This  constraint results in the following function to describe the intercept of the \MStarMuv\ relation,  
$ M_\textsc{\MStarMuv}^0(z)$, 
evaluated at $\Muv=-20$ and $z\le 3.75$:

\begin{equation}\label{eq:MstarMuveqn}
\begin{aligned}
& M_\textsc{\MStarMuv}^0(z) = a \, (z-3.75)^2 + [b - 2\,a\,(z-3.75)] \, (z-3.75) + c \\
\end{aligned}
\end{equation}
where $a=0.12$, $b=0.08$, and $c=9.41$.
This evolution of the \MStarMuv\ intercept is shown as the black solid curve
in Figure~\ref{MstarMuvevol}, along with the observed data (blue points). The resulting linear \MStarMuv\ relations in each redshift bin, according to the smoothly evolving intercept function defined in Equation \ref{eq:MstarMuveqn}, are also shown as black solid lines in Figure~\ref{MstarMuvdist}.
At $z>3.75$, the \MStarMuv\ intercept evaluated at $\Muv=-20$ evolves approximately
linearly with the following form:

\begin{equation}\label{eq:MstarMuveqn2}
\begin{aligned}
M_\textsc{\MStarMuv}^0(3.75<z<8) &\,=\, -0.12\times z+9.88 \\  
M_\textsc{\MStarMuv}^0(z\ge8) &\,=\,  8.92 \\  
\end{aligned}
\end{equation}

We use Equation \ref{eq:MstarMuveqn} to assign \Muv\ values to galaxies of a given stellar mass and redshift at $z\le3.75$, and Equation \ref{eq:MstarMuveqn2} to assign \Muv\ values at $z>3.75$, assigned randomly within the scatter observed in 3D-HST data in Figure~\ref{MstarMuvdist}.  
We have characterized this scatter in both stellar mass, \Muv, and redshift, and find that the scatter in \Muv\ is remarkably constant in both stellar mass and redshift, with an average value of $\sigma_{uv}\sim$0.7 magnitudes. 
Therefore, at all redshifts, we assign \Muv\ values randomly according to this relation and that assumed Gaussian scatter $\sigma_{uv}$. This direct assignment of \Muv\ applies only to low-redshift low-mass star-forming mock galaxies ($z\le4$ and have $\LMstarMsun \le8$), or high-redshift star-forming galaxies at $z>4$.  Massive, low-redshift mock galaxies (with $\LMstarMsun > 8$ and $z\le4$) are assigned \Muv\ values according to their matched 3D-HST realization.
The \MStarMuv\ relation and scatter as described here are used as priors
 when drawing realizations from fits to 3D-HST galaxies to ensure a smooth transition in the mock catalog \MStarMuv\ relation at $\LMstarMsun=8$.
 This procedure is detailed in Section~\ref{highmassgal}.

\subsubsection{Theoretical limits on the mass to light ratios of galaxies}

As mass to light ratios continue to decrease with increasing redshift and decreasing stellar mass according to our model, the mass to light ratios approach a theoretical limit of the stellar population models we generate with \beagle.
This limit is set by the UV luminosities of individual massive stars, and represents the minimum mass to light ratio possible for an instantaneous burst of star-formation for any given IMF (with no dust attenuation, the lowest metallicity, and corresponding nebular continuum emission). Any IMF choice will result in such a limit in the possible \Muv\ given a stellar mass, with more top-heavy or bottom-light IMFs allowing for brighter limiting \Muv\ and bottom heavy IMFs producing fainter limiting \Muv. 
The Chabrier IMF that we use in this work is
relatively bottom-light and has a larger mass to light ratio parameter space than a more bottom heavy IMF \citep[e.g.][]{Kroupa2001, Salpeter1955}.
For a Chabrier IMF with our assumed high-mass cutoff of 100 \Msun,
 we find that the stellar plus nebular continuum emission results in a theoretical mass to light ratio given by $\Muv \approx -2.45\,\LMstarMsun - 1.3$.

\begin{figure}
\begin{center}
\includegraphics[width=.5\textwidth]{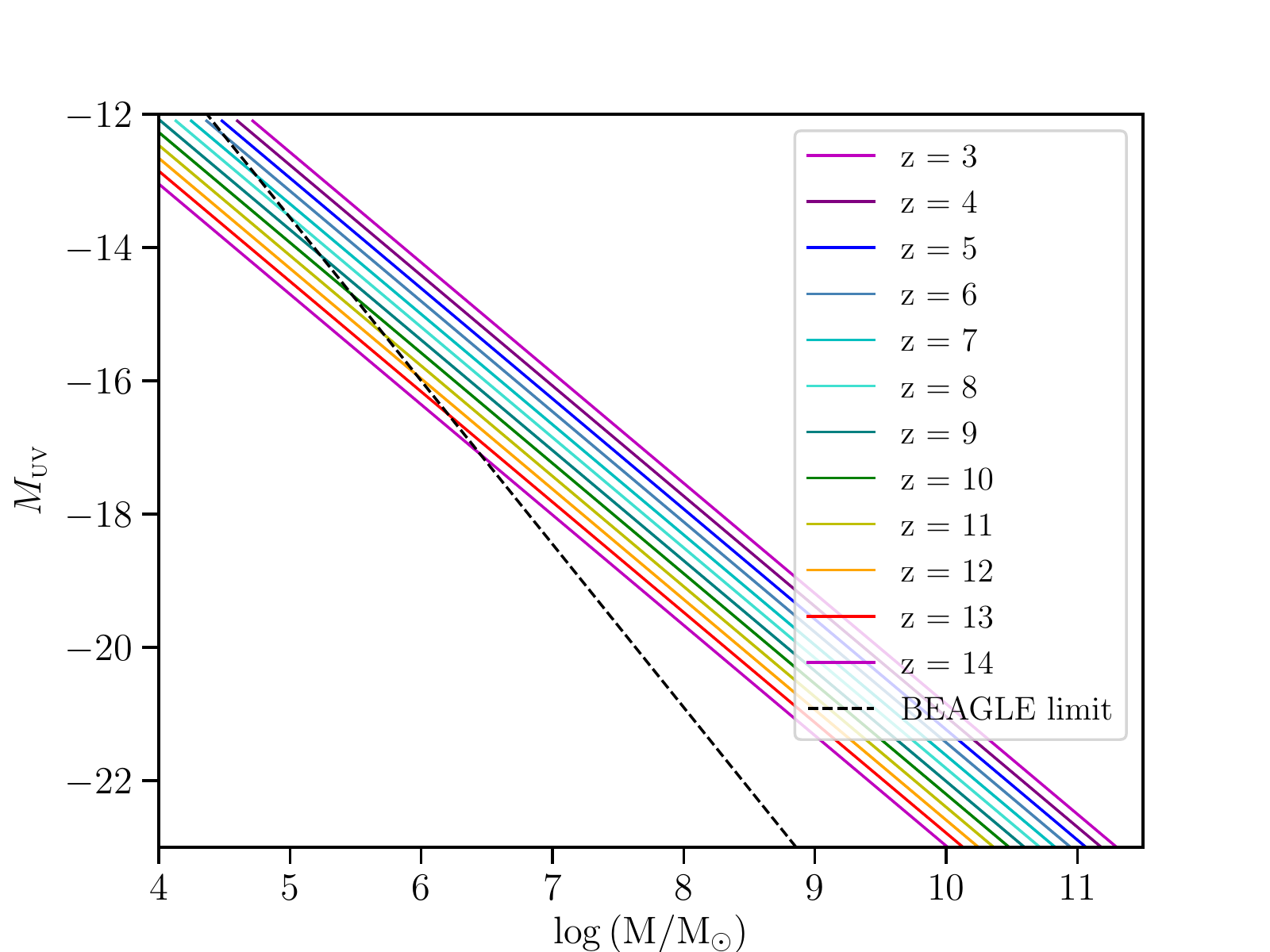}
\caption{Demonstration of the evolution of the mean \MStarMuv\ relation with redshift, (if mass to light ratios were allowed to continue decreasing above $z>8$), in comparison to the mass-to-light ratio limit imposed by the stellar population modeling with \beagle\ (which implicitly assumes a Chabrier IMF with a high mass cut-off of 100 \Msun.) 
In our model, all $z>8$ galaxies follow the \MStarMuv\ relation at $z=8$ to avoid a scenario where significant numbers of galaxies exceed the theoretical mass-to-light ratio limit set by \beagle.  }
\label{MstarMuvLimit}
\end{center}
\end{figure}

To accommodate this theoretical limit in the \MStarMuv\ evolution of our phenomenological model, we truncate the Gaussian distribution that we use to assign \Muv\ values.  For galaxies at the detection limit of future blank surveys (e.g. apparent magnitude $m_{app} \sim 31$) this truncation has a negligible effect on the overall \MStarMuv\ distribution at $z<4$. At $z>4$, we find that scatter using the truncated Gaussian at the limit changes the overall shape of the \MStarMuv\ distribution by steepening the low-mass end. The effect becomes significant by $z\sim8$. We therefore additionally halt the redshift evolution of the mean \MStarMuv\ relation parametrized in Equation \ref{eq:MstarMuveqn2} at $z=8$.
A demonstration of the \beagle\ mass to light ratio 
limit is shown in Figure~\ref{MstarMuvLimit},  compared
with the projected evolution of the mean \MStarMuv\ relation
allowed to evolve past z$\sim$8.

Although we make every effort to choose reasonable constraints where available, the overall shape of the \MStarMuv\ distribution at $z>4$ is still an extrapolation that impacts various evolutionary relations at high-redshift in the model, including the UV luminosity function and specific star-formation rate (sSFR).
We will discuss these issues in depth in Section~\ref{characterization}.


\subsection{UV continuum slope -\Muv\ relationship}\label{betamuv}

The spectral slope of the UV-continuum (\Betauv\, where $f_{\lambda}\propto \lambda^{\beta}$) of galaxies is sensitive to the properties of stellar populations (e.g. metallicity and age), star-formation history, and dust attenuation. Population studies of star-forming galaxies indicate well-characterized relationships between $\beta$ and 
UV luminosity. Bright, massive galaxies tend to have red
(i.e., shallower) UV continua, which likely owes to a combination of old stars, higher stellar metallicities, and a larger dust content. 
The bluer (i.e. steeper) UV continua of lower luminosity galaxies are often associated with younger, less metal-rich stellar populations,
and less dust attenuation \citep[e.g.][]{Stanway2005,Labbe2007, Bouwens2009, Labbe2010beta, Bouwens2012beta, Rogers2013, Rogers2014}.
The detailed relations between \Betauv\ and \Muv, scatter, and evolution out to redshift $z\sim8$ are still areas of active research, but many studies are consistent with 
a linear relationship between \Betauv\ and \Muv\,
\citep[e.g.][]{Bouwens2012beta, Kurczynski2014,Bouwens2014beta,Alavi2014,Rogers2014}
with an average evolutionary trend towards bluer \Betauv\ with increasing redshift \citep[e.g.][]{Labbe2007, Bouwens2009, Finkelstein2012beta, Castellano2012, Wilkins2011,Bouwens2014beta}.

\subsubsection{Mean \MuvBeta\ relation across cosmic time}\label{betamean}

We use a compilation of measured \MuvBeta\ relations and their scatter across redshifts to assign the rest-frame UV SEDs of mock galaxies. Following several studies (see previous Section), we model the average \MuvBeta\ relation with a linear function, where the slope $d\beta(z)/d\Muv$ and intercept $\beta(\Muv = -19.5, z)$ of the function vary with redshift. We consider sets of \MuvBeta\ relations at $1\le z \le 8$ obtained from HST/ACS \citep{Kurczynski2014, Alavi2014, Mehta2017} and HST/WFC3 imaging \citep{Bouwens2009, Bouwens2014beta}. The relationships describing \MuvBeta\ at the high redshifts
we model
are broadly consistent with fits measured in other studies \citep[e.g.][]{Bouwens2012beta, Finkelstein2015, Rogers2014}.

\begin{figure}
\begin{center}
\includegraphics[width=0.5\textwidth]{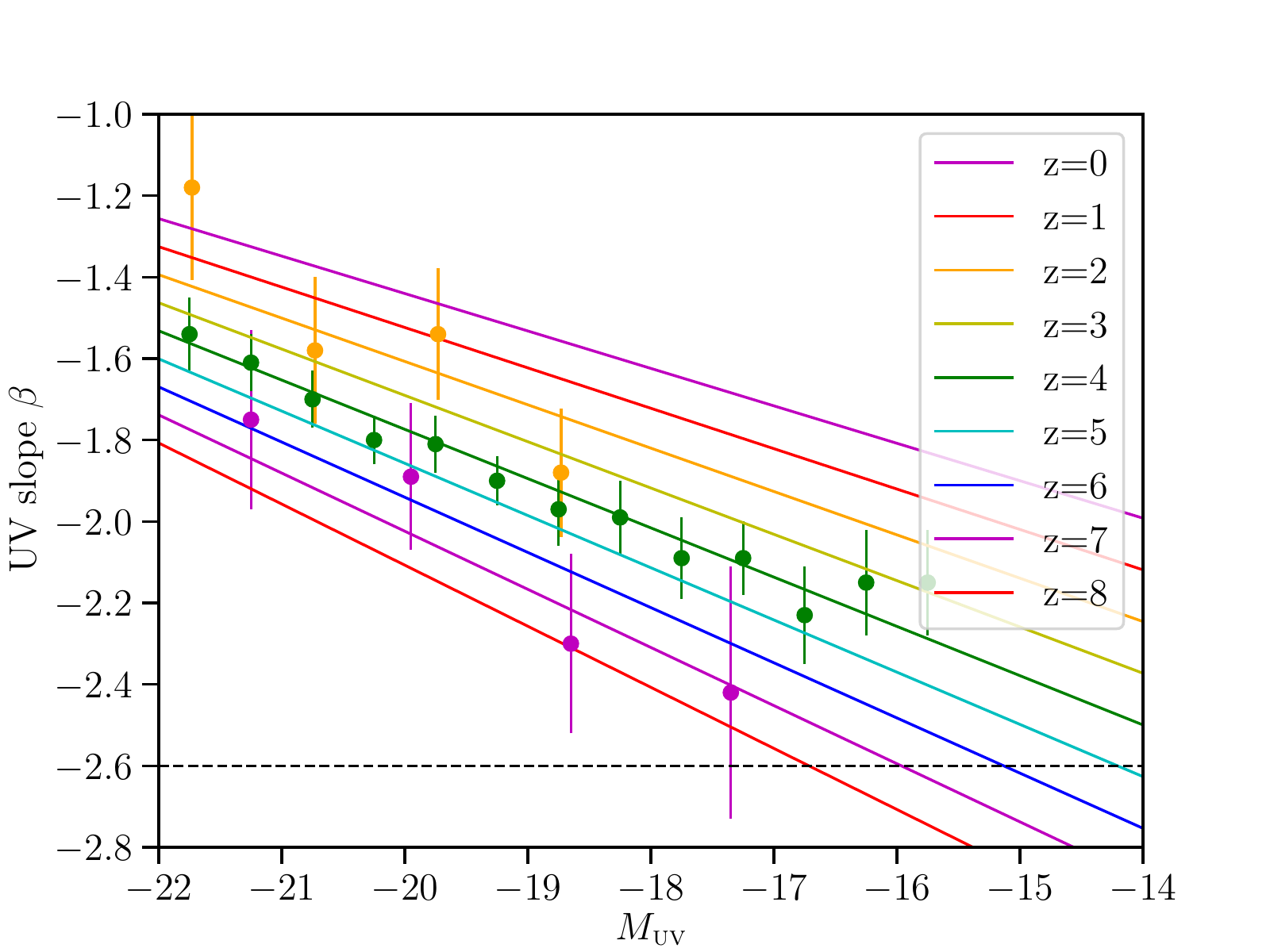}
\caption{Redshift evolution of the mean \MuvBeta\ relation. 
Intrinsic scatter is included as described in Section~\ref{betascatter}. Points indicate a selection of binned measurements from \citet[][circles]{Bouwens2009,Bouwens2014beta} at similar redshifts of the colored lines (orange at $z\sim2.5$; green at $z\sim4$; magenta at $z\sim7$). The black dashed line indicates the theoretical limit in the \beagle\ models as discussed in Section~\ref{sec:betalim}.  }
\end{center}
\label{betaevol}
\end{figure}

We find that the slope of the \MuvBeta\ relation shows little evolution at redshifts $1\le z \le 8$, as already found in previous works \citep[e.g.][]{Bouwens2012beta, Kurczynski2014,Bouwens2014beta}. The intercept of the relation increases significantly from redshift $z\sim1-8$,
reflecting the evolutionary trend that galaxies have older ages and higher metallicities at later cosmic times \citep[e.g.][]{Labbe2007}.
We perform least-squares linear fits to measurements of both 
$d\beta(z)/d\Muv$ and $\beta(\Muv = -19.5,z)$ and their errors from the literature to produce a mean relation that smoothly evolves with
redshift (see Figure~\ref{betaevol}), described by
\begin{equation}\label{meanbeta}
\begin{aligned}
d\beta(z)/d\Muv = -0.09 \, z-0.007 \\
\beta(\Muv=-19.5,z) = -1.49 \, z-0.09.
\end{aligned}
\end{equation}
For mock galaxies at $z\le1$, we extrapolate this relationship to lower redshifts to assign $\beta$ values. 
At $z>8$ we use the \MuvBeta\ relationship at $z=8$ to assign $\beta$ to mock galaxies. There exist several motivations to curb the evolution of \MuvBeta, with the foremost being the existence of a theoretical limit on the steepness of the UV spectrum emitted by non-Pop. III stars (see discussion in Section~\ref{sec:betalim} below).
Further, the data do not yet constrain evolutionary trends at the highest redshifts currently accessible (\range{z}{5}{8}).
While evolutionary trends with redshift are observed in most analyses \citep{Finkelstein2012beta,Bouwens2014beta}, 
the combination of high-redshift color selections with flux boosting from noise in the filters used to measure $\beta$ are still likely causing statistical studies to be biased  against redder $\beta$ measurements at $z>5$ \citep{Dunlop2012, Rogers2013}.
At the very highest redshifts currently accessible (\range{z}{7}{10})
the data do not provide strong 
evidence for or against any evolutionary trend in $\beta$ with redshift or \Muv\ \citep{Dunlop2013,Wilkins2016}, although the dynamic range in \Muv\ is relatively small at such early times. 
While evolution cannot be excluded by current data, deep imaging surveys with {\it JWST} will enable more robust characterization of the evolution beyond $z\sim8$.

In our model, all mock galaxies are assigned a $\beta$ according to the mean relationship and intrinsic, photometric bias-corrected
scatter described in Section~\ref{betascatter}. We now detail the
theoretical predictions for the UV continuum slope of our
combined stellar population and photoionization model.

\subsubsection{Theoretical limits on UV-continuum slopes}\label{sec:betalim}

\begin{figure}
\centering
\resizebox{\hsize}{!}{\includegraphics{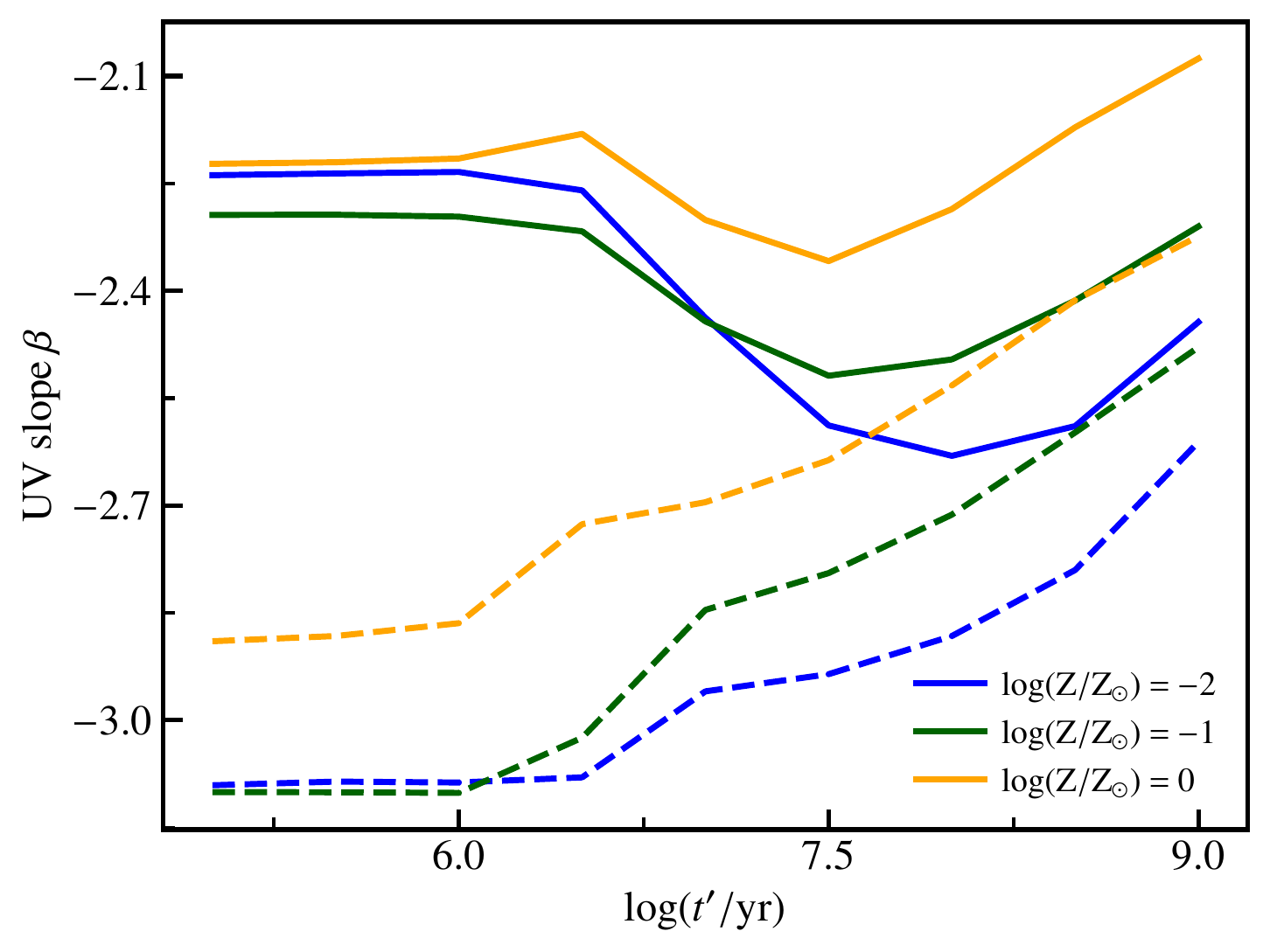}} 
\caption{UV-continuum slopes predicted by the spectral evolution model adopted in this work. We show model predictions for a constant SFH of different ages and three metallicities, $\log{\Z/\Zsun}=-2$ (blue), -1 (green), and 0 (orange). Dashed lines indicate predictions for stellar emission only, while solid lines for stellar and nebular continuum emission. In this latter case, we consider a photoionization model with ionization parameter $\logUs=-2.5$ and depletion factor $\xid=0.3$. }
\label{fig:UV_slope_model}
\end{figure}

Model predictions on the shape of the UV-continuum emission of galaxies depend on the assumed properties of stellar populations and ISM (gas and dust).
Young, massive stars show blue UV continua, which become bluer with decreasing metallicity.
The effects of stellar age and metallicity on the UV-continuum emission over time in our spectral evolution model are illustrated in Fig.~\ref{fig:UV_slope_model}, which shows  that the bluest UV spectra are obtained for very young ($\lesssim 1$ Myr) stellar populations with sub-solar metallicities (dashed lines).
Dust reddens the emitted stellar spectrum, and leads to a relation between
the attenuation suffered by a galaxy and its UV slope \Betauv\ \citep{Meurer1999}. 
Recombination-continuum from ionized hydrogen also reddens the UV continuum emission emerging from a galaxy. This effect is shown by the solid lines in Fig.~\ref{fig:UV_slope_model}, which illustrates how our combined stellar population $+$ photoionization model predicts redder \Betauv\ slopes than a model accounting only for stellar emission.

In order to avoid unphysical values for the \Betauv\ slopes associated to our mock galaxies through the redshift-dependent \MuvBeta\ relation presented in Section~\ref{betamean} above, we impose a limit $\Betauv_\txn{min}=-2.6$ for the bluest possible value. 
This limit corresponds, approximately, to the bluest \Betauv\ value obtained with \beagle\ for a model with constant SFH, sub-solar metallicity, low depletion factor ($\xid=0.1$), and low ionization parameter ($\logUs=-3.5$). We note that models with non-zero escape fraction of H-ionizing photons can reach bluer values.      

\subsubsection{Scatter in the \MuvBeta\ relation across cosmic time}\label{betascatter}

The scatter in the \MuvBeta\ distribution discussed in Section~\ref{betamean} encodes the intrinsic diversity in age, metallicity, and dust attenuation of the galaxy population at fixed redshift and UV luminosity. We aim to assign $\beta$ values to mock galaxies following the intrinsic scatter (i.e., corrected for photometric biases) of the \MuvBeta\ distributions over cosmic time.
Much effort has been put into characterizing the scatter, and how it might change with redshift, UV luminosity, or other galaxy properties \citep[e.g.][]{Bouwens2012beta}.
The intrinsic scatter in $\beta$ at fixed UV luminosity, $\sigma_{\beta}$, is surprisingly uniform across all redshifts from \range{z}{1}{6}, and relatively independent of UV luminosity with values between \range{\sigma_{\beta}}{0.3}{0.4} \citep{Bouwens2012beta,Bouwens2014beta, Kurczynski2014,Mehta2017}. 
We note that there is some evidence for the intrinsic scatter of the 
distribution increasing with UV luminosity, such that populations of brighter galaxies will have larger intrinsic scatter in $\beta$ \citep{Rogers2014}. This evidence comes from a careful analysis at $z\sim5$ only, however, and such luminosity dependence in the $\sigma_{\beta}$ is not characterized sufficiently across cosmic time to be incorporated in our model. We correspondingly adopt an intrinsic scatter of $\sigma_{\beta}\sim0.35$ in our model uniformly across all redshifts and UV luminosities.

To include intrinsic scatter $\sigma_{\beta}\sim0.35$ in our model, we assign $\beta$ values to mock galaxies according to a Gaussian distribution with a mean defined at a given redshift and \Muv\ according to Equation \ref{meanbeta} with $\sigma_{\beta}=0.35$. 
To avoid values of $\beta$ bluer than the theoretical limits described in the previous section (for which we would not be able to associate a \beagle\ spectrum), we truncate the distribution at $\beta=-2.6$. With this truncated Gaussian scatter, at the very highest redshifts and faintest \Muv\ the mean value of $\beta$ reddens slightly so as to cause a mild flattening of the linear relations shown in Figure~\ref{betaevol}. However, we accept this feature as more favorable  than artificially fixing to the bluest value of $\beta$ and reducing the galaxy diversity in the mock. In addition, it mimics the behavior of \MuvBeta\ seen in some studies that indicate an apparent flattening of the linear relation at faint luminosities \citep[e.g. $\Muv \gtrsim -19$;][]{Bouwens2014beta}. Future measurements from {\it JWST} imaging and spectroscopy will help to inform our stellar population synthesis models as we uncover the full range and distribution of UV continuum slopes in the early universe.

As described further in the following section, when possible we use 3D-HST galaxies to provide the constraints on the shape of the SED for mock galaxies. However, when this is not possible, we use the $\beta$ slope that is assigned to each galaxy to match to a parent catalog of SEDs produced by \beagle\ (see Section~\ref{BEAGLE_SF_grid}).  This ensures that our catalog will follow observed trends in \MuvBeta.

\subsection{Assigning Galaxy SEDs and spectroscopic properties}
\label{section:SF_SED_parent_catalog}

We assign a set of spectral properties to each mock galaxy, 
allowing us to provide filter photometry as well as a full spectrum for each object. The general method is to produce a parent catalog of spectra that can be matched to galaxies in the mock. 
Where possible we produce this parent catalog from the results of SED fitting to galaxies in the 3D-HST catalog, allowing the observed photometry to provide the diversity of observed SEDs at given stellar mass and redshift (see Section~\ref{BEAGLE_general}). We limit the use of these empirical SEDs to $z\le4$ galaxies, as beyond this redshift the rest-frame optical is only sampled by the Spitzer IRAC bands where the poor resolution leads to a significant confusion of sources at faint magnitudes. 

For galaxies at redshifts $z>4$, or $z\leq4$ but below the mass completeness limits of 3D-HST, we rely on extrapolations of observed relationships between \MStarMuv\ and \MuvBeta\ to provide constraints on the galaxy SEDs.  These constraints are used to match to a parent catalog, produced using \beagle\ in mock catalog mode.  When generating this parent catalog we have the full parameter space of the stellar and nebular templates to choose from (described in Section~\ref{BEAGLE_general}) and so we use observed trends in galaxy physical parameters, albeit with large scatter, to restrict this parameter space.  Specifically we use three observed relations: $\Mstar-Z-\sfr$, where \sfr\ is the SFR of the object (the `fundamental metallicity relation'); $\sfr-Z-\tauV$, 
to provide physically motivated constraints on dust attenuation (where \tauV is the effective V-band optical depth); and $Z-\logUs$.

\subsubsection{SED fitting to 3D-HST catalogs}
\label{section:3DHST_SEDfitting}

The photometric catalogs produced by the 3D-HST team \citep{Skelton2014} are selected from the noise-equalized combination of HST/WFC3 $J_{125}$, ${JH_{140}}$ and $H_{160}$ images 
taken from an extensive set of publicly available imaging data over 5 fields (AEGIS, COSMOS, GOODS-North, GOODS-South, and the UKIDSS UDS) covering $\simeq900$ arcmin$^2$.  From these catalogs we use the data in the deeper regions of the CANDELS (\citealt{Grogin2011}, \citealt{Koekemoer2011}) GOODS-South and GOODS-North fields to provide a parent catalog of redshift and mass-dependent SEDs that can be assigned to our mock catalog galaxies.

We use version 4.1 of the 3D-HST photometric catalogs \citep{Momcheva2016}.  These catalogs include selection from mosaics that include HUDF-09 (11563; PI:Illingworth) and HUDF-12 (12498; PI: Ellis) WFC3 imaging in the HUDF and parallels (11563; PI: Illingworth) that was performed as part of release 3.0\footnote{as outlined in 3D-HST v3.0  release documentation http://monoceros.astro.yale.edu/RELEASE\_V3.0/Photometry/\\3dhst\_v3.0\_readme.pdf}.  The catalogs do not include deeper HUDF ACS imaging of \cite{Beckwith2006} and so ACS photometry across the GOODS-South is at the depths of the original GOODS imaging \citep{Giavalisco2004}.

From the GOODS-South and GOODS-North 3D-HST catalogs we fit to the broad-band HST fluxes ($B_{435}$,  $V_{606}$, $i_{775}$, $z_{850}$, $J_{125}$, $JH_{140}$ and $H_{160}$), as well as the \textit{Spitzer}/IRAC Channel 1 (3.6$\mu$m) and Channel 2 (4.5$\mu$m) imaging from SEDS \citep{Ashby2013} to provide constraints in the rest-frame optical at high redshifts.
We also use a subset of the ground-based filters that required small photometric zeropoint corrections in the SED fitting analysis of \cite{Skelton2014} compared to the $H_{160}$ band (see their Table~11). 
In the GOODS-South field we use photometry from VLT/ISAAC $J$, $H$, and $K_s$ band imaging from the ESO/GOODS and FIREWORKS surveys \citep{Wuyts2008,Retzlaff2010},
and VLT/VIMOS U-band imaging from the ESO/GOODS survey \citep{Nonino2009}.
In the GOODS-North field we use Subaru/MOIRCS imaging in J, H and Ks bands from the MODS survey \citep{Kajisawa2011}\footnote{We choose not to fit using the KPNO U-band data in GN as the imaging is significantly shallower than the VLT/VIMOS U-band imaging in GS, and a large zeropoint offset was measured in \cite{Skelton2014}}.
We do not apply the zeropoint offsets reported in \cite{Skelton2014}, Table~11 after verifying that applying these corrections does not improve the accuracy of photometric redshifts output by \beagle.

We fit the broad-band photometry using \beagle\ (see Section~\ref{BEAGLE_general} for details on the model).
We use a delayed star formation history $\psi(\t) \propto \t \exp{(-\t/\tausfr})$, where \tausfr\ is the star formation timescale and \t\ the age of the galaxy, taken to lie between $10^7$ yr and the maximum time allowed since the onset of star formation at the galaxy redshift.  This parameterization gives a star formation history that rises at early times and declines exponentially at later times.  This star formation history is shown to better reproduce the colors and mass-to-light ratios of galaxies in the smoothed particle hydrodynamics (SPH) simulations of \cite{Simha2014} than the widely-used exponentially decreasing star-formation histories.  Additionally, simulations have been shown to predict that high-redshift galaxies have rising star-formation histories \citep{Finlator2011}, a scenario that is naturally achieved using this parameterization.
To ensure that galaxies are not fitted with models that are older than the age of the Universe, we set an upper limit of $\zformm=15$ to the redshift of onset of star formation. We further employ a weakly informative Gaussian prior on $\log \t$, with mean at $\log (\t/{\rm yr}) = 9.3$ and $\sigma=0.7$.\footnote{
Since the galaxy age is only weakly constrained by broad-band data alone, the resulting stellar masses are sensitive to the choice of the age prior \citep[e.g.][]{Pacifici2015}. Empirically, we find that adopting a uniform prior on $\log \t$ overweights young ages, therefore leading to underestimated stellar masses with respect to those derived by \citet{Pacifici2015}.} We approximate the distributions of stellar and interstellar metallicities in a galaxy with a single metallicity $\Zism=\Z$.
We use an exponential prior for $\hat\tau_V$ and fix $\mu=0.4$.  The free parameters in the model fitting are summarized in Table~\ref{tab:BEAGLE_fitting_parameters}, and see Section~\ref{BEAGLE_general} for a general overview of the individual parameters.

\begin{table}

\centering
\caption{Parameters allowed to vary in the \beagle\ fitting to galaxies in the 3D-HST catalog with their priors.} 
\begin{tabular}{ C{0.24\columnwidth-2\tabcolsep} C{0.33\columnwidth-2\tabcolsep} C{0.43\columnwidth-2\tabcolsep} } 

\toprule

 \multicolumn{1}{c}{Parameter}	  & \multicolumn{1}{c}{Prior} & \multicolumn{1}{c}{Description} \\

\midrule

 $z$ & Uniform $\in [0,15]$ & redshift \\
 
 $\log(\M_{tot}/\Msun)^{a}$ & Uniform $\in [7,13]$ & Integrated SFH \\

 $\log(\t/\yr)$ & Gaussian	$\mathcal{N}(9.3;\,0.7)$ truncated $\in [7, 10.15]$  & Age of oldest stars in the galaxy \\

 $\log (\tausfr / \yr)$ & Uniform $\in [7,12]$ & Timescale of star formation \\

$\log(\Z / \Zsun)$  &Uniform $\in [-2.2,0.24]$ & Stellar (and interstellar) metallicity (\Z=\Zism)\\

 \tauV & Exponential exp(-\tauV) truncated $\in [0, 4]$ & $V$-band attenuation optical depth \\

\mud & Fixed 0.4 & Fraction of attenuation arising in the diffuse ISM \\

\logUs &  Uniform $\in [-4,-1]$  & Effective gas ionization parameter \\
 
\xid & Fixed 0.3 & Dust-to-metal mass ratio \\
 
\bottomrule
\end{tabular}
\raggedright $^a$\beagle\ samples over the integral of the past star formation history of the galaxy ($\M_{tot}$).  It returns the stellar mass (\Mstar), which accounts for the mass returned by evolved stars to the ISM.
\label{tab:BEAGLE_fitting_parameters}
\end{table}

\subsubsection{Parent SED catalog of galaxies based on 3D-HST catalog for $z\le4$, $\log (\Mstar/\Msun) > 8$}\label{highmassgal}

To generate our parent SED catalog for high mass galaxies at $z<4$, we use
fits to the broad-band photometry of
star-forming galaxies in the 3D-HST catalog using \beagle.
Since T14 use rest-frame $U-V$ vs $V-J$ colors to separate galaxies into star-forming and quiescent galaxies before measuring the type-dependent mass functions, we select star-forming galaxies from the 3D-HST catalog using a similar star-forming/quiescent (SF/Q) classification scheme.  Specifically, we select objects in the star-forming region of $U-V$ vs $V-J$ color space as defined by \cite{Whitaker2011} (see their Figure~17 and Equations 14 and 25) using the reported $U$, $V$ and $J$-band absolute rest-frame magnitudes supplied in the 3D-HST catalog.

As described in Section~\ref{BEAGLE_general}, \beagle\ uses \multinest\ \citep{Feroz2009} to sample the parameter space, and records the associated
SEDs and \multinest\ weights.
This information can be used to produce samples drawn from the corresponding
posterior probability distribution, and we use these samples to 
populate a parent catalog with a range of statistically acceptable SED fits (plus associated physical parameters) for each object.

At low masses, constraints on the rest-frame UV, metallicity and dust attenuation can suffer from poor photometric constraints.
For each of these parameters we therefore impose conditional priors on \MStarMuv, $\Mstar-\Z-\sfr$ and $\sfr-\Z-\tauV$.  These are additional priors to those already set in the SED-fitting (listed in Table~\ref{tab:BEAGLE_fitting_parameters}) that we must apply after the fact, as \beagle\ does not accept conditional priors.  Luckily it is relatively simple to apply these priors as one effectively needs to adjust the \multinest\ weights derived from the initial \beagle\ fitting.  This method was presented in \cite{Chevallard2017} (see their Section~2.3, where they describe drawing SEDs that match the observed main sequence of star-forming galaxies).
Essentially we perform weighted draws from the \multinest\ output for each galaxy fitted to in the 3D-HST catalog such that the probability of a galaxy entering into the parent catalog follows: 

\begin{equation}\label{eq:3dHSTsamplingPrior}
\begin{aligned}
\conditional{\Muv,\Z,\tauV}{\Mstar,\sfr,z}  \propto \\ 
\conditional{\Muv}{\Mstar,z} \, \conditional{\Z}{\Mstar,\sfr} \, \conditional{\tauV}{\Z,\sfr}\,\conditional{\thetab}{\Db}
\end{aligned}
\end{equation}

\noindent
where \conditional{\thetab}{\Db} is the posterior probability of a set of parameters sampled over in the fitting ($\thetab = [z$, $\M_{tot}$, $\t$, $\tausfr$, $\Z$, $\tauV$, $\mu$, $\logUs$, $\xid]$) 
given the data, $\Db$.
The shape of the prior in \MStarMuv\ is set by the observational 
constraints detailed in Section~\ref{MstarMuv}, 
and for $\conditional{\Z}{\Mstar,\sfr}$ and $\conditional{\tauV}{\Z,\sfr}$ we use the same priors imposed when constructing the parent catalog not based on 3D-HST photometry (as described in the following sub-section, specifically equations \ref{eq:SF_M_Z_SFR} - \ref{eq:students-t} describe the prior in $\conditional{\Z}{\Mstar,\sfr}$ and equations \ref{eq:tauVperp} - \ref{eq:tauV_AA_scatter} describe the prior in $\conditional{\tauV}{\Z,\sfr}$)
\footnote{Please note that, although we impose a prior on the $\Z-\logUs$ relation in the parent catalog \textit{not} produced using SED fits to galaxies in the 3D-HST catalog, we do not apply the same prior here.  }  
The prior on \MStarMuv\ ensures a smooth transition in observables between galaxies assigned from the 3D-HST catalog constraints
and those at lower masses or higher redshift that lie
outside the parameter space covered by the observed galaxies. 
For objects in the 3D-HST catalog with firm \Muv\ constraints the prior changes the sampling very little, allowing the final mock \MStarMuv\ relation to display the wide
diversity of SEDs
seen in the observed population.

\begin{figure}
\begin{center}
\includegraphics[width=0.5\textwidth]{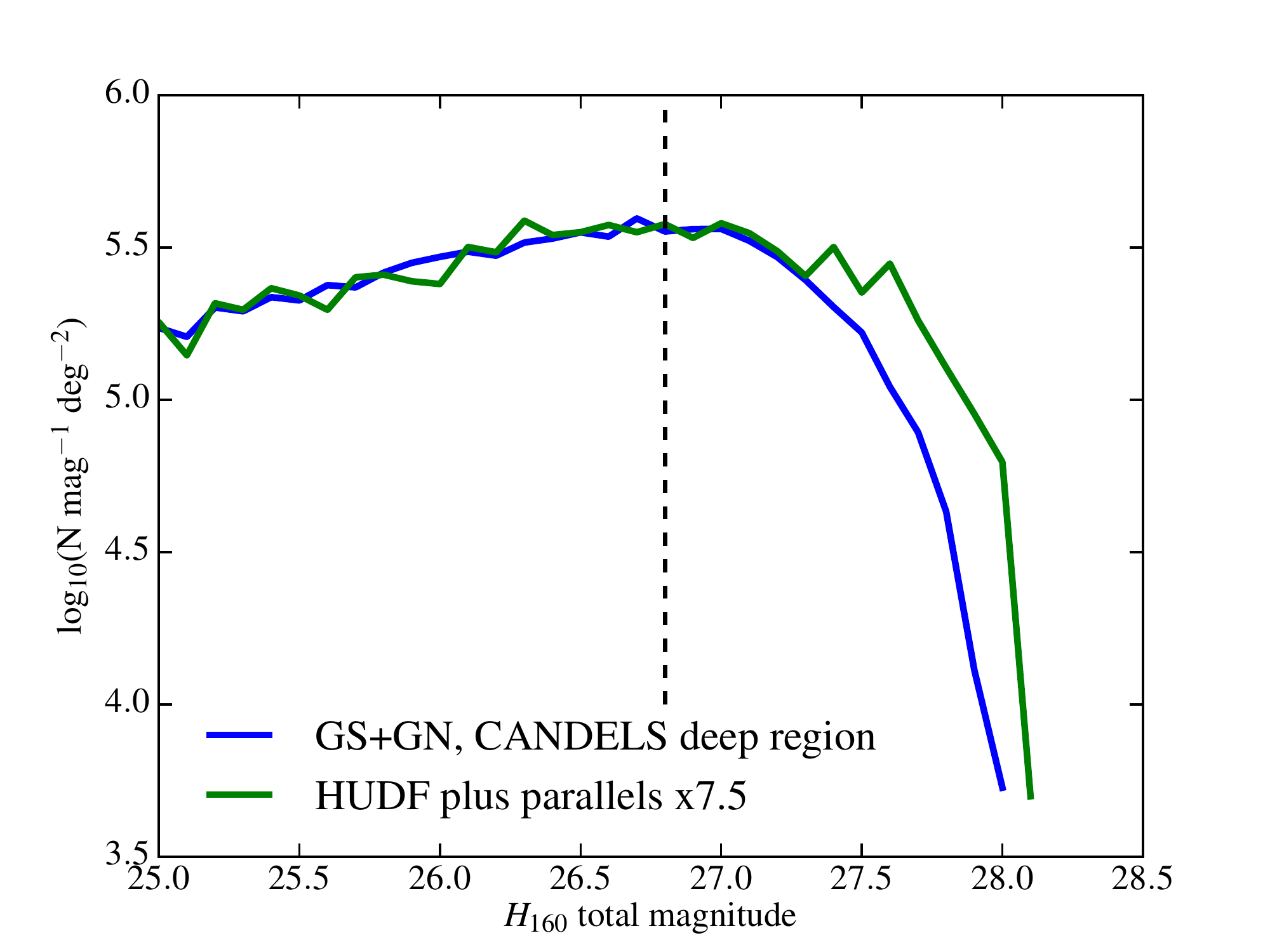}
\caption{Number density of galaxies as a function of total $H_{160}$ AB magnitude for the combined GOODS-S and GOODS-N CANDELS deep regions (blue line) and HUDF and parallel fields (green line).  The number counts for the HUDF and parallel fields have been scaled by a factor of 7.5 to match the number counts of the GOODS-S+GOODS-N region (see text for details). The vertical dashed line shows the conservative limit we assign such that objects brighter than this limit are only included in the parent catalog if they are in the GOODS-S+GOODS-N region, and objects fainter than this limit are from the HUDF and parallel fields only.}
\label{relativeDepths}
\end{center}
\end{figure}

When drawing SEDs from the fits to 3D-HST galaxies, we account for the varying
sensitivity limits of the adopted datasets.  
The HUDF field and parallels provide the faintest $H_{160}$ objects in the catalogs, while the sensitivity in CANDELS varies between the Deep and Wide regions.
To avoid favoring SEDs from the numerous $H_{160}$-bright objects found in shallow, wider survey areas, we limit the shallower area to that in the CANDELS deep regions within GOODS-N and GOODS-S.  Figure~\ref{relativeDepths} displays the residual difference in depths between the deeper HUDF and parallel fields and the CANDELS deep regions of GOODS-S and GOODS-N (plotted together).   The number counts at given total $H_{160}$ magnitude start to turn over at  $H_{160}\sim27$ for the GOODS-N+GOODS-S region, while the number counts for the HUDF and parallels remain flat to $H_{160}\sim27.5$.  We apply a conservative cut of $H_{160}=26.8$ above which, only objects in the deeper HUDF and parallels are included in the parent catalog and below which, only objects from GOODS-N+GOODS-S are included.  We also require that the transition in number counts across this magnitude limit is smooth, ensuring that the distribution of $H_{160}$ magnitudes at a given mass is not more heavily weighted by the more numerous objects in the shallower region.  The ratio between cumulative number counts at $H_{160}<26.7$ for the two regions is equal to 7.5, which essentially accounts for the difference in area between the two regions (this factor has already been applied to the number counts in the HUDF+parallels in the figure). 
We therefore draw 45 realizations per object from the HUDF+parallel regions, while only 5 random draws are included in the parent catalog for each object in the shallower region.

\subsubsection{Parent catalog for galaxies at $z\le4$ and with $\log(\Mstar/\Msun) < 8$, and for all $z>4$ galaxies}
\label{BEAGLE_SF_grid}

For mock galaxies outside of the stellar mass and redshift range of
3D-HST galaxies, we generate a parent catalog by 
using \beagle\ to produce SEDs covering a wide range of parameter values
(e.g., $z$, \Mstar, $\tau$, \tauV, \logUs, \Z, and \t). 
We impose constraints 
to avoid populating the parent catalog with objects in
unphysical regions of parameter space, such as
galaxies with high metallicity and high \logUs\ or high-mass galaxies with extremely small metallicities.
To constrain these parameters, we utilize the distributions of
$\Mstar-\Z-\sfr$ and $\Z-\logUs$ inferred from observations.

We use delayed SFHs to produce the SEDs, where the $\tau$ and \t\ values are chosen to ensure that the galaxies would be classified as star-forming, with $\log(\ssfr/\yrInv)>-10$.  Here we allow $\log(\t/\yr)$ to vary between 6 and the age of the Universe.  The lower limit in age is lower than that introduced in the prior on $\log(\t/\yr)$ used in the fitting to objects in the 3D-HST catalog (see Table~\ref{tab:BEAGLE_fitting_parameters}) as this parent catalog is going to be used to match to lower mass/higher redshift objects.
\beagle\ provides the stellar masses accounting for mass returned to
the ISM as stars evolve and die, as well as the current SFR which can be used to assign metallicity, ionisation parameter and V-band optical depth due to attenuation by dust.

To constrain the metallicities of galaxies, we use the fundamental
metallicity relation between $\Mstar-\Z-\sfr$ measured by \cite{Hunt2016} from a compilation of $\sim1000$ galaxies covering a wide range in \sfr\ and stellar mass, with oxygen abundance estimates derived from consistent calibrations, and, crucially, covering a wide range in redshifts up to $z\sim3.7$. The 
redshift distribution of \cite{Hunt2016}  can be viewed in 
their Figure~1, which can be compared with the
fundamental metallicity relation measured in \citet{Mannucci2010} from galaxies 
in the Sloan Digital Sky Survey for objects with redshifts between 0.07 and 0.3. 
A fit to this fundamental metallicity relation is given by

\begin{equation}\label{eq:SF_M_Z_SFR}
\begin{aligned}
\logOH = &-0.14\, \log(\sfr/\MsunyrInv) + \\
&0.37\, \LMstarMsun + 4.82.\\
\end{aligned}
\end{equation}

\noindent

This relation is a fit to the gas-phase oxygen abundance, while we require a relation based on the nebular metallicity $\Zism=\Z$ associated with our models. From the grid of $\xid=0.3$ models used here, we infer the approximate relation $\logOH \simeq \log(\Zism/\Zsun) + 8.7$. While this approximation is not suitable to determine accurately \logOH\ from the output metallicity for individual nebular models (see \citealt{Gutkin2016}, Table~2 for values of \logOH\ for different model metallicities), the errors it generates are much smaller than the scatter we introduce below in the fundamental metallicity relation.

It is important to highlight here that the \cite{Hunt2016} relation only models a linear dependence between oxygen abundance and stellar mass, whereas we know the mass-metallicity relation is \textit{not} linear at high stellar masses (e.g. \citealt{Tremonti2004}).  In fact the upper limit in metallicity of $\Z/\Zsun=0.24$ (also the upper limit of the prior in metallicity employed when fitting to objects in the 3DHST catalogs) introduces a turnover in the catalog mass-metallicity relation at low redshifts and high masses (see Section~\ref{MZR_comparison}).

We wish to instill a broad diversity in our parent catalog SEDs and avoid
an over-representation of unphysical parameters in our resulting mock catalogs.
We therefore apply a broad scatter to the fundamental metallicity
relation, and do not attempt to
predict the form of the $\Mstar-\Z$ relation or $\Mstar-\Z-\sfr$ plane to high redshifts. The broad range of spectral parameters will enable investigations 
of selection effects in future observations, especially for redshift and magnitude
regimes where current measurements cannot yet reach.
We characterize the scatter with a Student's-t distribution:

\begin{equation}\label{eq:students-t}
\begin{aligned}
f(x) = \frac{\Gamma(\frac{\nu+1}{2})}{\sqrt{\nu\pi}\Gamma(\frac{\nu}{2})}\Big(1+\frac{x^2}{\nu}\Big)^{-\frac{\nu+1}{2}}\,,
\end{aligned}
\end{equation}
where $\nu$ is the number of degrees of freedom and $\Gamma$ is the gamma function and:

\begin{equation}
\begin{aligned}
x = \frac{\log(\Zism/\Zsun) - \log(\Zbarism/\Zsun)}{\sigma_{x}}\,.
\end{aligned}
\end{equation}
We set $\sigma_x = 0.3$ and $\nu=3$,  where $\nu=3$ has been chosen to provide a distribution with more weight in the tails compared to a Gaussian.

To constrain the ionization parameter of these galaxies we use a linear fit between metallicity and \logUs\ measurements at low redshift from \citet[][see their figure~2]{Carton2017}: 
\begin{equation}\label{eq:SF_Z_logU}
\begin{aligned}
\logUs(\Z) = -0.8\, \log(\Z/\Zsun) - 3.58\,.
\end{aligned}
\end{equation}
We again use the Student's-t distribution
with 3 degrees of freedom to introduce scatter in this relation.

We account for dust attenuation by using an approach commonly featured
in semi-analytic models of galaxy formation \citep[e.g.][]{Guiderdoni1987, DeLucia2007,Fontanot2009}. Following \citet[][their equation~6]{Devriendt1999}, we estimate the $V$-band, face-on attenuation optical depth \tauVperp\ using the relation
\begin{equation}\label{eq:tauVperp}
\tauVperp = \left(\frac{\Zism}{\Zsun}\right)^{1.6} \, \left( \frac{N_\textsc{h}}{2.1 \times 10^{21} \, \txn{cm}^{-2}} \right ) \,
\end{equation}
where \Zism\ is the interstellar metallicity and $N_\textsc{h}$ the mean hydrogen  column density. As in \citet{Devriendt1999}, we compute $N_\textsc{h}$ from the (cold) gas fraction
\begin{equation}\label{eq:N_H}
N_\textsc{h} = 6.8 \times 10^{21} \, \frac{\Mgas}{\Mstar+\Mgas} \, ,
\end{equation}
where the cold gas mass \Mgas\ is computed by inverting the Schmidt-Kennicutt relation \citep{Schmidt1959, Kennicutt1998}. In practice, we consider the size and SFR of each galaxy to compute the SFR density $\SigmaSfr = \sfr/(\pi \, r^2)$, in units of $\Msun \, \yrInv \, \txn{pc}^{-2}$, where $r$ is the galaxy effective radius. We can then compute the cold gas surface density \SigmaGas\ from the Schmidt-Kennicutt relation, and estimate
the cold gas mass as $\Mgas=\SigmaGas \, \pi \, r^2$.

Equation~\eqref{eq:tauVperp} provides us with the face-on attenuation optical depth, from which we can derive an angle-averaged attenuation optical depth by assuming a spatial distribution of dust and stars. Following \citet{Devriendt1999}, we approximate our galaxies as oblate ellipsoids where dust and stars are homogeneously mixed. The $V$-band attenuation optical depth averaged over all galaxy inclinations $i$ can then be written as
\begin{equation}\label{eq:tauV_AA}
\tauVAA = -2.5 \, \log \left ( \frac{a_\textsc{v}}{1-\omega_\textsc{v} + \omega_\textsc{v}\,a_\textsc{v}} \right ) \, \frac{1}{1.086} \, ,
\end{equation}
where $\omega_\textsc{v}=0.87$ is the albedo at 5500 \AA\ for dust grains with properties as in the Small Magellanic Cloud \citep{Pei1992}, and $a_\textsc{v}$ is computed as 
\begin{equation*}
a_\textsc{v} = \frac{3}{4\,\tauVperpP} \, \left [ 1 - \frac{1}{2\,\hat{\tau}_{\textsc{v}\,\perp}^{\prime \, 2}} + \left( \frac{1}{\tauVperpP} + \frac{1}{2\,\hat{\tau}_{\textsc{v}\,\perp}^{\prime \, 2}} \right )\, \exp(-2\,\tauVperpP) \right ] \, ,
\end{equation*}
where $\hat{\tau}_{\textsc{v}\,\perp}^{\prime} = 2.62 \, \tauVperp$.

Equations~\eqref{eq:tauVperp}--\eqref{eq:tauV_AA} enable us to associate to each mock galaxy a physically-motivated value for the angle-averaged $V$-band attenuation optical depth, which depends on the galaxy SFR, size, and metallicity. We then account for the effect of galaxy inclination on the $V$-band attenuation optical depth \tauV\ by randomly drawing 
\tauV\ from a Gaussian distribution centered at \tauVAA\ and truncated at $\tauV=0$.  The width of the Gaussian used to draw the \tauV\ value is chosen to be dependent on \tauVAA\ and \Z, according to:
\begin{equation}\label{eq:tauV_AA_scatter}
\sigma = 0.2 + \Z/\Zsun - \tauVAA
\end{equation}
This function ensures that at low metallicities, there is a smaller scatter in \tauV, limiting range of dust attenuation in the regime where we expect low dust-to-gas ratios, while the minimum value of 0.2 prevents the values of \tauV\ being too constrained at the lowest metallicities.  We note that non-negligible attenuation by dust even at very low metallicities is not unreasonable, as even if they have low gas-to-dust ratios they may be gas-rich, allowing for non-negligible dust-to-stellar mass ratios (e.g. \citealt{DaCunha2010}). 
Our choice of the negative dependence of $\sigma$ on \tauVAA\ mimics results obtained from radiative transfer calculations of dust attenuation in galaxies \citep[e.g.][]{Tuffs2004, Pierini2004}, which show that galaxies with low angle-averaged attenuation optical depths $\tauVAA \lesssim 0.1$ exhibit a larger fractional range of inclination-dependent attenuations than galaxies with larger \tauVAA.

\subsubsection{Matching mock galaxies to the parent catalog}
\label{section:matching_sf_mock_to_parent_cat}

To assign SEDs from the parent catalog to mock galaxies, we find the closest
match between the mock galaxy parameters and the physical
parameters of the parent catalog galaxies.
Which properties are matched differ when assigning SEDs drawn from
fits to 3D-HST galaxies or from the wide grid of galaxies produced by \beagle.  
For those objects with $z<4$ and stellar mass higher than $\log(\M/\Msun) = 8$ or the mass completeness limit of the 3D-HST catalog ($\log(\M/\Msun) > 0.7z+0.63$)\footnote{The approximate mass completeness limits are estimated in bins of redshift and fit with a linear relation. Specifically, we randomly sample 100 SEDs from the posterior probability distribution of each galaxy in the HUDF portion of the 3D-HST catalog.   These SEDs are binned in redshift and mass and the completeness limit calculated as the mass at which 95\% of the SEDs are brighter than 27.6 in $H_{160}$ (the magnitude at which the number counts in the UDF portion of the field start to turn over, see Figure \ref{relativeDepths}). }, red vertical lines shown in figure \ref{MstarMuvdist}), we find the closest match in \Mstar\ and redshift, as we rely on observed broad-band photometry to constrain the expected SED shape of an object at a given stellar mass.  
For objects at $z>4$ or with lower stellar masses at $z\le4$, the expected SED shape is based on extrapolations of the \MStarMuv\ and \MuvBeta\ relations, as described in Sections~\ref{MstarMuv} and \ref{betamuv}, respectively.  These relations are used to assign \Muv\ and $\beta$ values to each galaxy in the star-forming galaxy mock catalog.  Each mock object is then assigned an SED based on the closest match in redshift, stellar mass, \Muv\ and UV-slope in the parent catalog.

\begin{figure*}
\begin{center}
\includegraphics[width=.9\textwidth]{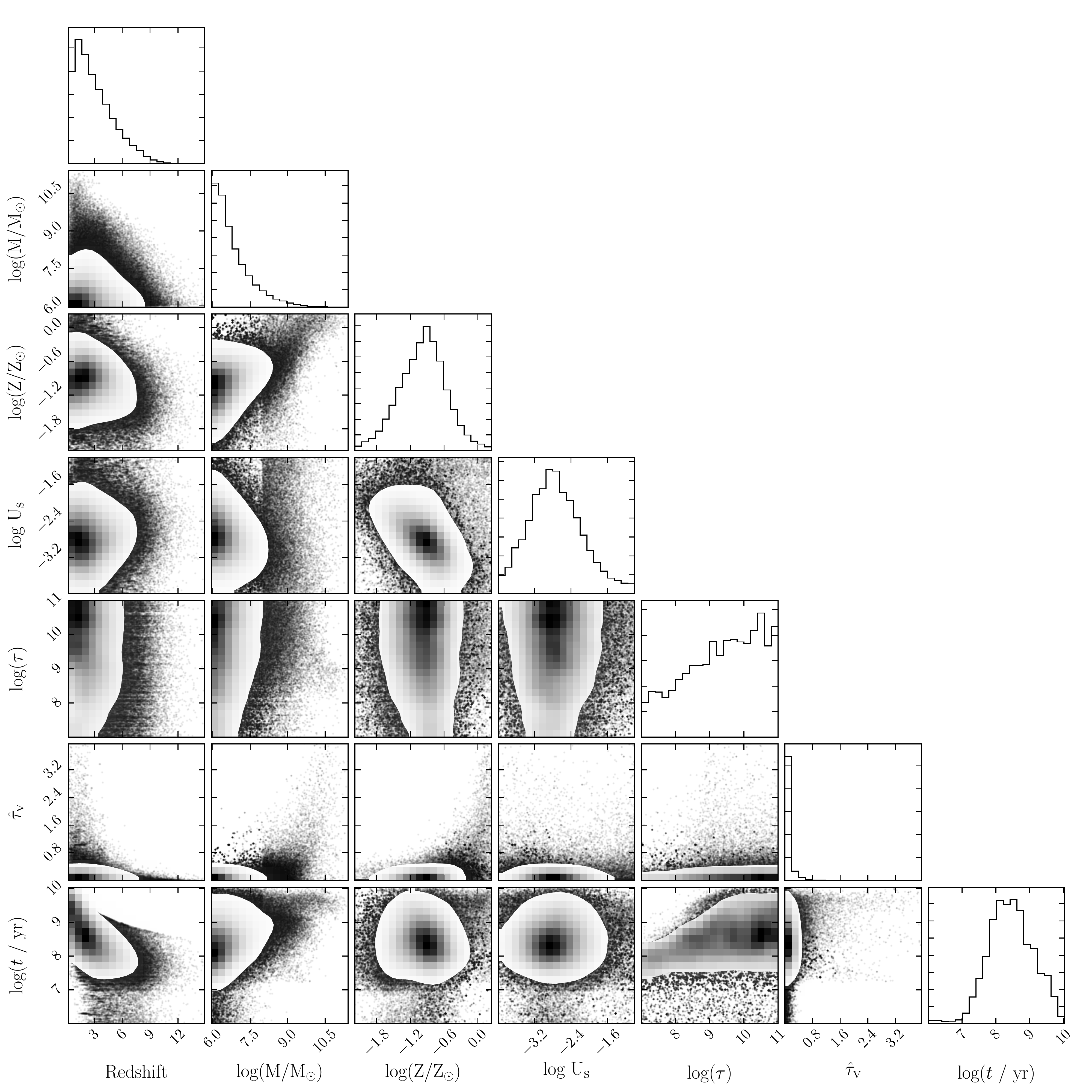}
\caption{The parameters assigned to all star-forming galaxies in the mock catalog.  Where the point density becomes high, a two-dimensional histogram indicates the density of points by the level of shading.  The plot was made using the python package \textit{corner} \citep{corner}}. 
\label{fig:physicalParametersMock}
\end{center}
\end{figure*}

For all matches between mock galaxies and parent catalog it is then possible to shift the redshift of the SED to the exact redshift of the mock galaxy.  Figure~\ref{fig:physicalParametersMock} displays the distributions of physical parameters assigned to all star-forming galaxies in the mock catalog.
      
\section{Generating Quiescent galaxies across cosmic time}\label{QG_MF}

\subsection{Quiescent galaxy counts}

\begin{figure*}
\begin{center}
\includegraphics[width=0.7\textwidth,trim=0 150 0 0,clip]{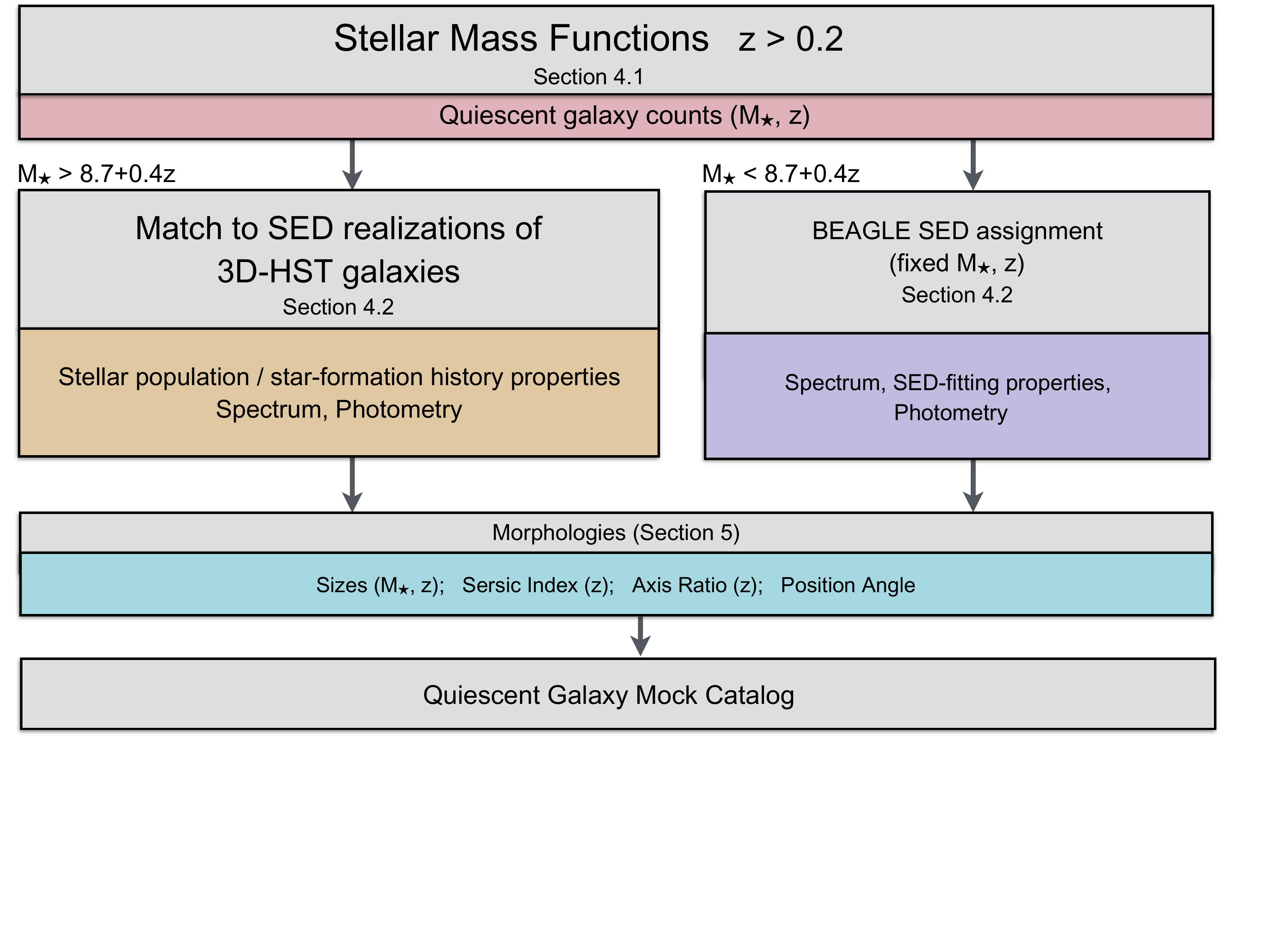}
\caption{Diagram summarizing the procedures for generating the quiescent galaxies. \Mstar\ is defined as \LMstarMsun. Gray boxes indicate the empirical relationships, distributions, or data on which mock galaxy properties are based, and colored boxes indicate the mock galaxy property which is generated in that step. All quiescent mock galaxies are generated following these procedures.  }
\label{flowchart_qg}
\end{center}
\end{figure*}

Our model for the redshift evolution of the stellar mass function of
quiescent galaxies is based on observed stellar mass functions that have been measured in the redshift range $0.2\le z \le3.0$ by T14. These authors use the redshift-dependent $UVJ$ color selection from \citet{Whitaker2011} to select quiescent galaxies from the ZFOURGE medium-band photometric survey \citep{Straatman2016} and data from the CANDELS survey (see Section~\ref{evolvemassfn}). T14 find that the quiescent galaxy stellar mass function is best fitted by a
double-Schechter function (equation~\ref{eq:DoubleSchechter} above) in the redshift range $0.2 \le z \le 1.5$, and by a single-Schechter function at higher redshifts, $1.5 \le z \le 3.0$. The differing functional forms were chosen to match the observed upturn in the stellar mass function below $\LMstarMsun \le 9.5$ at $z\le1$, a result in line with observations by \citet{Santini2012, Muzzin2013mf, Ilbert2013}.

In this work, we use both the observed binned stellar mass functions and the fitted Schechter parameters in bins of redshift from T14 to construct a continuous model for the redshift evolution of the quiescent stellar mass function. 
To produce smooth evolution at all redshifts, 
we choose to adopt the double-Schechter function description for the mass function at all redshifts, even at $z>1.5$ where observations find consistency with
a single Schechter function. 
This double-Schechter function has five parameters, $\MFMstar$, $\alphaOne$, $\phiStarOne$, $\alphaTwo$, and $\phiStarTwo$. We fit the double-Schechter parameters from T14 at $z < 1.5$, but substituted the single-Schechter parameters ($M^*$, $\alpha$, and $\phi^*$) for the double-Schechter parameters ($\MFMstar$, $\alphaOne$, $\phiStarOne$) at $z > 1.5$.  In a double-Schechter function, $\alphaTwo$ and $\phiStarTwo$ control the slope and overall normalization at low-masses.

We choose to fix \MFMstar\
to a value of $\LMstarMsun \sim 10.6$ at all redshifts, owing to the lack of significant shift in the observed evolution in \MFMstar\ 
with redshift from the T14 observations, and we fit the $\alphaOne$ and $\phiStarOne$ evolution with a cubic function with redshift. We fit a linear function to the $\alphaTwo$ evolution and a quadratic function to the $\phiStarTwo$ evolution. 

Given the chosen forms for the evolution in the parameters, in order to prevent the cubic functions from diverging at low-redshift and high-redshift 
we stop the evolution for some of the parameters. For $\alphaOne$, $\phiStarOne$, $\alphaTwo$, and $\phiStarTwo$ we stop the redshift evolution at $z < 0.5$, $z < 0.75$, $z < 0.5$, and $z < 0.5$, respectively. 
We require $\phiStarOne$ to stay constant below $z < 0.75$ because the cubic fit
to the T14 Schechter parameters would otherwise over-predict by an order of magnitude the number of objects as compared to observations at low redshift. We fix $\alphaOne$ at $z > 1.75$ to the value of the cubic fit at that redshift, and at $z > 1.88$ we impose a linear decline on the evolution of $\phiStarOne$ to extrapolate the quiescent galaxy counts to $z>3$, beyond the range probed by the observations. 
This projected decline in $\phiStarOne$ predicts a number density of 2.16$\pm$ 0.8 $\times 10^{-5}$ Mpc$^{-3}$  quiescent galaxies at $3.4\le z\le4.2$, consistent with the number density of quiescent galaxies identified in \citet[][1.8$\pm$0.8$\times$10$^{-5}$ Mpc$^{-3}$]{Straatman2014, Straatman2015} to their stellar mass limit of $\LMstarMsun \sim 10.6$ and the ZFOURGE survey area of $\sim$363 arcmin$^2$. Thus our extrapolation is in broad agreement with the few constraints available on counts of quiescent galaxies at $z>3.5$.

\begin{table*}
\caption{Quiescent Mass Function double-Schechter parameters (Figure~\ref{fig:quiescentwithTomczak}) and their evolution.   The parameter is shown on the left column, the functional form of the parameter (with associated constants) on the right column,  and the redshift-range of that functional form in the middle column.  }
\label{tab:quiescentwithtomczak} 
\begin{center}

\renewcommand{\arraystretch}{1.1}
\begin{tabular}{|l|l|l|}
\hline

\textbf{Parameter}  & \textbf{Redshift range} & \textbf{Functional Form} \\ 
\hline \hline

\MFMstar & $z\geq0.2$ & $\MFMstar = 10.617$ \\ \hline
\multirow{4}{*}{\alphaOne} & $z<0.5$ & $\alphaOne = -0.225$\\ \cline{2-3}
                    & \multirow{2}{*} {$0.5\leq{z}<1.75$} &  $\alphaOne=b_0\times z^3+b_1\times z^2+b_2\times z+b_3$\\ 
                    &                               & $b_0 =0.43 $, $b_1=-2.33$, $b_2=3.49$, $b_3=-1.44$\\ \cline{2-3}
                    & $z\geq1.75$ & $\alphaOne=-0.150$\\ \hline
\multirow{5}{*}{\phiStarOne} & $z<0.75$ & $\log(\phiStarOne)=-2.67$\\ \cline{2-3}
                       &\multirow{2}{*} {$0.75\leq z<1.877$} & $\log(\phiStarOne)=c_0\times z^3+c_1\times z^2+c_2\times z + c_3$ \\ 
                       &                                   & $c_0=-0.35$, $c_1=1.92$, $c_2 = -3.77$, $c_3=-0.78$\\ \cline{2-3}
                       & \multirow{2}{*} {$z\geq1.877$} &  $\log(\phiStarOne)=c_4\times z+c_5$ \\
                       &                     & $c_4=-=0.43$, $c_5=-2.59$ \\ \hline
\multirow{3}{*}{\alphaTwo} & $z<0.5$ & $\alphaTwo=-1.83$ \\ \cline{2-3}
                    & \multirow{2}{*} {$z\geq0.5$} & $\alphaTwo = d_0\times z + d_1$ \\ 
                    &                & $d_0=1.15$, $d_1=-2.41$  \\ \hline
\multirow{3}{*}{\phiStarTwo} & $z<0.5$ & $\log(\phiStarTwo)=-4.71$ \\ \cline{2-3}
                        &\multirow{2}{*} { $z\geq0.5$} & $\log(\phiStarTwo)=e_0 \times z^2 + e_1 \times z + e_2$ \\ 
                        &                 & $e_0=-0.59$, $e_1=1.93$, $e_2=-5.52$ \\ \hline

\end{tabular}

\renewcommand{\arraystretch}{1}
\end{center}
\end{table*}

\begin{figure}
\includegraphics[width=0.5\textwidth]{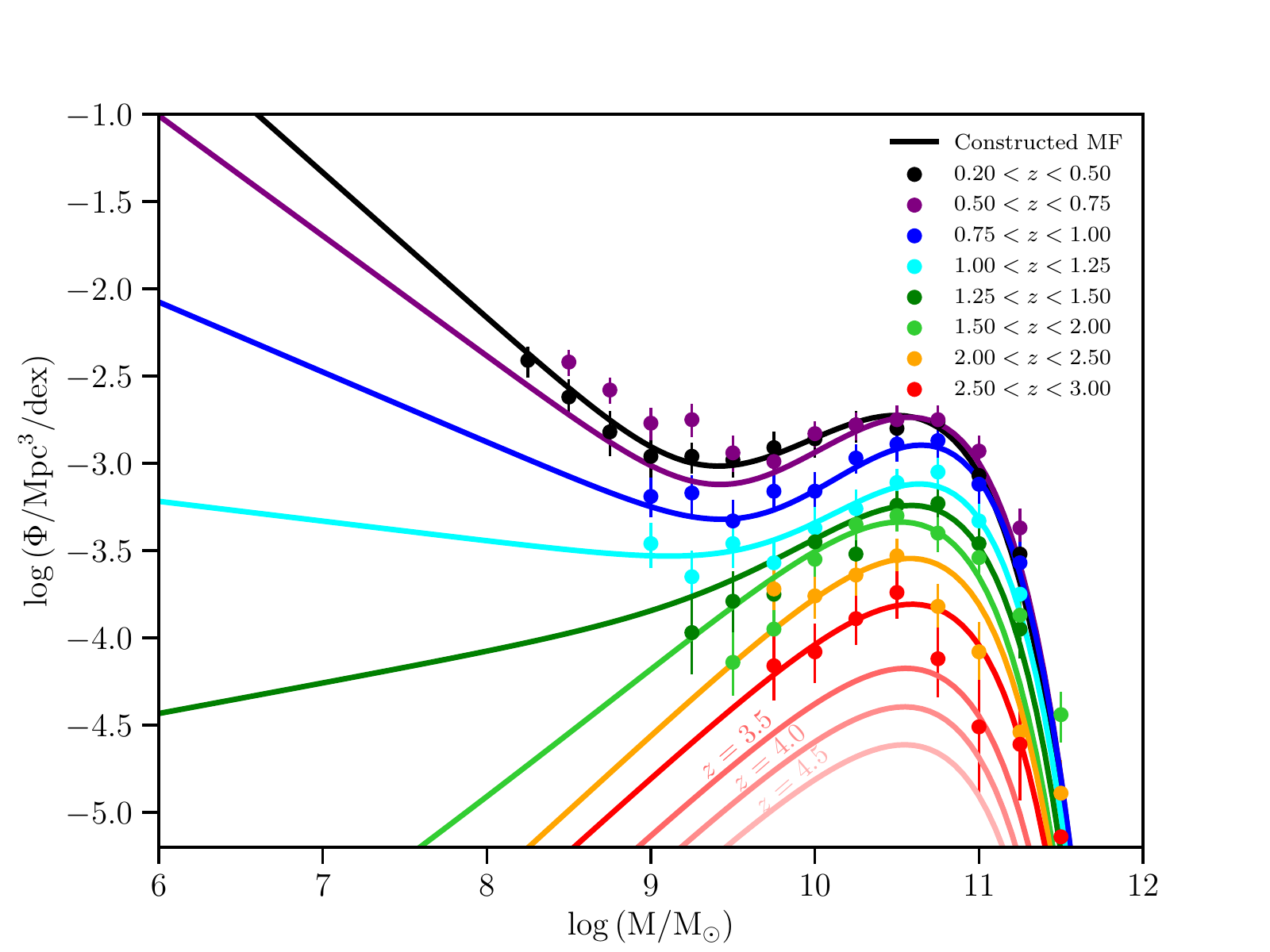}
\caption {Evolution of the quiescent galaxy stellar mass function (dashed), plotted with the observations from T14 (circles). The parameters of this fit to the MF evolution are given in Table~\ref{tab:quiescentwithtomczak}.  }
\label{fig:quiescentwithTomczak}
\end{figure}

The resulting quiescent galaxy stellar mass function is compared with
the T14 data in Figure~\ref{fig:quiescentwithTomczak}, and we provide the detailed functional fit parameters in Table~\ref{tab:quiescentwithtomczak}.  
Our model for the stellar mass function evolution broadly agrees within the uncertainties of the binned stellar mass function observations from T14. We note that the observations imply more low-mass [$\LMstarMsun \le 9.5$] quiescent galaxies at $0.5 \le z \le 0.75$ than below $z\le0.5$. The continuously varying stellar mass function evolution model we describe here monotonically increases in counts with decreasing redshift,  
and therefore it does not replicate this rise and fall of the low-mass end that is seen in the observations (Figure~\ref{fig:quiescentwithTomczak}).

\subsection{Quiescent galaxy SEDs}
\label{Q_SEDs}

We apply the same SED assignment technique for quiescent galaxies as that for star-forming galaxies, described in  Section~\ref{section:SF_SED_parent_catalog}, whereby we produce a large parent catalog of SEDs and associated physical parameters that can then be used to associate an SED with each mock galaxy. Where possible, we use observed galaxies to guide the allowed range of mock galaxy  SEDs at a given stellar mass. Outside of the parameter space covered by 3D-HST objects we produce theoretical mock SEDs using \beagle.

To produce a parent catalog from observed objects, we use \beagle\ fits to 3D-HST galaxies classified as quiescent using the rest-frame $U$, $V$ and $J$ band absolute magnitudes and the \cite{Whitaker2011} $U-V$ vs. $V-J$ color criteria.  Following the method described in Section~\ref{highmassgal}, we draw SEDs (and associated physical parameters) from the \beagle\ fits to produce the parent catalog.   Specifically, the stellar mass and associated 68\% central credible interval are estimated for each galaxy, and the SEDs are randomly drawn from the BEAGLE output files if their stellar mass lies within this interval, without further weighting based on other physical parameters.

Uncertainties associated with the star-forming/quiescent separation have to be dealt with carefully when producing this parent catalog.  More details are given in Appendix \ref{app:robust_Q_determination}, where we define a redshift-dependent $H_{160}$ magnitude cut of $H_{160} < 24.5+z$, brighter than which quiescent objects from the 3D-HST catalog can be used to provide SEDs for the parent catalog.  This limit lies well above the sensitivity limit of the CANDELS deep region of the 3D-HST catalog (and so we do not encounter the situation we found for the star-forming galaxies where the faintest objects were only drawn from the smaller area HUDF portion of the catalog).   
We therefore skip the areal correction 
described in Section~\ref{highmassgal}, and populate the 3D-HST parent catalog of quiescent galaxies with 5 random draws from the \beagle\ fit output for each quiescent galaxy in the 3D-HST catalog satisfying the $H_{160}<24.5+z$ criterion. 
The $H_{160}$ limit translates to an approximate mass limit of $\log(\M/\Msun) > 8.7+0.4z$.  Thus only objects in the mock catalog with $\LMstarMsun > 8.7+0.4z$ are paired with the closest match in redshift and stellar mass among parent catalog galaxies.

For objects with $\LMstarMsun < 8.7+0.4z$, 
we produce a parent catalog of SEDs and associated physical parameters using \beagle.  For quiescent galaxies, we do not need to assign nebular \HII-region parameters (e.g. \xid, \logUs), and we neglect dust attenuation. We therefore vary only the galaxy age \t, star formation timescale \tausfr, and metallicity \Z\ to generate SEDs for the parent galaxy catalog.  Using a delayed SFH (Section~\ref{section:3DHST_SEDfitting}) with $\log(\tausfr) < 1.11\times\log(\t)-2.02$ ensures that the specific star formation rate of objects be less than $\log(\ssfr/{\rm yr}^{-1})=-10$.  The parameters \t\ and \tausfr\ are assigned to each mock catalog galaxy from uniform distributions.
The parameter \t\ is
allowed to vary between 30\,Myr and the age of the Universe at the redshift of the object. We allow \tausfr\ to vary 
between 10\,Myr and the maximum value required to produce $\log(\ssfr/{\rm yr}^{-1})<-10$.  

Measuring the stellar metallicities of quiescent galaxies and their evolution is technically challenging, requiring deep rest-frame optical spectra to measure stellar absorption-line indices.  
({\it JWST} will provide new opportunities to probe stellar metallicities of quiescent galaxies at high-redshift). However, with existing data,  most stellar metallicity measurements for quiescent galaxies exist only for high mass field galaxies ($\LMstarMsun\gtrsim9.5$) to moderate redshifts (e.g. \citealt{Gallazzi2006}, $z\sim0.1$ and \citealt{Gallazzi2014}, $z\sim0.7$) or cluster galaxies (e.g. \citealt{Sanchez-Blazquez2009}, \citealt{Jorgensen2013}).   The highest mass galaxies in our mock have their physical properties assigned from SED fits to 3D-HST galaxies, and so, with the lack of current constraints on the stellar metallicities of low-mass and high-redshift quiescent galaxies,  
we assign metallicities with a uniform weight between the limits of our templates ($-2.2<\log(\Z/\Z_{\odot})<0.24$).

After each galaxy in the mock catalog is assigned \t, \tausfr, and \Z\ values,
\beagle\ computes the fraction of mass $1-\Mstar/\M_{tot}$ 
returned by evolved stars to the ISM, and hence the scaling required to generate an SED with the corresponding stellar mass assigned to the mock catalog object.  \beagle\ then generates
the SEDs in ``mock mode'' using the assigned parameters.
As with the star-forming catalog, the SEDs assigned to each of the quiescent mock catalog galaxies are used to generate NIRCam filter fluxes.

\section{Galaxy morphologies from $0.2\le z\le 15$}\label{morphs}

The evolution of galaxy morphologies with cosmic time represents one of many key insights into the galaxy formation process that \JWST\ will provide.
Perhaps more practically,
galaxy shapes and light distributions affect detectability and measurability of other galaxy properties, and fully understanding these systematics in future {\it JWST} data will be important for characterizing uncertainties.
Anticipated applications of this mock catalog include {\it JWST} NIRCam image and NIRSpec spectroscopic simulations, as well as NIRSpec MSA slit assignment. Therefore, we assign simple morphologies to mock galaxies to enable these types of analyses. All mock galaxies are modeled as simple Sersic profiles \citep{Sersic1968}, and follow the redshift evolution of the relevant morphological parameters that has been characterized using deep extragalactic surveys with {\it HST}.  In the following sections, we describe our method for producing continuous evolutionary models and realistic distributions for galaxy sizes, shapes, and light profiles. 
Below, we describe the procedure for assigning half-light radii to both star-forming and quiescent galaxies at $z\le4$ where WFC3 has provided accurate rest-frame optical morphologies, and at $z > 4$ where HST has
characterized rest-frame UV morphologies,
and shapes, light profiles, and orientations for star-forming galaxies.

\subsection{Galaxy Sizes at $z<4$}

\begin{figure*}
\begin{center}
\includegraphics[width=0.48\textwidth]{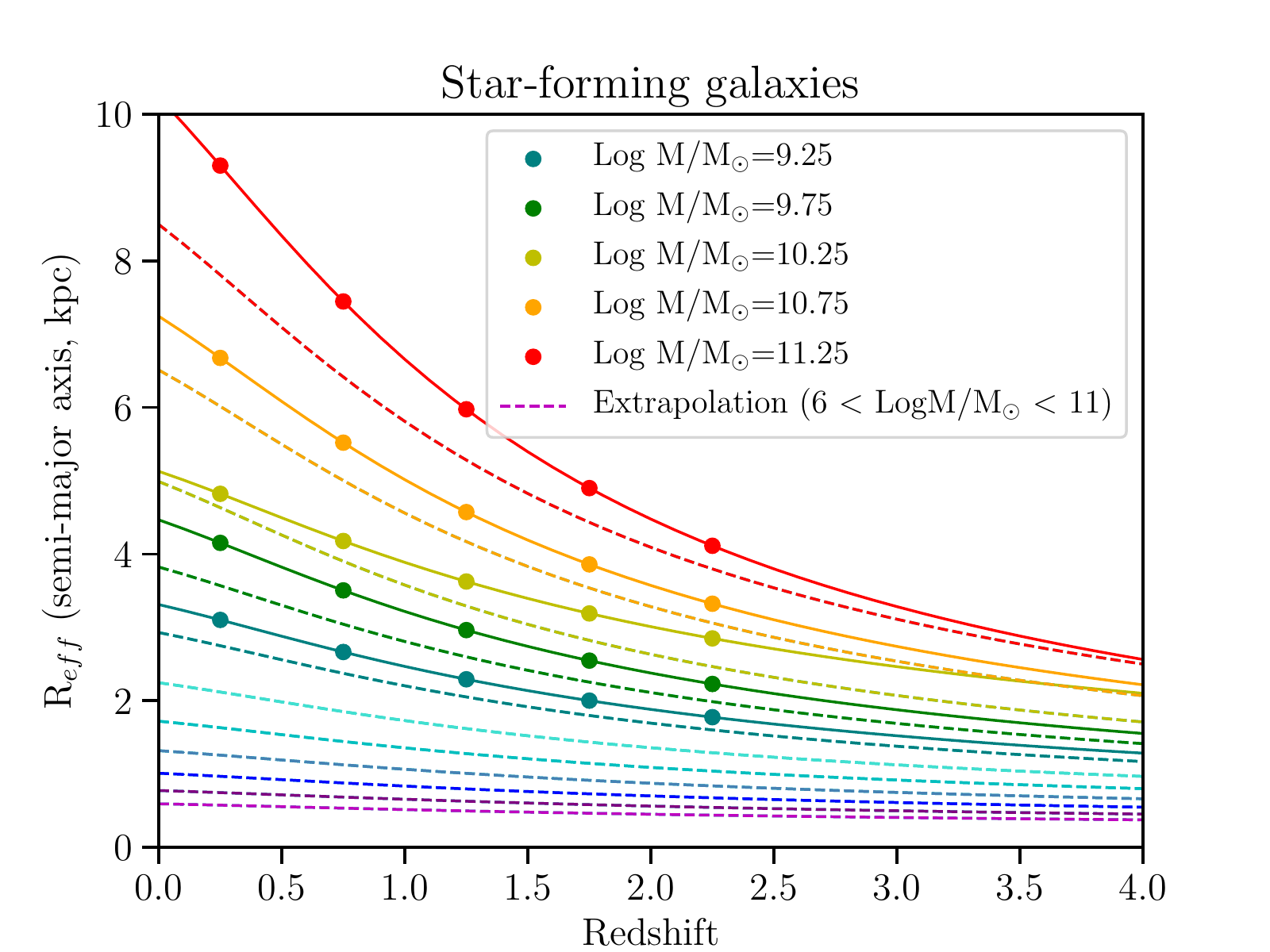}
\includegraphics[width=0.48\textwidth]{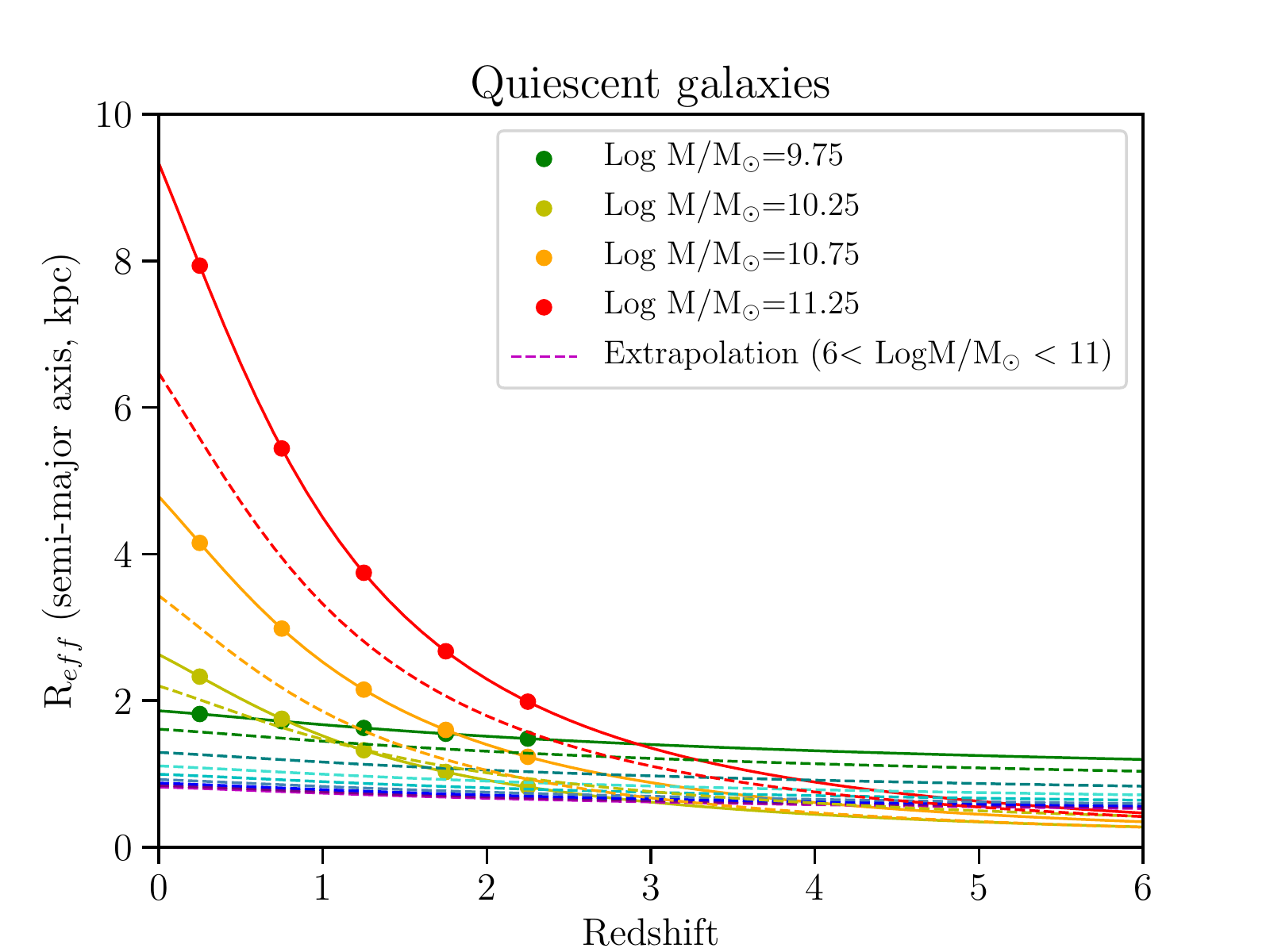}
\caption{Model describing the variation of average \Reffmaj\ for star-forming galaxies (left) and quiescent galaxies (right) as a function of both stellar mass and redshift. Points with solid lines indicate the observations in stellar mass bins and best-fit redshift evolution as published in \citet{vanderWel2014}. Dashed lines illustrate the behavior of our continuously varying model with stellar mass, as defined in equations \ref{eqn_sfg_rmaj} and \ref{eqn_etg_rmaj}. Dashed lines go from $\LMstarMsun=6$ (magenta) to $\LMstarMsun=11$ (red) in increments of $\Delta \LMstarMsun=0.5$.}
\end{center}
\label{lowz_Rmaj_evol}
\end{figure*}

We aim to generate a continuously evolving model in stellar mass, UV luminosity, 
and redshift, using the observed size-mass relationships that have been measured for 
galaxies at $0\le z\le4$. For this purpose, we use the relationships measured in \citet{vanderWel2014} for both star-forming and quiescent galaxies 
using CANDELS data, and extrapolate their behavior down to \LMstarMsun$\sim$6. 
These relationships have also been shown to agree with 
the measured size-mass relation of local galaxies in SDSS \citep{Shen2003, Guo2009}.

\citet{vanderWel2014} parametrize the redshift evolution of the half-light semi-major radius in kpc, \Reffmaj, as a power-law function of the Hubble parameter at a given redshift, H(z), in bins of stellar mass. The parametrization has the form
\begin{equation}
 R_{\textrm{eff,maj}}(z) = B_{H} \Big{(}\frac{H(z)}{H_{0}}\Big{)}^{\beta_{H}} 
 \end{equation}
 where both $B_{H}$ and $\beta_{H}$ (respectively, the coefficient and power-law slope of the redshift evolution)  vary with stellar mass. To generate a smoothly evolving model, we generalize this evolution of $\Reffmaj(z)$ between stellar mass bins by fitting both $B_{H}$ and $\beta_{H}$ as functions of stellar mass, to produce one smooth function in both redshift and stellar mass.

The behavior of $B_{H}$ and $\beta_{H}$ with stellar mass differs between star-forming galaxies and quiescent, and therefore we model the stellar mass dependence differently between the two samples. For star-forming galaxies, both $B_{H}$ and $\beta_{H}$ appear linear with $\Mstar = \LMstarMsun$. The best-fitting linear relationships are 

\begin{equation}\label{eqn_sfg_rmaj}
\begin{aligned}
B_{H}(\Mstar)=&\,\,0.23\Mstar-1.61\\ 
\beta_{H}(\Mstar)=&-0.08\Mstar+0.25,
\end{aligned}
\end{equation}
\noindent
and the shape of the resulting function $\Reffmaj(z,\Mstar)$ for star-forming galaxies is shown in the left panel of Figure~\ref{lowz_Rmaj_evol}. We extrapolate the relationship out to $z\sim4$, and assign sizes to all star-forming galaxies in this redshift range using this method.

For quiescent galaxies, the behavior of both $B_{H}$ and $\beta_{H}$ with stellar mass are not linear. We find that $B_{H}$ (\Mstar) is well described by either a quadratic or exponentially declining function of mass, 
however we choose to parametrize $B_{H}$ using the exponential to avoid the undesirable quadratic feature that galaxy size increases unphysically to low masses.
We find that $\beta_{H}$ is well fit by an exponentially increasing function with decreasing stellar mass; however, increasingly large values of this power-law exponent at low masses produce unphysical size evolution at low mass. Therefore, we fix the value of $\beta_{H}$ below $\LMstarMsun<9.75$.  The relationships we use for quiescent galaxies are:
\begin{equation}\label{eqn_etg_rmaj}
	\begin{aligned}
B_{H}(\Mstar)=& 3.8\mathrm{e}{-4} e^{0.71\scMstar}-0.11 &(\textrm{at all masses})\\
	\beta_{H}(\Mstar) =& 1.38\mathrm{e}{12} e^{-2.87\scMstar} - 1.21   &(\Mstar \geq 9.75)\\
	\beta	_{H}(\Mstar) =& -0.19     &(\Mstar < 9.75).
	\end{aligned}
\end{equation} 
\noindent

The resulting function $\Reffmaj(z,\Mstar)$ for quiescent galaxies is shown in Figure~\ref{lowz_Rmaj_evol}.
The significant size evolution among massive quiescent galaxies (\LMstarMsun$\gtrsim$10) is in agreement with other studies \citep[e.g.][]{Cassata2013}. The flattening of the quiescent galaxy size-mass relation evolution at lower stellar mass is consistent with the expectations of environmental effects due to satellite quenching \citep[e.g.][]{vanderWel2010, Kawinwanichakij2017} and the observations that quiescent galaxies in high-density regions typically have larger sizes \citep[e.g.][]{Cooper2012, Delaye2014}.

As we have outlined in Section~\ref{QG_MF}, our model extrapolates the evolution of the stellar mass function for quiescent galaxies at $z>4$, and therefore mock catalogs will contain samples of quiescent galaxies at redshifts beyond where current surveys can identify them or measure their morphologies. Although the expected number of quiescent galaxies at $z>4$ in mock surveys would be small, we estimate that mock catalogs of comparable area to one GOODS field ($\sim$150 arcmin$^{2}$) will contain quiescent galaxies out to $z\sim$6.
Therefore we note here that sizes for such objects come from an extrapolation of the relationship presented in Equation \ref{eqn_etg_rmaj}  
which can be calculated for arbitrarily large redshift. We show the extrapolation out to $z\sim6$ in the right panel of Figure~\ref{lowz_Rmaj_evol}. 

The data presented in \citet{vanderWel2014} indicate that the scatter in galaxy sizes within quiescent and star-forming galaxy samples is approximately uniform in both redshift and stellar mass. The size distributions within redshift and mass bins for quiescent galaxies are log-normal with $\sigma=0.16$ dex, 
whereas the sizes of star-forming galaxies follow a skewed distribution better fit by a Gauss-Hermite polynomial expansion in $\Logten \Reffmaj$  \citep{vanderMarelFranx1993} with fixed dispersion of 0.18 dex and skewness $h3 = -0.15$. The skewed distribution is motivated by the observation that the star-forming galaxy size distribution has a tail of small-sized galaxies \citep[e.g.][]{Williams2014,Williams2015,vanDokkum2015}. The distribution has the form 

\begin{equation}
\begin{aligned}
P(\Rmock|M*,z) \propto&\, e^{-\frac{1}{2}y^{2}(1+h3(\frac{1}{\sqrt{6}}(2\sqrt{2}y^{3}-3\sqrt(2)y)}\\
 y=& \frac{1}{\sigma} (\Logten \Rmock - \Logten\Reffmaj).
\end{aligned}
 \end{equation}
\noindent

To assign sizes to mock galaxies of each type, we draw a size \Rmock\ from random distributions of these forms, where \Reffmaj\ is the median size in kpc 
for each mock galaxy's redshift and stellar mass from equations \ref{eqn_sfg_rmaj} and \ref{eqn_etg_rmaj} (Figure~\ref{lowz_Rmaj_evol}).

\subsection{Star-forming galaxy sizes at $z>4$}

At $z>3.5$ the WFC3 $H$-band moves out of the rest-frame optical.
At higher redshifts, the current best imaging data set for morphological measures, the CANDELS $H$-band imaging, will then become a probe
of the rest-frame UV morphology of high-redshift galaxies. 
For mock star-forming galaxies at $z>4$ we therefore assign morphologies according to the redshift evolution of the observed M$_{UV}$-size relationships presented in \citep{Shibuya2015}. (As mentioned in the last section, quiescent galaxies at $z>4$ follow the extrapolated relation presented in Equation \ref{eqn_etg_rmaj}). The size evolution in that work is parametrized by circularized half-light radius, defined as  
\begin{equation}
\Reffcirc = \Reffmaj\sqrt{b/a} 
\end{equation}
\noindent
where $b/a$ is the axis ratio (ratio of the semi-minor to semi-major axis size). \citet{Shibuya2015} find that median \Reffcirc\ of galaxies at $z>4$ is correlated with UV luminosity, and at fixed UV luminosity, galaxy size significantly decreases with increasing redshift. Sizes at $z>4$ in \citet{Shibuya2015} are measured from imaging at $1500<\lambda_{\textrm{rest}}<3000$\AA.
We generate sizes according to their size parameterizations with UV luminosity:
\begin{equation}
\Reffcirc = r_{o}\Big{(}\frac{L_{UV}}{L_{o}}\Big{)}^{0.27}
\end{equation}
\noindent
 where $r_{o}=6.9(1+z)^{-1.2}$ represents the effective radius in kpc  at a characteristic UV luminosity L$_{o}$ (corresponding to $\Muv = -21$). This relation is generally consistent with other measurements at $z\sim6-7$ \citep{Ono2013, Grazian2012, Oesch2010, CurtisLake2016, Bowler2017, Kawamata2015,Holwerda2015}. 
 The observed distribution of sizes is log-normal at all redshifts, with a mean size according to the above parametrization and a constant scatter
 with redshift and luminosity, $\sigma_{ln \sc{R}_\txn{eff,circ}}$= 0.5.

\subsection{Axis Ratios, Sersic Indices, and position angles across cosmic time}

\begin{figure*}
\begin{center}
\includegraphics[width=0.9\textwidth]{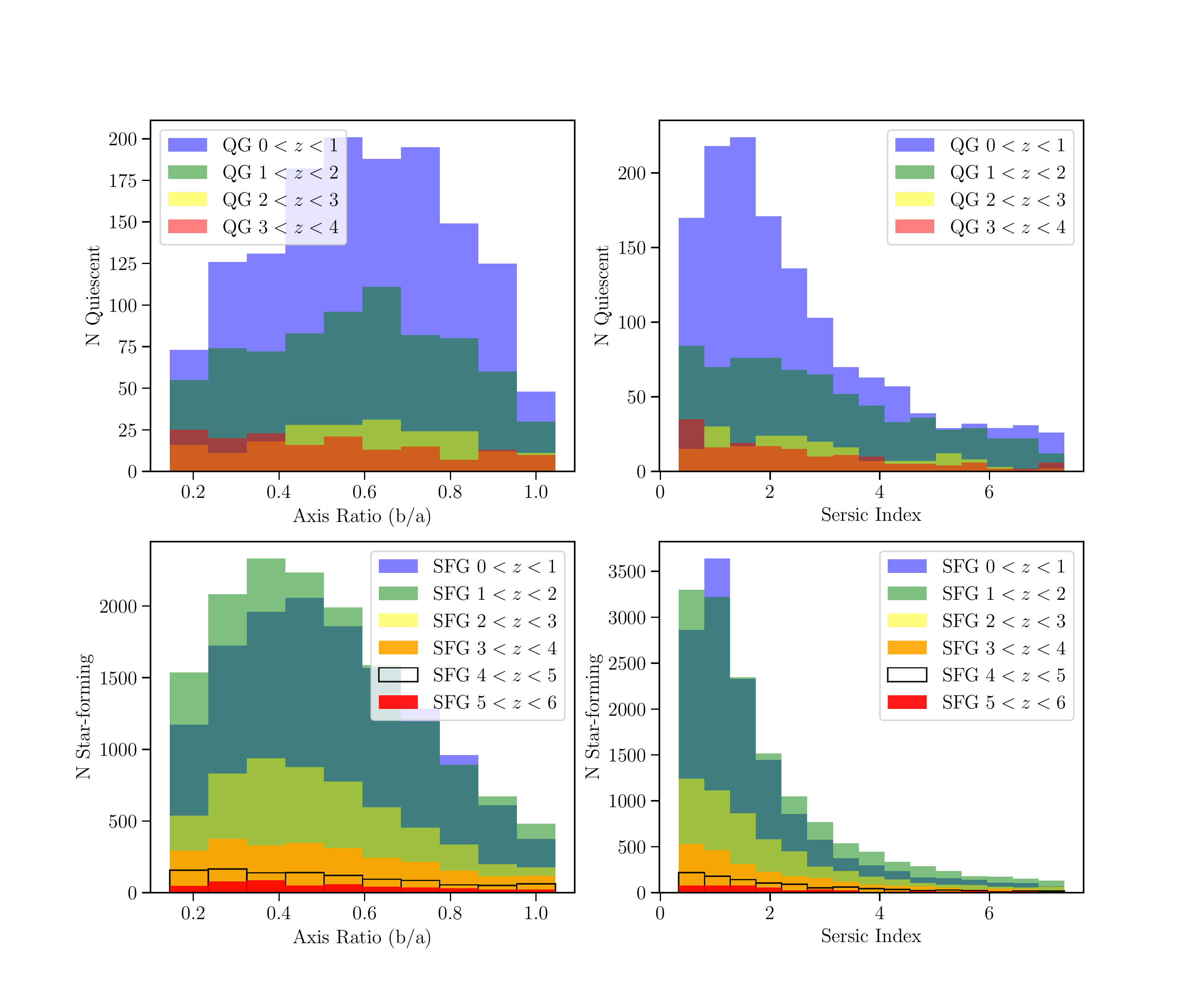}
\caption{Observed distributions of axis ratios (left panels) and Sersic index (right panels) for observed 3D-HST galaxies as measured by \citet{vanderWel2012}. Only galaxies with robust measurements (``flag = 0'') are plotted. 
Top row shows the distributions for quiescent galaxies split by redshift, and bottom panels show distributions for star-forming galaxies split by redshift. The distributions are used to randomly assign mock galaxies axis ratio and sersic indices as a function of star-forming class and redshift. }
\end{center}
\label{axissersic}
\end{figure*}

In addition to sizes, galaxy shapes are characterized by the following additional properties: projected axis ratio (semi-minor half-light size / semi-major half-light size; $b/a$), Sersic index $n$ (setting the overall concentration of the light profile), and position angle on the sky. The distribution of observed axis ratios and Sersic indices are known to correlate with star-formation \citep[e.g.][]{Franx2008,Bell2012, Mortlock2013}. 
Star-forming galaxies often exhibit extended, disk-like light profiles characterized by $n<2$ and lower values of $b/a$, whereas quiescent galaxies exhibit centrally concentrated light profiles with $n>2$ and higher $b/a$.
These properties are known to evolve within each star-formation sub-class, in that the $n$ and $b/a$ of quiescent galaxies tend to decrease with increasing redshift 
\citep[e.g.][]{vanderWel2011}, while at high-redshift star-forming galaxies exhibit more clumpy concentrated morphologies in contrast to the extended disks seen at low-redshift \citep[e.g.][]{Elmegreen2009, ForsterSchreiber2011,Lee2013, Conselice2014, Guo2015}. 
To capture the differing evolution within each class of galaxy, we use the observed axis ratios and Sersic indices that have been measured using GALFIT \citep{Peng2002} from the deep HST/WFC3 F160W CANDELS imaging in both GOODS-North and GOODS-South \citep{vanderWel2012}. To produce the observed distribution of these parameters
for each galaxy class and binned in redshift ($\Delta z=1$),
we match the real CANDELS galaxies in those morphology catalogs with robust morphology measurements (``flag = 0'') to the photometric redshifts and rest-frame UVJ colors published by 3D-HST \citep{Skelton2014}.
The resulting evolution with redshift of the distribution of parameters for star-forming and quiescent galaxies are shown in Figure~\ref{axissersic}.

We assign axis ratios and Sersic indices to mock star-forming and quiescent galaxies at a given redshift by generating random variates from those redshift-dependent distributions. We choose to draw directly from the binned distribution, rather than fit with an assumed functional shape for the evolution of the distributions. We first bin axis ratio measurements in bins of $\Delta(b/a)=0.1$ and Sersic
index in bins of $\Delta n=0.5$, 
and treat these distributions as probability distribution functions. Then we draw random variates from these distributions to assign to mock galaxies, assuming uniform probability within each bin. Quiescent galaxies at $z>4$ are assigned morphologies according to the $3 \le z \le 4$ distributions, and star-forming galaxies at $z>6$ follow the $5 \le z\le 6$ distributions.

The resulting axis ratio for each mock galaxy is used to convert the semi-major axis at $z\le 4$ into both semi-minor axis and circularized half-light radius, and to convert circularized half-light radius into semi-major and minor axes at $z>4$. Position angles for all galaxies are assigned randomly from a uniform distribution.

\section{Mock galaxy properties}\label{characterization}

To assess the performance and possible limitations of our phenomenological model, 
we compare mock galaxy properties, distributions, and relations to observations that were not used to inform our methodology. For this purpose we use a single realization 
(i.e. a JAGUAR mock catalog) on an area of $11\times11$ square arcminutes containing both star-forming and quiescent galaxies with $\LMstarMsun = 6 - 12$ and at $z = 0.2 - 15$.\footnote{
Available at http://fenrir.as.arizona.edu/jaguar/} In the following sections, we compare the mock galaxies from this realization to the redshift evolution of observed quantities including galaxy UV luminosity functions, 
star-formation rate densities and average specific star-formation rates, 
the mass--metallicity relation, emission line diagnostic diagrams, and observed infrared galaxy colors.

\subsection{UV Luminosity Function Evolution}\label{sect:LFevol}

\begin{table*}
\begin{center}
\caption{Stepwise average binned luminosity functions from the literature in the redshift bins presented in Figure~\ref{fig:binnedLF}  }
\begin{tabular}{ccccccccc}
 \hline
\Muv\ & z$\sim$ 0.5 & z$\sim$ 0.8 & z$\sim$ 1.25 & z$\sim$ 1.75 & z$\sim$ 2.25 & z$\sim$ 2.75 & z$\sim$ 3.75 & \\
\hline
 & Log$\Phi^{a}$ & Log$\Phi$ & Log$\Phi$ & Log$\Phi$ & Log$\Phi$ & Log$\Phi$ & Log$\Phi$  \\
\hline
-22.75  &  -14.59   &  -12.02 $\pm$ 1.34  &  -10.38 $\pm$ 0.88  &  -9.06 $\pm$ 0.21  &  -6.55 $\pm$ 0.70  &  -6.27 $\pm$ 0.30  &  -6.44 $\pm$ 0.17 \\ 
-22.25  &  -10.42  &  -8.96 $\pm$ 1.10  &  -7.91 $\pm$ 0.69  &  -6.92 $\pm$ 0.17  &  -5.34 $\pm$ 0.53  &  -5.15 $\pm$ 0.24  &  -5.16 $\pm$ 0.11 \\ 
-21.75  &  -7.74   &  -6.98 $\pm$ 0.85  &  -6.27 $\pm$ 0.50  &  -5.35 $\pm$ 0.15  &  -4.49 $\pm$ 0.38  &  -4.35 $\pm$ 0.19  &  -4.28 $\pm$ 0.07 \\ 
-21.25  &  -6.02   &  -5.68 $\pm$ 0.62  &  -5.15 $\pm$ 0.34  &  -4.26 $\pm$ 0.15  &  -3.88 $\pm$ 0.27  &  -3.77 $\pm$ 0.17  &  -3.66 $\pm$ 0.05 \\ 
-20.75  &  -4.88   &  -4.78 $\pm$ 0.43  &  -4.37 $\pm$ 0.21  &  -3.54 $\pm$ 0.16  &  -3.43 $\pm$ 0.18  &  -3.33 $\pm$ 0.17  &  -3.23 $\pm$ 0.05 \\ 
-20.25  &  -4.13   &  -4.11 $\pm$ 0.27  &  -3.78 $\pm$ 0.11  &  -3.07 $\pm$ 0.17  &  -3.08 $\pm$ 0.12  &  -3.01 $\pm$ 0.17  &  -2.92 $\pm$ 0.05 \\ 
-19.75  &  -3.61   &  -3.58 $\pm$ 0.16  &  -3.34 $\pm$ 0.04  &  -2.76 $\pm$ 0.17  &  -2.8 $\pm$ 0.09  &  -2.76 $\pm$ 0.17  &  -2.69 $\pm$ 0.05 \\ 
-19.25  &  -3.25   &  -3.17 $\pm$ 0.08  &  -3.0 $\pm$ 0.03  &  -2.54 $\pm$ 0.16  &  -2.59 $\pm$ 0.10  &  -2.56 $\pm$ 0.17  &  -2.51 $\pm$ 0.06 \\ 
-18.75  &  -2.98   &  -2.86 $\pm$ 0.03  &  -2.75 $\pm$ 0.05  &  -2.39 $\pm$ 0.14  &  -2.41 $\pm$ 0.12  &  -2.39 $\pm$ 0.17  &  -2.36 $\pm$ 0.07 \\ 
-18.25  &  -2.76   &  -2.63 $\pm$ 0.01  &  -2.54 $\pm$ 0.06  &  -2.27 $\pm$ 0.12  &  -2.27 $\pm$ 0.15  &  -2.23 $\pm$ 0.17  &  -2.22 $\pm$ 0.08 \\ 
-17.75  &  -2.59   &  -2.45 $\pm$ 0.01  &  -2.37 $\pm$ 0.07  &  -2.16 $\pm$ 0.10  &  -2.14 $\pm$ 0.17  &  -2.09 $\pm$ 0.18  &  -2.10 $\pm$ 0.10 \\ 
-17.25  &  -2.44   &  -2.3 $\pm$ 0.00  &  -2.23 $\pm$ 0.09  &  -2.06 $\pm$ 0.08  &  -2.03 $\pm$ 0.21  &  -1.95 $\pm$ 0.20  &  -1.99 $\pm$ 0.12 \\ 
 \hline
 \label{table:binnedLF}
\end{tabular}

\end{center}
 $^a$ The $z\sim0.5$ binned luminosity function only includes one measurement from the literature, therefore the observed scatter in that bin is zero.
\end{table*}

We compare the UV luminosity function at $z \le 4$ computed from the mock catalog with measurements from the literature, as this enables us to test the adopted evolutionary model of both the star-forming galaxy stellar mass function (Section~\ref{evolvemassfn}) and the \MStarMuv\ relation (Section~\ref{MstarMuv}).

To compare with observations of the UV luminosity function, we use the compilation of literature measurements analyzed in \citet{Parsa2016} from $0.4<z<4$.
\citet{Parsa2016} provide a compilation of Schechter function parameters with errorbars but do not quantify
the covariance(s) 
 between these parameters, which are known to be strong. 
Because of the degeneracies among the parameters,  we 
avoid comparing directly to individually measured Schechter functions, and rather convert these into step-wise binned luminosity functions with equal magnitudes and bin widths.
We then average the binned luminosity functions to produce a mean step-wise luminosity function at each redshift. We quantify the scatter in the literature as the standard deviation of the galaxy counts in each luminosity bin, divided by the square root of the number of measured luminosity functions contributing to the average. The binned averages and scatter from the literature are presented in Table~\ref{table:binnedLF}.

\begin{figure*}
\begin{center}
\includegraphics[width=1\textwidth]{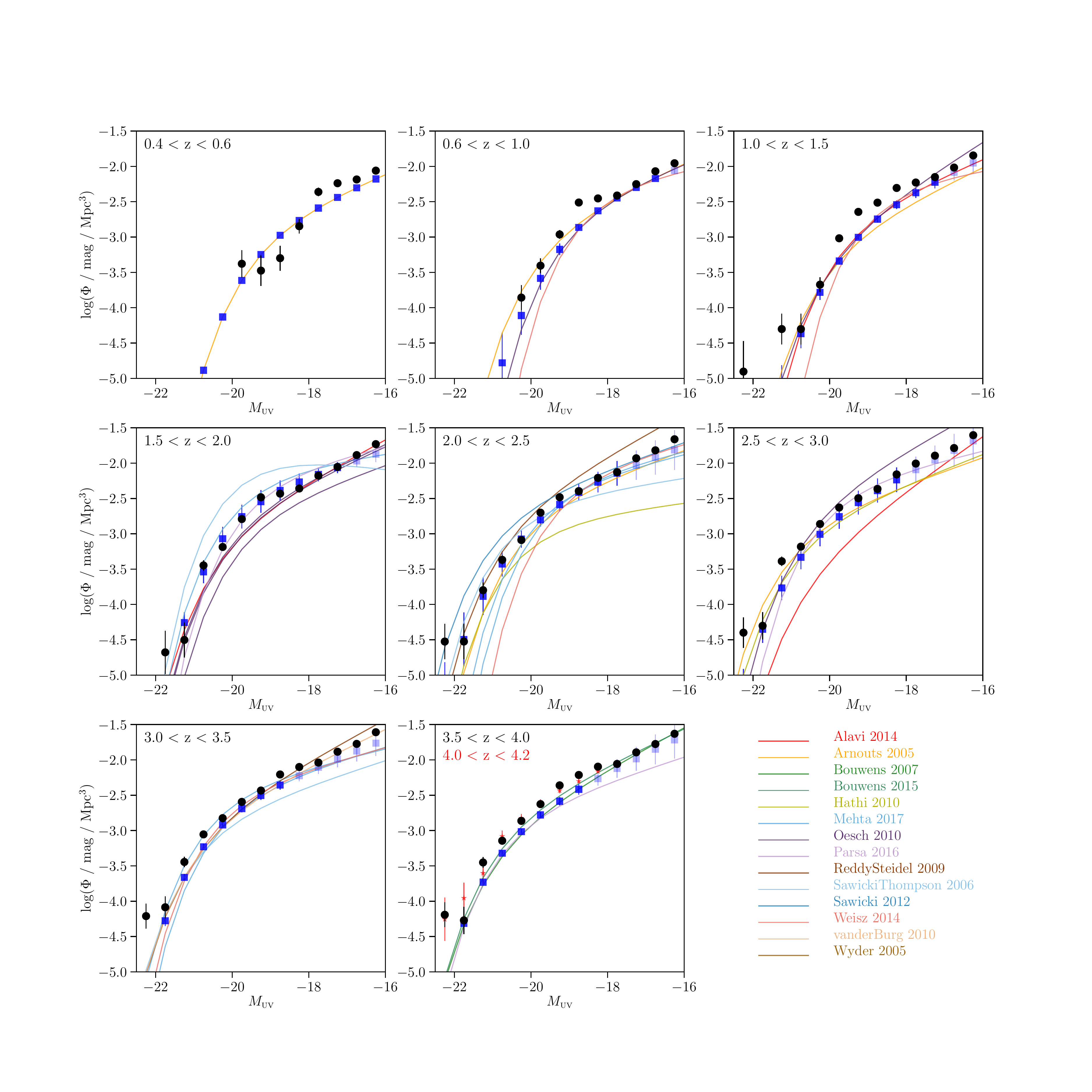}
\caption{Comparison of the $z<4$ UV luminosity functions of mock and observed galaxies. Black points indicate mock galaxies, where the error bars represent Poisson errors. Colored lines indicate UV luminosity functions from the literature \citep{Wyder2005, Arnouts2005, Hathi2010,Bouwens2007,Bouwens2015LF,Alavi2014,Mehta2017,Oesch2010,ReddySteidel2009,Sawicki2006,Sawicki2012,Weisz2014,vanderBurg2010}, while the average (step-wise) luminosity functions from the literature is shown by the blue squares (blue error-bars indicate the scatter of the observations as described in the text). Observed points below the incompleteness limit are indicated by light-blue points. Red stars in the highest redshift bin indicate the luminosity function measured from mock galaxies at $4<z<4.2$.}
\label{fig:binnedLF}
\end{center}
\end{figure*}

Figure~\ref{fig:binnedLF} shows the comparison of the $0.4<z<4$ mock galaxy UV luminosity functions (black circles) with the average literature measurements (blue squares).\footnote{We exclude mock galaxies at $z<0.4$, as the volume probed in the realization is small, and the empirical constraints on \MStarMuv\ and observed UV luminosity functions are less robust.} At $z>1.5$, the mock catalog exhibits excellent agreement with the observations, while at $z<1.5$ the agreement is less robust. We note, however, that the mock catalog never overpredicts the measured number density of galaxies by more than $\sim 0.36$ dex (at $\Muv\sim-19$ at $0.6<z<1$ and $1<z<1.5$).
This overprediction is likely caused by the poor observational constraints on the rest-frame UV emission of galaxies at $z<1.5$,  where photometry in the 3D-HST catalog no-longer provides coverage at 1500\AA\ (rest wavelength).
 This affects both the characterization of the \MStarMuv\ relation, which we use to directly assign \Muv\ values to low-mass mock galaxies at these redshifts, as well as the \Muv\ values of high-mass galaxies that are derived directly from fits to the 3D-HST photometry.
The same uncertainties affect the observed UV luminosity functions compiled from the literature between $z\sim 0$, where GALEX data are available, and $z\sim 1$, where \HST\ near-UV bands probe the rest-frame 1500 \AA\ flux \citep[e.g.][]{Oesch2010, Windhorst2011, Teplitz2013, Rafelski2015}.

\subsection{Star-formation rate density evolution}\label{CSFRD}

\begin{figure}
\includegraphics[width=0.5\textwidth]{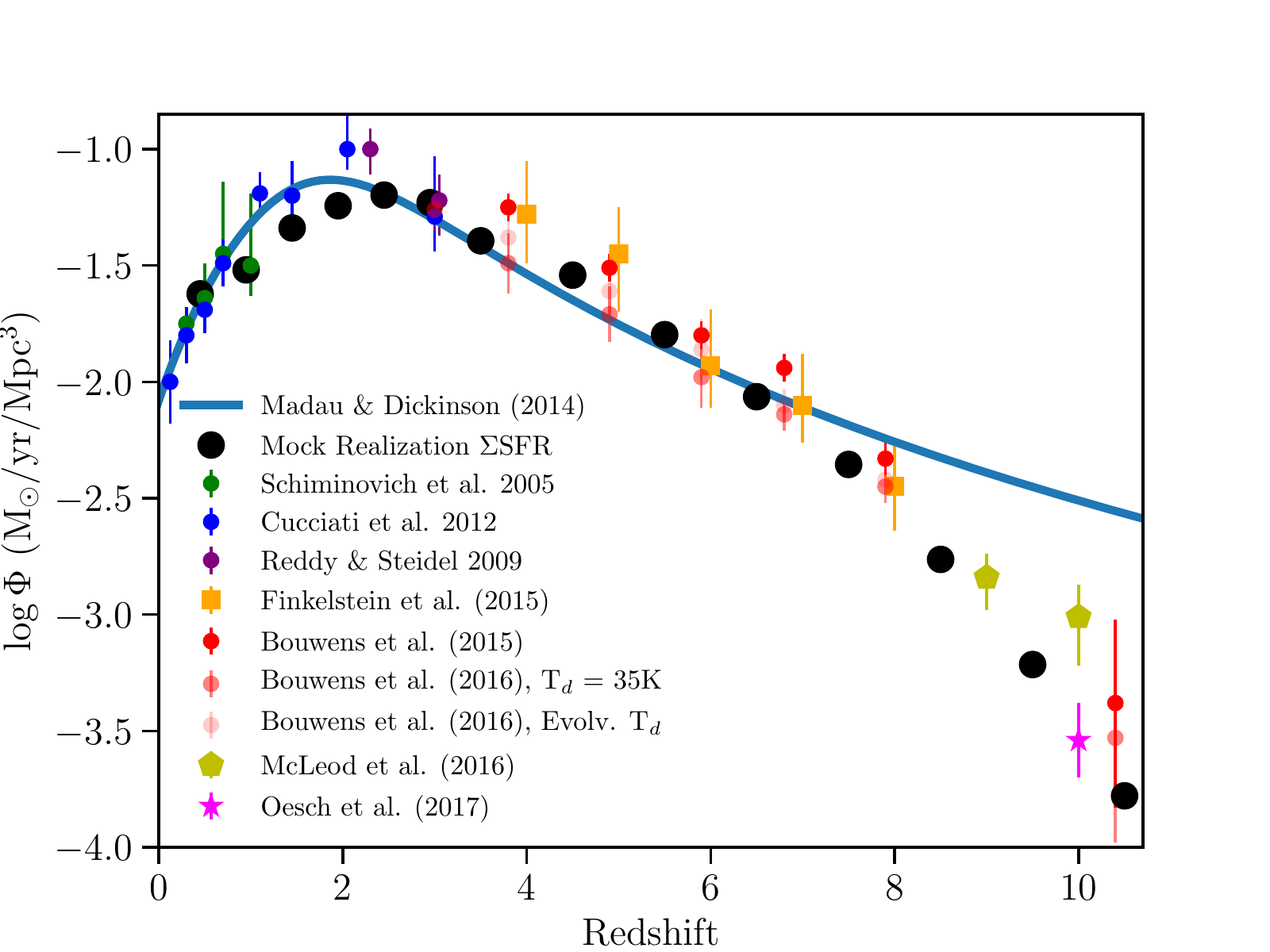}
\caption{Evolution of the cosmic star formation rate density of the Universe as compiled by \citet{MadauDickinson2014} (blue curve) and as computed from our mock catalogue (black points). Colored points indicate individual measurements from the literature.}
\label{fig:SFRD}
\end{figure}

In Figure \ref{fig:SFRD}, we compare the cosmic star-formation rate density (CSFRD) of mock galaxies with the average CSFRD evolution presented in \citet[][based on a uniform analysis of luminosity function measurements in the literature]{MadauDickinson2014}, converted to a Chabrier IMF. To approximate the same UV luminosity limits imposed by \citet{MadauDickinson2014}, we estimate the limiting \Muv\ corresponding to $0.03 \, L^\ast$ in each redshift bin.
These values are taken directly from the Schechter UV luminosity function parameter \LFMstar 
 which is used to estimate the equivalent limit in \Muv\ from $0<z<10$. The equivalent of $0.03 \, L^\ast$, as published in the following studies, corresponds to $\Muv \sim -14.5$ at $1<z$ \citep{Cucciati2012}; $\Muv \sim -15.5$ at $1<z<2$ \citep{Cucciati2012}; $\Muv \sim -16.89$ at $1.9<z<2.7$ \citep{ReddySteidel2009}; and $\Muv \sim -17$ at $z>2.7$ \citep{Bouwens2015LF, Finkelstein2015, Bouwens2016LF,Oesch2017}.

To calculate the CSFRD of mock galaxies in each redshift bin, we identify all mock galaxies above the limiting \Muv\ corresponding to $0.03 \, L^\ast$ and take the sum of their star-formation rates (averaged over the past 100 Myr) per comoving volume. We find that the evolution of the CSFRD estimated from the mock catalog qualitatively reproduces the overall shape of the observed CSFRD; 
however, at $4<z<6$ mock galaxies  show a slight excess with respect to the results of 
\citet{MadauDickinson2014}, who derive their $z>4$ points exclusively from the UV luminosity functions of \citet[][]{Bouwens2012beta, Bouwens2012LF}. \citet{Bouwens2012LF} provided measurements of the $z\sim 6$ luminosity function, along with a compilation of earlier measurements from at $z\sim 4-5$ and $z\sim 7-8$ \citep{Bouwens2007, Bouwens2011}. 
However, our model is based on the UV luminosity function of \citet{Bouwens2015LF}, who find higher number counts than in previous works, in particular at the bright end at $\Muv<-18$ and $z\sim 4-5$, and at $\Muv<-20$ and $z>6$, which likely explains our excess relative to \citet{MadauDickinson2014}. At $1<z<3$, the CSFRD measured from mock galaxies is slightly lower than the one of \citet{MadauDickinson2014}. A reason for this may be that our phenomenological model does not explicitly include extremely dust-obscured  galaxies 
which may be missing from the {\it HST}-selected samples that we use to characterize the stellar mass function evolution, whereas \citet{MadauDickinson2014} do incorporate far-infrared measurements of dust-obscured objects.  
To validate this explanation, we include individual measurements of the CSFRD from UV-selected samples at $z<3$ and their uncertainties \citep{Schiminovich2005,Cucciati2012,ReddySteidel2009}. These indicate a better agreement with our mock galaxies, within the scatter of binned observations,  than implied by the \citet{MadauDickinson2014} curve.

We additionally plot measurements of the CSFRD at $z>4$ from \citet[][converted to Chabrier IMF when necessary]{Bouwens2015LF, Finkelstein2015, Bouwens2016irxb, McLeod2016, Oesch2017}. 
The CSFRD evolution of mock galaxies at $z>4$ is in better agreement with these later measurements.  The \Muv\ integration limits in these studies match those used to estimate the CSFRD from the mock catalog with the exception of \citet{McLeod2016}, whose integration limit is $\Muv \sim -17.7$. The CSFRD of \citet{McLeod2016} would be higher if measured to the same limiting \Muv\ used for the mock catalog.
The higher UV counts (and thus CSFRD) measured in \citet{McLeod2016} have been interpreted as evidence for a slower rate of decline in galaxy number counts with redshift, in contrast to the "accelerated'' evolution that is seen in, e.g., \citet[][]{Bouwens2015LF,Oesch2017}. Since our model follows the more rapidly evolving luminosity function of \citet{Oesch2017}, the CSFRD in the mock follows more closely the CSFRD measured in that work. 
We find that our mock realization generally reproduces the $z>4$ evolution of the CSFRD based on the latest results from current extragalactic surveys, and in particular those we have used to develop our model for the evolution of the stellar mass function.

\subsection{Evolution of the average specific star-formation rate}

\begin{figure}
\includegraphics[width=0.5\textwidth]{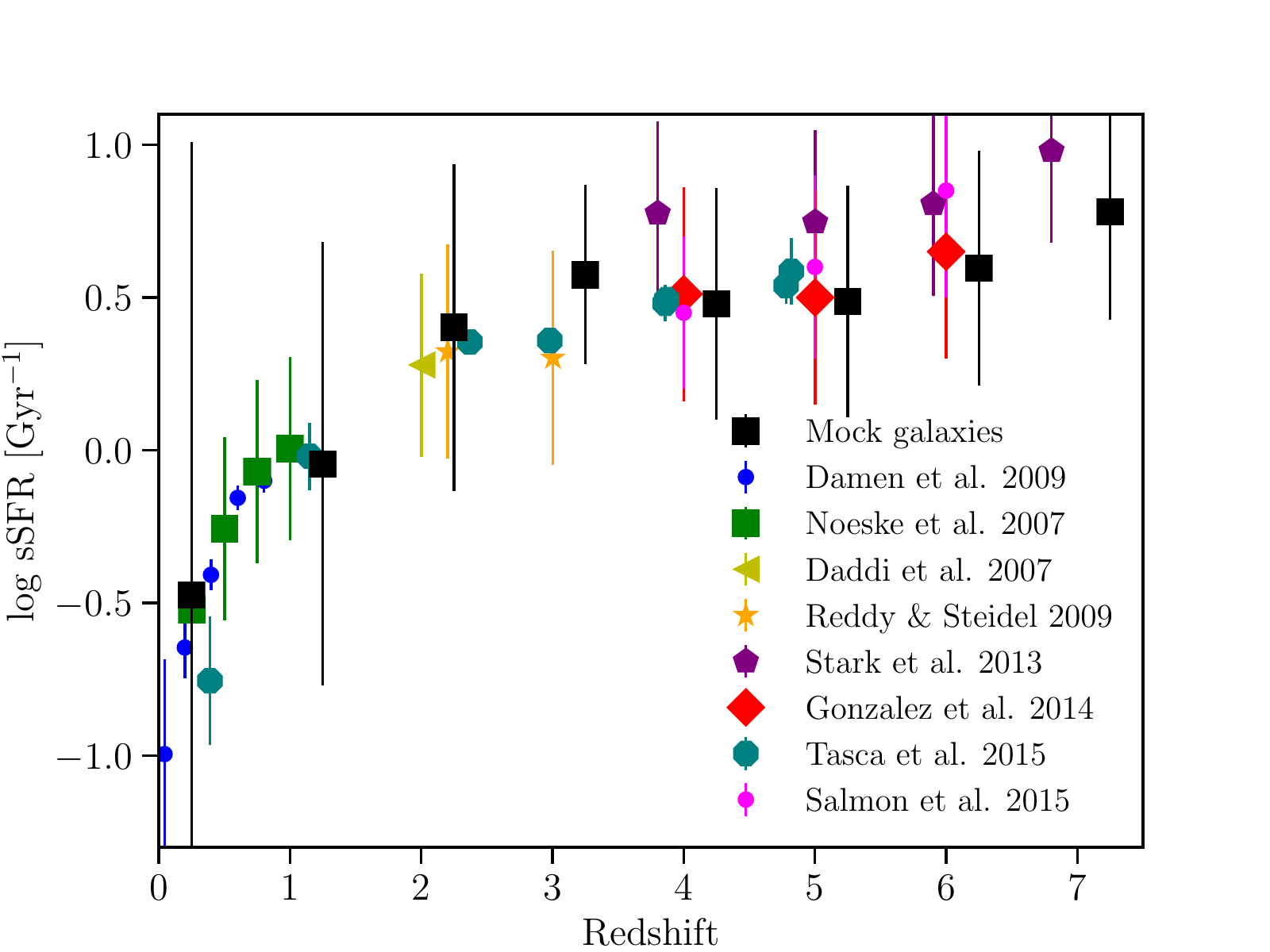}
\caption{Redshift evolution of the (median) specific star formation rate of mock galaxies with $8.8<(\log\Mstar/\Msun)<10$ (black squares) compared to measurements from the literature (colored points).} 
\label{fig:ssfr}
\end{figure}

We compare the redshift evolution of the specific star-formation rate (sSFR) from the mock catalog with literature measurements at $0<z<7$, the highest redshift at which observational constraints are available. We consider the median sSFR of galaxies with masses $8.8<\LMstarMsun<10$ as computed by \citet{Noeske2007, Damen2009, Daddi2007,ReddySteidel2009,Stark2013,Gonzalez2014,Tasca2015, Salmon2015}. 
To calculate the median sSFR of mock galaxies, we use the \beagle-assigned SFR averaged over the last 100 Myr divided by stellar mass of mock galaxies in the same stellar mass range as the observations. We plot this comparison in Figure \ref{fig:ssfr}, and show that the sSFR values for our mock are in excellent agreement with the observations within their  
uncertainties.

As discussed in Section~\ref{MstarMuv}, the sSFR of galaxies at $z\gtrsim 4$ and the redshift dependence of the average sSFR are directly related to the extrapolation we adopted for the redshift evolution and normalization of the \MStarMuv\ relation. 
This is in contrast to mock galaxy sSFRs at $z<4$, where the \MStarMuv\ relation is based directly on more reliable
stellar mass measurements across a range of UV luminosities.
Therefore, the agreement between the mock galaxy sSFRs with measurements in the literature at $z>4$ shown in Figure~\ref{fig:ssfr} is a validation of the adopted extrapolations of the \MStarMuv\ relation at high redshift. We note, however, that the evolution of the sSFR of galaxies at $z>4$ is still a matter of active research; most studies support an increasing average sSFR rather than a plateau at $z>4$ \citep[see discussion in][]{Stark2016}. An accurate characterization of the sSFR evolution at $z>4$ will await improved measurements that will be made possible by \JWST.

\subsection{{Evolution of the Mass-Metallicity Relation}}
\label{MZR_comparison}

The mass-metallicity relation of galaxies is known to evolve, such that galaxies at higher redshifts are more metal-poor at a given stellar mass \citep[e.g.][]{Maiolino2008}. When assigning SEDs (and associated physical properties) to each mock galaxy, we use a parent catalog that has a built-in redshift-independent prior linking $\Mstar-\Z-\sfr$. based on the fundamental metallicity relation as measured by \cite{Hunt2016}. The redshift-evolving \MStarMuv\ relation, however, produces a redshift evolution of the $\Mstar-\sfr$ relation which, in turn, should generate an evolving mass-metallicity relation. 

\begin{figure}
\includegraphics[width=0.5\textwidth]{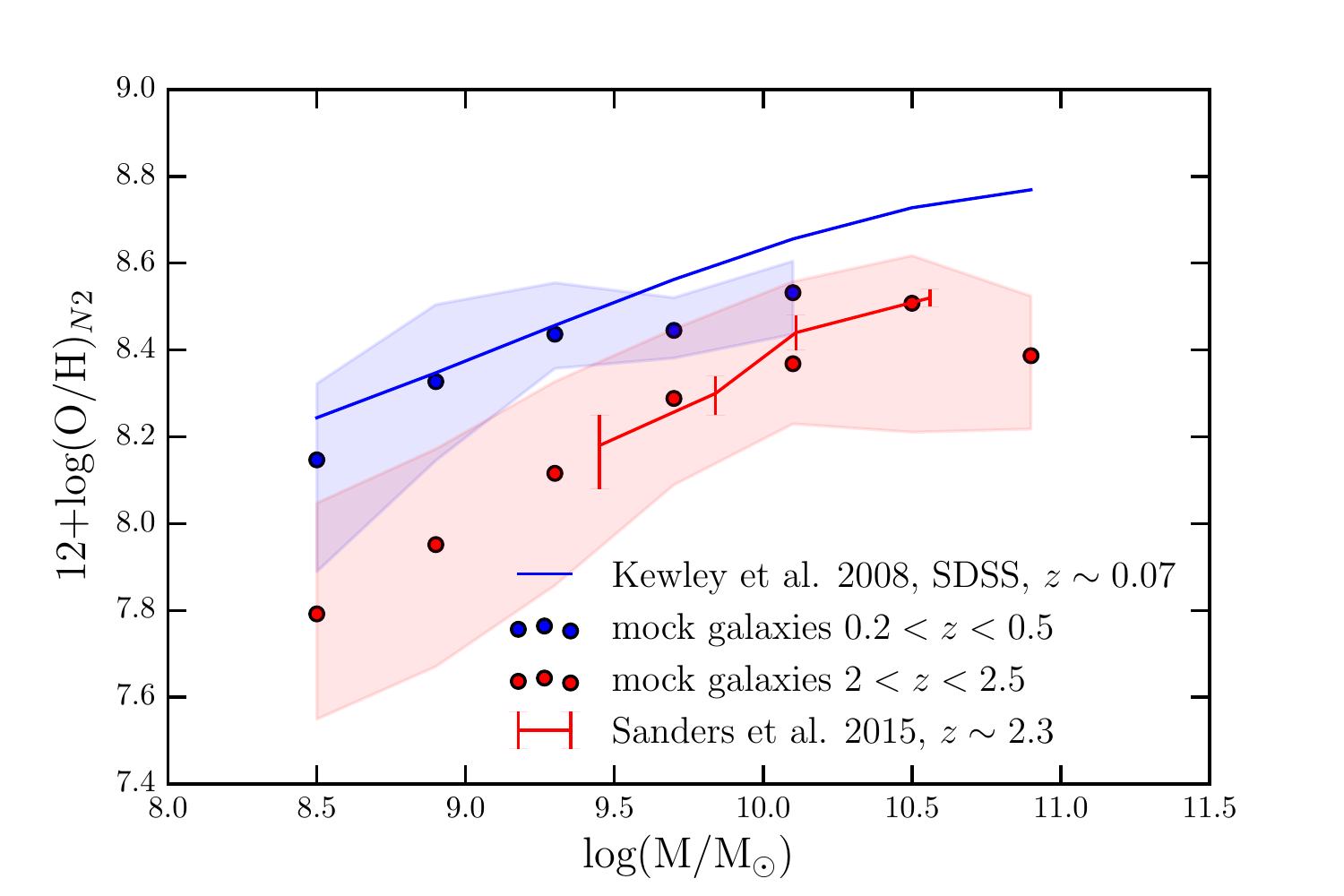}
\caption{Mass-metallicity relation of mock galaxies compared to observations in two bins of redshift, $0.2<z<0.5$ (blue) and $2<z<2.5$ (red).  The points show the median values of metallicity  for mock galaxies in those redshift ranges, estimated using the NII calibration (see text for details), while the shaded regions encompass the 25\% and 75\% percentiles for mock galaxies. Only bins containing at least 5 galaxies are plotted. The blue solid line indicates the observed relation of \cite{Kewley2008} ($z\sim 0.07$), while the red line and error bars show the observations of \cite{Sanders2015} ($z\sim 2.3$).}
\label{fig:MZR}
\end{figure}

Figure~\ref{fig:MZR} shows the mass-metallicity relation in two redshift bins, $0.2 \le z \le 0.5$, and $2 \le z \le 2.5$. Given that different metallicity indicators provide systematically different metallicity estimates \citep{Kewley2008}, we choose to estimate the mock galaxy metallicities using a metallicity calibration often used in the literature, based on the ratio between the lines $[\txn{N}\textsc{ii}]\lambda 6584$ and $\txn{H}\alpha$ (the \cite{Pettini2004} N2 metallicity calibration):
\begin{equation}
\logOH_{N2} = 8.90 + 0.57 \times \log ([\txn{N}\textsc{ii}]\lambda 6584/\txn{H}\alpha)
\end{equation}\noindent
This allows us to compare directly to observationally-derived mass-metallicity relations. 

Figure~\ref{fig:MZR} indicates that the mass-metallicity relation derived from the mock catalog does show a turn-over at high stellar masses, as seen in the observations, despite the \cite{Hunt2016} fundamental metallicity relation only being linearly dependent on $\log(\Z/\Zsun)$. The turnover is caused by the presence of a maximum metallicity $\log(\Z/\Zsun)=0.24$ that can be assigned to the mock galaxies (see Sections~\ref{highmassgal} and \ref{BEAGLE_SF_grid}). Figure~\ref{fig:MZR} demonstrates a good agreement among our mock galaxies and the mass-metallicity relation of \cite{Kewley2008} (at $z\sim0.07$) and of \cite{Sanders2015} (at $z\sim2.3$). As discussed above, the evolution of the mass-metallicity relation in our mock catalog is a result of the priors we impose on both $\Mstar-\Z-\sfr$ and \MStarMuv.

\subsection{Emission line diagnostic diagrams}
\label{BPT_comparison}

\begin{figure*}
\includegraphics[width=0.5\textwidth]{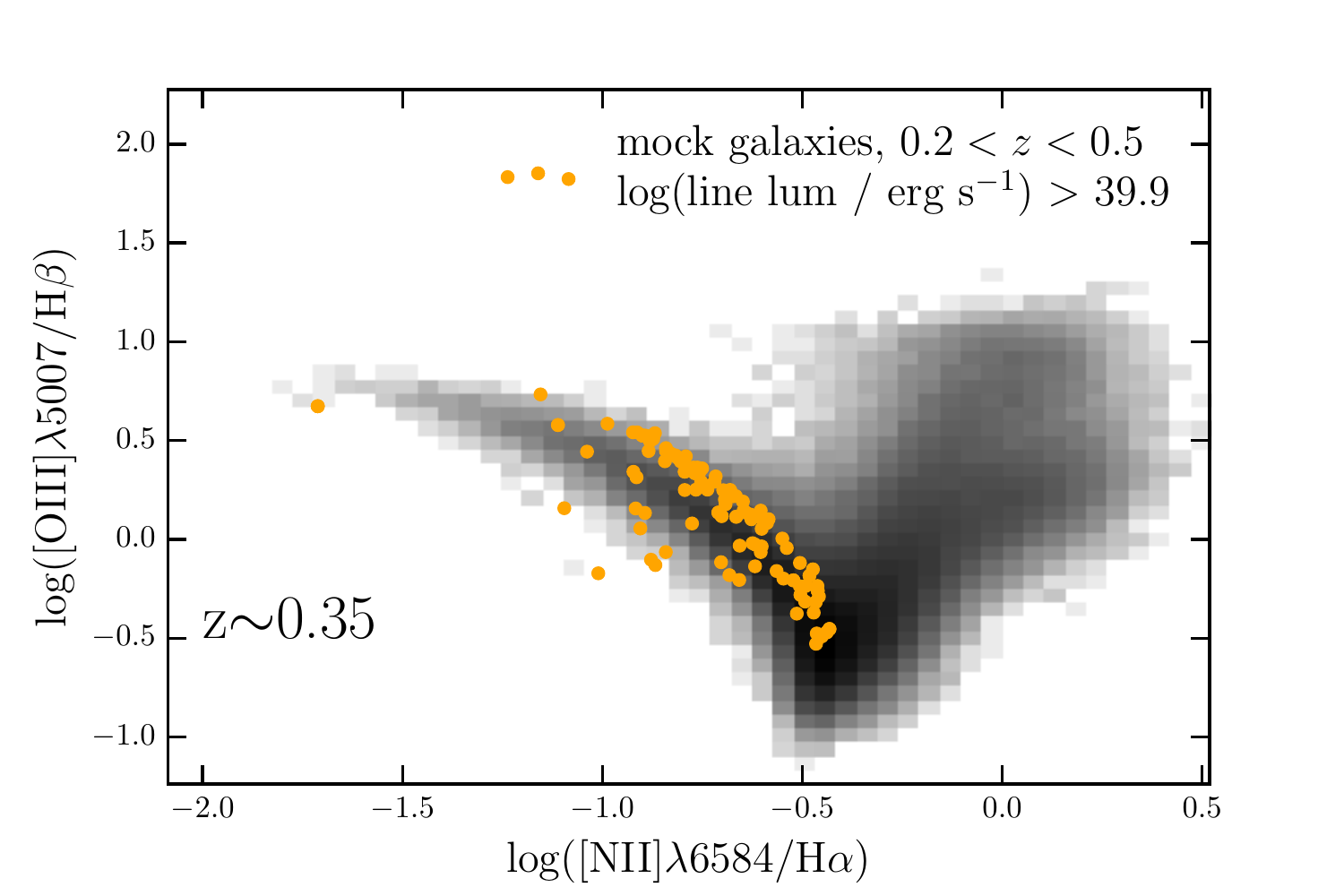}
\includegraphics[width=0.5\textwidth]{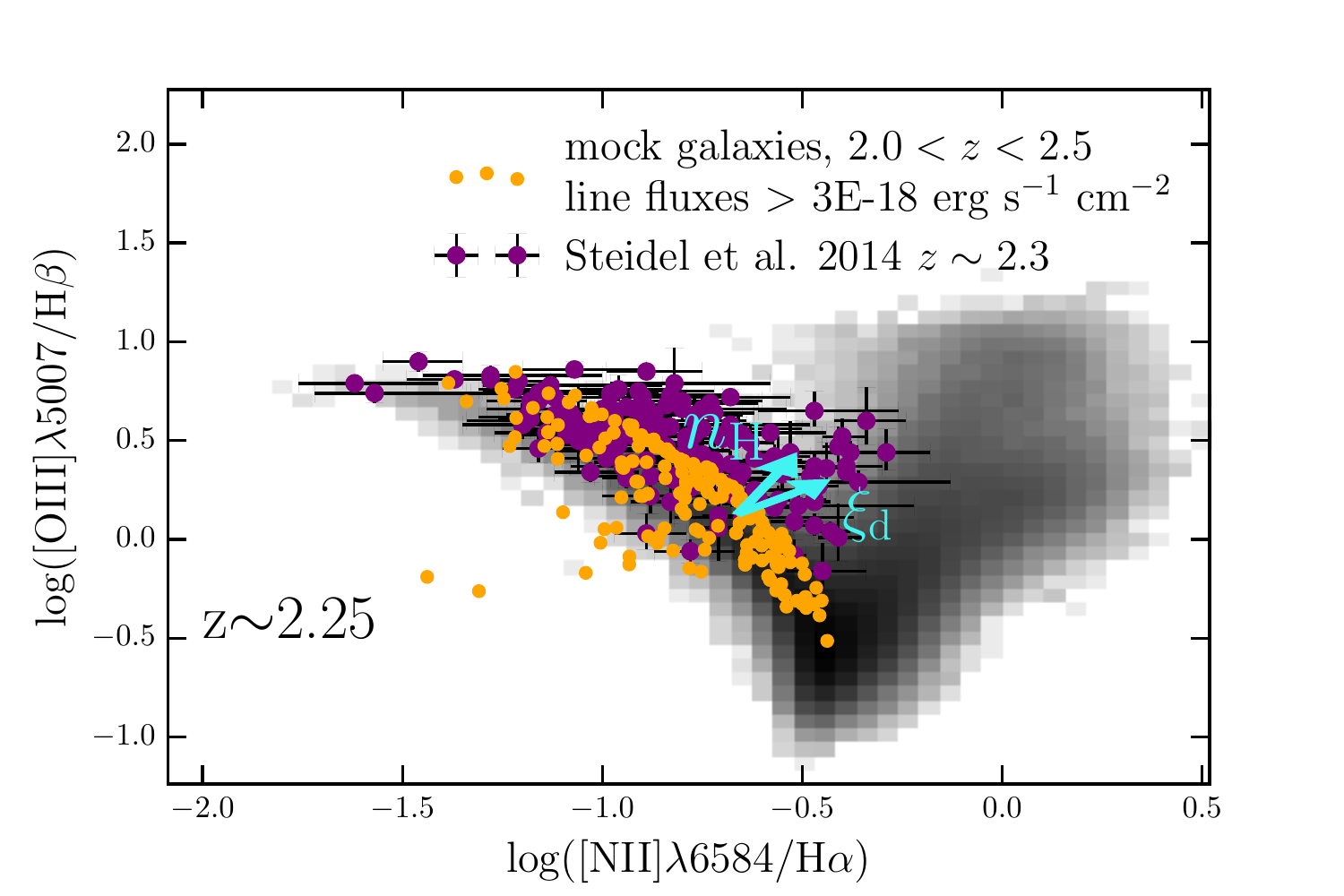}
\caption{The BPT diagram of mock galaxies compared to observations in two different redshift bins, $0.2<z<0.5$ in the right panel and $2<z<2.5$ in the left panel.  The mock galaxies are plotted as orange points in each panel, and we display the observations from SDSS as the grey 2D histogram.  Purple points and error-bars in the right panel are the observations of \cite{Steidel2014}, excluding those objects classified as AGN.  In the left panel, only mock galaxies with line luminosities above $\log(\textrm{line lum / erg s}^{-1})>39.9$ are plotted, to approximate the limits in the SDSS sample, while in the right panel a flux limit of $>3 \times 10^{-18}$ erg s $^{-1}$ was applied to mock galaxies plotted at $2<z<2.5$, chosen to mimic the flux limit of the \cite{Steidel2014} sample.  The two blue vectors plotted on the right panel display the direction that objects move in the diagram with changes to \nH\ (changing \nH\ from 100 to 1000 cm$^{-3}$) or \xid\ (changing \xid\ from 0.3 to 0.5).}
\label{fig:BPT}
\end{figure*}

Emission line diagnostic diagrams are commonly used in the literature to identify the ionization sources in galaxies and characterize their physical properties. \cite{Baldwin1981} (BPT hereafter) pioneered the use of the ratios of \OIIIHb\ and \NIIHa\ to separate star-forming galaxies and AGN-dominated galaxies into two distinct regions of the plot, with composite galaxies (those with both significant star-formation and AGN contribution)
spanning the region in between. Recent observations of rest-frame optical spectra at redshifts out to $z\sim2.5$ have revealed that the locus of star-forming galaxies is slightly shifted to higher \OIIIHb\ ratios, at fixed \NIIHa, than in the local Universe \citep[e.g.][]{Shapley2005, Brinchmann2008, Hainline2009, Kewley2013, Steidel2014, Kashino2017}.
This shift can been explained if the physical conditions of the ionized gas are evolving with redshift. 
For example, evolution in the ionization parameter \citep[e.g.][]{Kashino2017}, the incident ionizing spectrum of stars \citep[e.g.][]{Steidel2014, Steidel2016}, gas-phase nitrogen abundance \citep[e.g.][]{Masters2016}, and hydrogen density \citep[e.g.][]{Sanders2016} can all introduce a shift in observed \OIIIHb\ ratio.  However, it is also possible  that selection effects play a role \citep[e.g.][]{Juneau2014}. 
More likely, the observed evolution of the location of star-forming galaxies in the BPT diagram is caused by a combination of the effects above, but current data do not allow us to quantify their respective roles.

In Figure~\ref{fig:BPT} we plot the BPT diagram of mock galaxies at low ($0.2<z<0.5$) and high ($2<z<2.5$) redshift. We compare the mock galaxies with SDSS galaxies ($z\sim0.07$) and with a sample of star-forming galaxies from the KBSS survey \citep{Steidel2014} ($z\sim 2.5$). For mock galaxies at $0.2<z<0.5$, we adopt the lowest luminosity threshold of \cite{Juneau2014}, and only plot objects with log luminosities $>39.9$ erg s$^{-1}$ in the lines $[\txn{O}\textsc{iii}]\lambda5007$, $\txn{H}\beta$, $\txn{H}\alpha$, $[\txn{N}\textsc{ii}]\lambda6584$.
As Figure~\ref{fig:BPT} shows, the mock galaxies sit close to the main star-forming galaxy locus, despite being at higher redshift than the SDSS sample. At $2<z<2.5$, we instead apply a flux limit close to that of \cite{Steidel2014}, i.e. line flux $>3 \times 10^{-18}$ erg s$^{-1}$ cm$^{-2}$. At these higher redshifts, the mock galaxies cover similar regions to the \cite{Steidel2014} points at low \NIIHa, but not at high values of \NIIHa. This difference can be understood by appealing to the recent study of \cite{Hirschmann2017}, where they self-consistently couple the nebular emission models of \cite{Gutkin2016} for star forming galaxies and  narrow-line AGN-driven models of \cite{Feltre2016} to cosmological zoom-in hydrodynamic simulations. In their work, they consistently tie \logUs\ to the simulated galaxy properties, and fix \xid\ and \nH\ to the same values as fixed in this mock catalog.  They reproduce the observed evolution of the line ratios, i.e. higher \OIIIHb\ on average at higher redshifts, but their figure~12 indicates that covering the elevated \OIIIHb\ values at high \NIIHa\ requires either a higher value of \xid\ and/or of \nH, or some level of AGN activity.  We add to the the right panel of Figure~\ref{fig:BPT} two vectors indicating how variations of these two parameters modify the expected line ratios. Given that the possible reasons for the elevated \OIIIHb\ ratio is far from determined, we choose to supply line ratios for mock galaxies with different values of \xid\ and \nH. 
Since an elevated \xid\ ratio acts to increase the relative abundance of gas-phase nitrogen to oxygen (since oxygen is a refractory element, unlike nitrogen, and thus heavily depleted onto dust grains), we additionally provide line fluxes and EWs for a single mock realization with $\xid=0.5$ (where the fiducial mock has \xid=0.3) which will be consistent with elevated nitrogen abundance (whether physically we expect this to be due to dust depletion or the relative importance of primary vs. secondary nitrogen production in stars at different epochs, the effect on these two line ratios will be similar).  We also provide line fluxes and equivalent for galaxies in that realization with a higher hydrogen density.  The \cite{Gutkin2016} models are evaluated in unit steps in $\log(\nH)$, so we are able to supply this line information for $\nH=1000$ cm$^{-3}$ (where the fiducial mock has $\nH=100$ cm$^{-3}$).

\subsection{Rest-frame optical colors at $z\sim4-7$}

\begin{figure}
\includegraphics[width=0.5\textwidth]{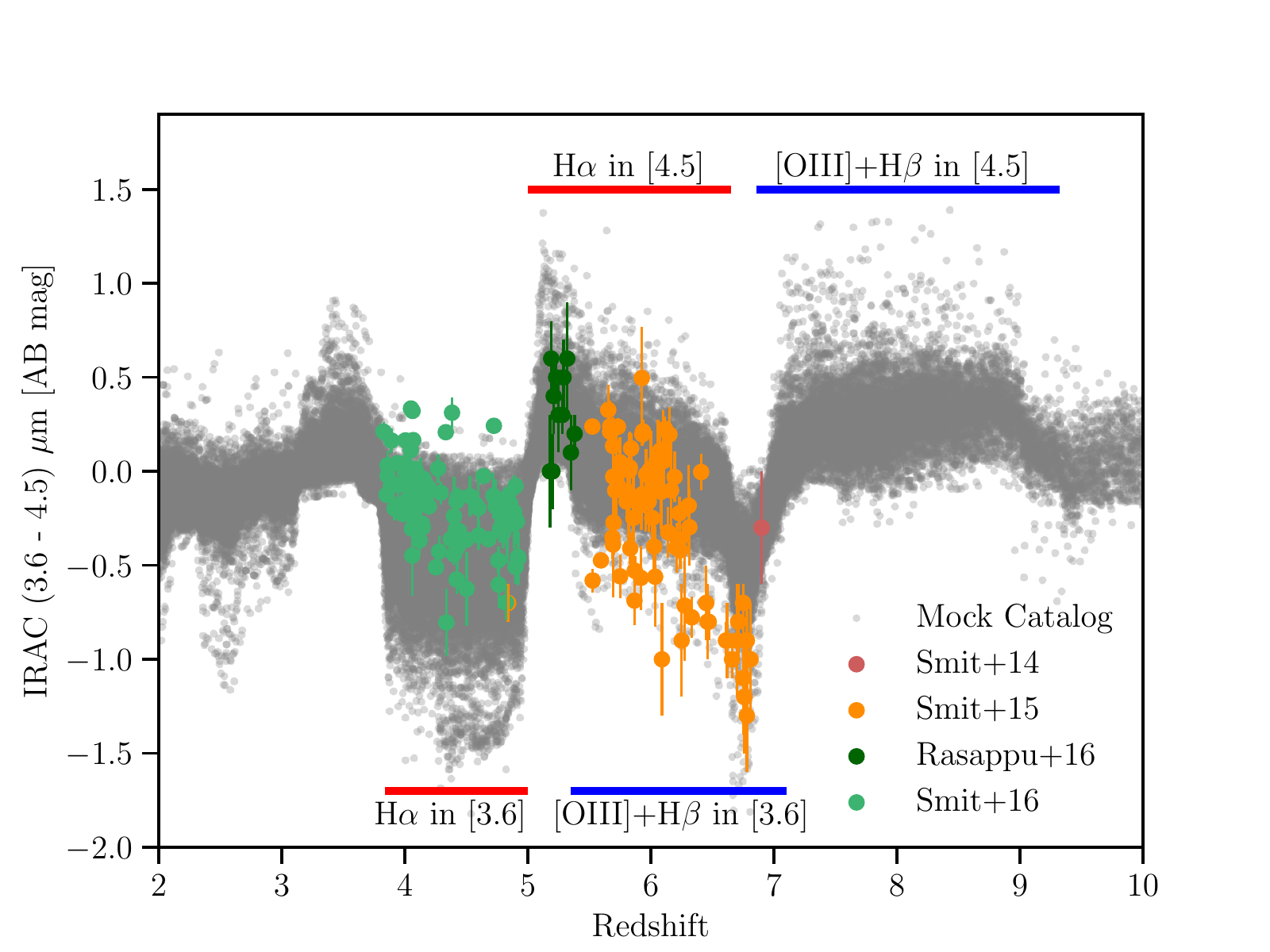}
\caption{ {\it Spitzer}/IRAC 3.6 and 4.5$\mu$m band colors for mock galaxies with \LMstarMsun$>$6 (grey points) compared to observations (colored points). Redshift ranges where H$\alpha$, [OIII] and H$\beta$ enter the 3.6 (4.5)$\mu$m bands are shown as the horizontal bars on the bottom (top) of the figure. Our model produces mock galaxies whose rest-frame optical emission line properties span the observed color excesses from current surveys.  }
\label{fig:wide_fluxexcess}
\end{figure}

Strong nebular emission lines are now known to contaminate the broad-band photometry of galaxies. In particular, colors from \Spitzer/IRAC have been extensively used to infer the EW of strong optical emission lines \citep[][]{Shim2011,Smit2014,Smit2015, Smit2016,Marmol-Queralto2016,Rasappu2016,Faisst2016,Castellano2017}.  
In Figure~\ref{fig:wide_fluxexcess}, we compare the IRAC 3.6 - 4.5 \micron\ color of mock galaxies with measurements at $z>4$ from \citet{Smit2014,Smit2015, Smit2016} and \citet{Rasappu2016}.The redshift evolution of the 3.6 - 4.5 \micron\ color predicted by the mock catalog, caused by the emission lines $\txn{H}\alpha$, $[\txn{O}\textsc{iii}]$, and $\txn{H}\beta$ entering and leaving the IRAC bands, is consistent with the observations. This suggests that the emission line strengths in the mock galaxies are compatible with the values inferred from IRAC observations. 
In Section~\ref{sec:nircammedflux} below, we will use the mock catalog to show how the sensitivity and wavelength coverage of \JWST/NIRCam will constrain emission line EWs with greater accuracy than existing \Spitzer\ data.

\begin{figure}
\includegraphics[width=0.5\textwidth]{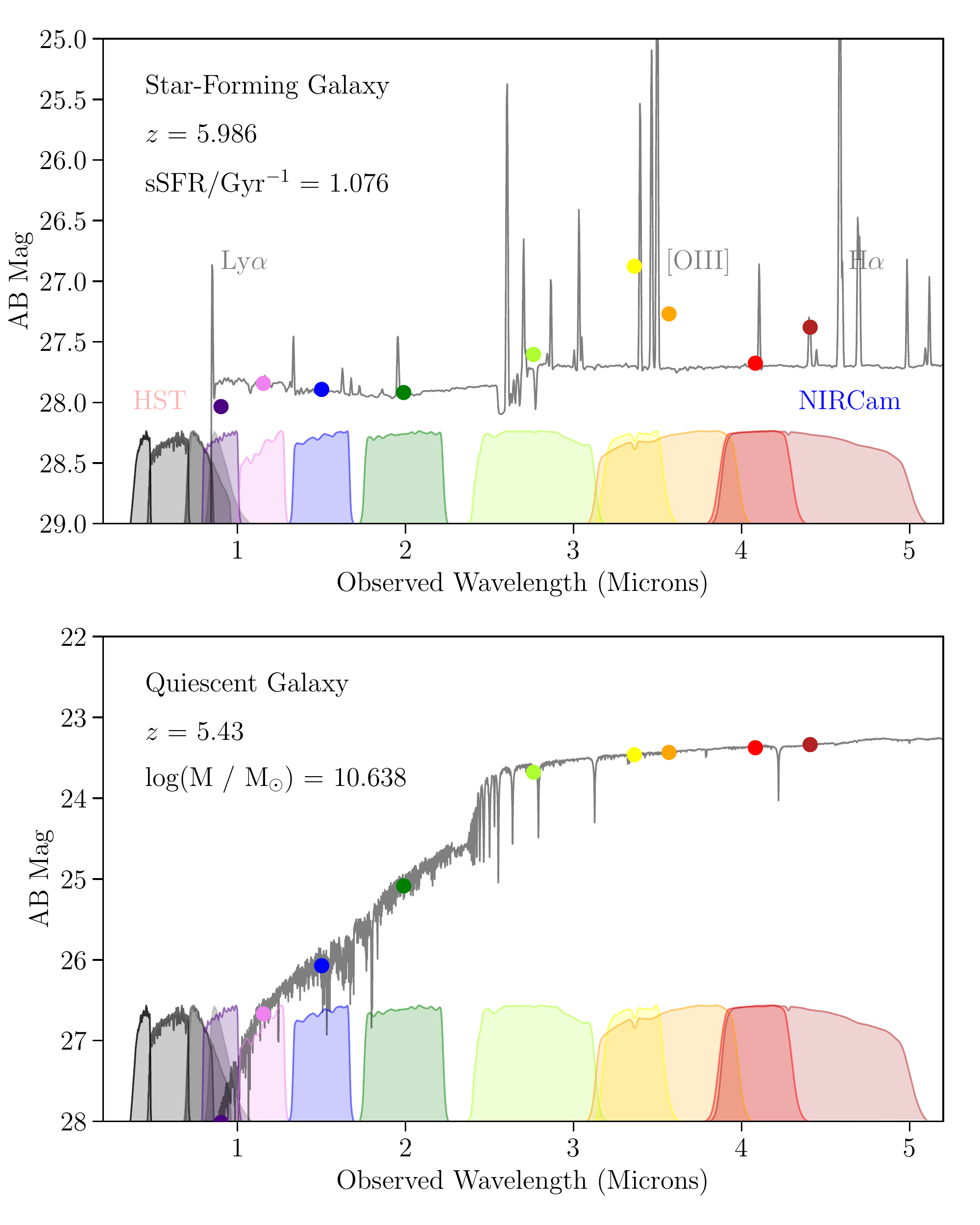}
\caption{\ Example SEDs and photometry for two objects in the mock catalog, with HST and NIRCam photometric filter transmission curves plotted below. The bands used are, from left to right HST ACS F435W, F606W, F750W, F814W, and F850LP (shades of grey), and NIRCam F070W, F090W, F115W, F150W, F200W, F277W, F335M, F356W, F410M, and F444W (rainbow). (Top) A star-forming galaxy at $z = 5.986$, with strong emission lines labeled. (Bottom) A quiescent galaxy at $z = 5.43$. The full SED of each galaxy, as well as line fluxes and rest-frame EWs, are provided in the mock catalog, which contains over 300,000 model SEDs across 11$\times$11 arcmin$^{2}$. }  
\label{fig:exampleseds}
\end{figure}

\section{Predictions for the first NIRCam/{\it JWST} imaging surveys}\label{results}
 
We demonstrate the predictive power of JAGUAR  
using the realization of the phenomenological model analyzed in Section~\ref{characterization} in order to make predictions for the initial deep extragalactic surveys that will be produced with {\it JWST}. 
The NIRCam and NIRSpec Science Teams have proposed a joint GTO program, the JADES survey, which will encompass both imaging and spectroscopy over 236 arcmin$^{2}$ 
in two well-studied fields, GOODS-S and GOODS-N. The NIRCam imaging aspect of the survey is a two-tiered strategy composed of a 46 arcmin$^{2}$ deep sub-survey, and a 190 arcmin$^{2}$ medium-depth sub-survey, with ten photometric bands between 0.7 and 5 \micron. The average point source detection limits for each subsurvey are summarized in Table~\ref{tab:nc}. For illustration we plot in Figure~\ref{fig:exampleseds} two example SEDs of mock high-redshift galaxies that would be detectable in the JADES survey.

\begin{figure*}
\includegraphics[width=1.0\textwidth]{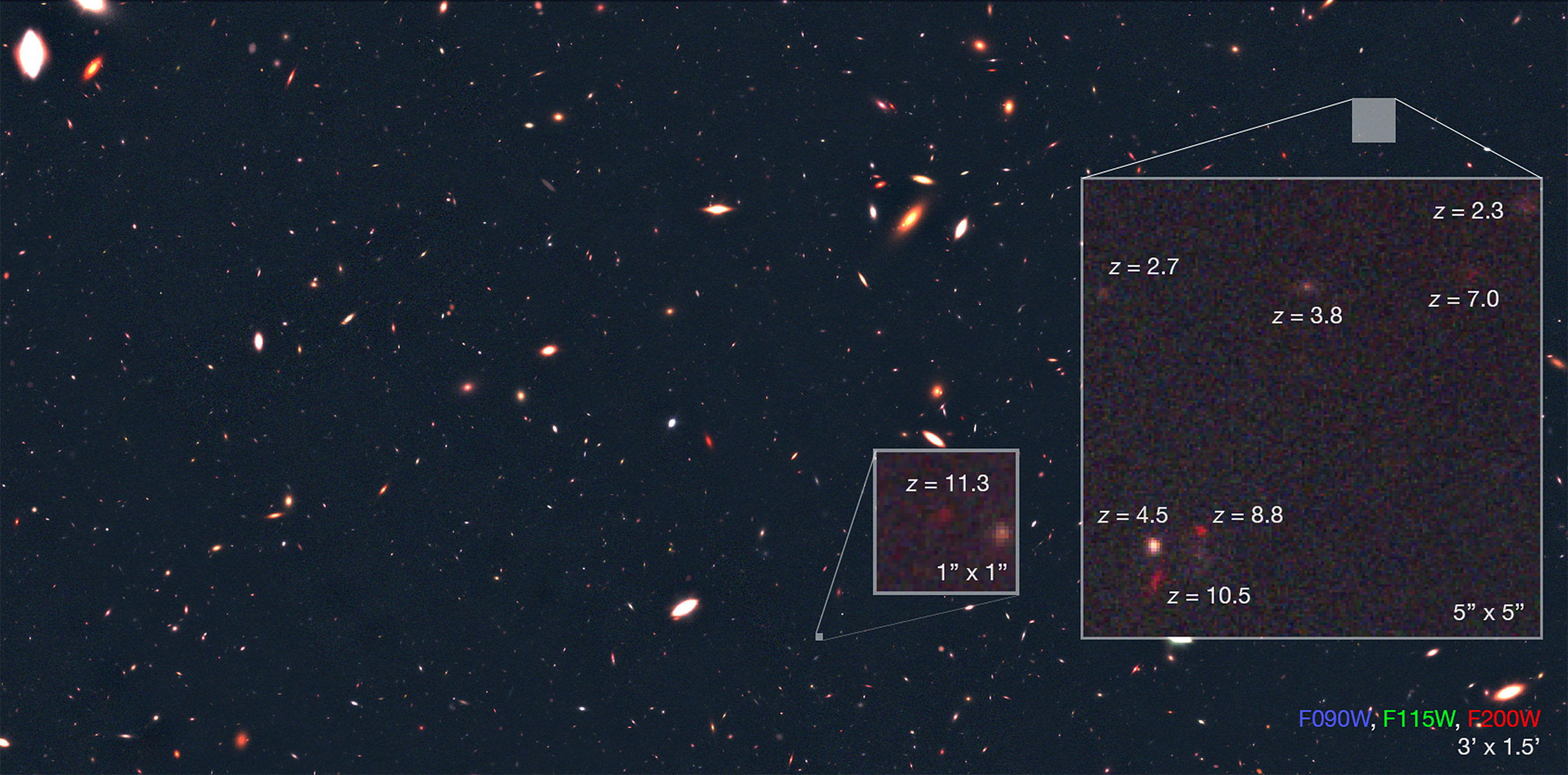}
\caption{ Simulated {\it JWST}/NIRCam mosaic generated using JAGUAR and the NIRCam image simulator \textit{Guitarra} (C. Willmer, in preparation), at the depth of the JADES Deep program. This image is focused on a region of 3' by 1.5', and is a composite of the F090W (blue), F115W (green), and F200W (red) filters.
The insets show a 5'' by 5' region with multiple high-redshift galaxies, and a 1'' by 1'' region focused on a galaxy at $z = 11.3$. }
\label{mockfig}
\end{figure*}

A visualization of the 
mock catalog can be seen in Figure \ref{mockfig}, which shows a simulated NIRCam mosaic of the JADES deep subsurvey (see Table \ref{tab:nc}). The images in the mosaic were generated using {\it Guitarra} (C. Willmer, in preparation), 
a ray-tracing image simulator specifically designed to create mock {\it JWST}/NIRCam scenes.
Photons in the simulated image that are associated with an object are added to the detector
pixels after being convolved with the mock galaxy Sersic model, the point spread function and the intra-pixel capacitance \citep{Rauscher2007}. The simulated image also includes the contribution due to the zodiacal and telescope background light, cosmic rays, read noise and the detector signatures as measured from ground-based data. 
The scenes are made using the same read-out patterns that will be used in flight, and are reduced using the pipeline that will be applied to the JWST data once these become available. The image shown in Figure \ref{mockfig}  
is a composite of a total of 648 images in F090W, F115W and F356W, combined using $swarp$ \citep{Bertin2002}, using the dither positions calculated by the JWST Astronomer's Proposal Tool (APT) for GTO proposal 1180.

\begin{table*}[ht]
\begin{center}
\begin{tabular}{|l|c|ccccc|ccccc|}
\hline
& Area & \multicolumn{10}{c|}{5$\sigma$ Point Source Magnitude (AB)} \\
Subsurvey & [arcmin$^{2}$] & F070W & F090W & F115W & F150W & F200W & F277W & F335M & F356W & F410M & F444W \\
\hline
Deep 	& 46  & ---     &  30.3 &  30.6 & 30.7 &  30.7 & 30.3 & 29.6 & 30.2 & 29.8 & 29.9 \\
Medium  &  190 & 28.8$^a$ & 29.4 & 29.6 & 29.7 & 29.8 & 29.4 & 28.8$^a$ & 29.4 & 28.9 & 29.1 \\

\hline
\end{tabular}
\end{center}
\caption{\label{tab:nc}Summary of NIRCam Imaging as part of JADES.  
The 5-$\sigma$ depth for a point source corresponding to the average exposure times.  We also include the area covered in square  arcminutes.  
$^a$The F070W and F335M areas of the Medium survey are only 93 square arcminutes. 
}
\end{table*}

\subsection{High-redshift galaxy counts}

\begin{figure}
\includegraphics[width=0.53\textwidth]{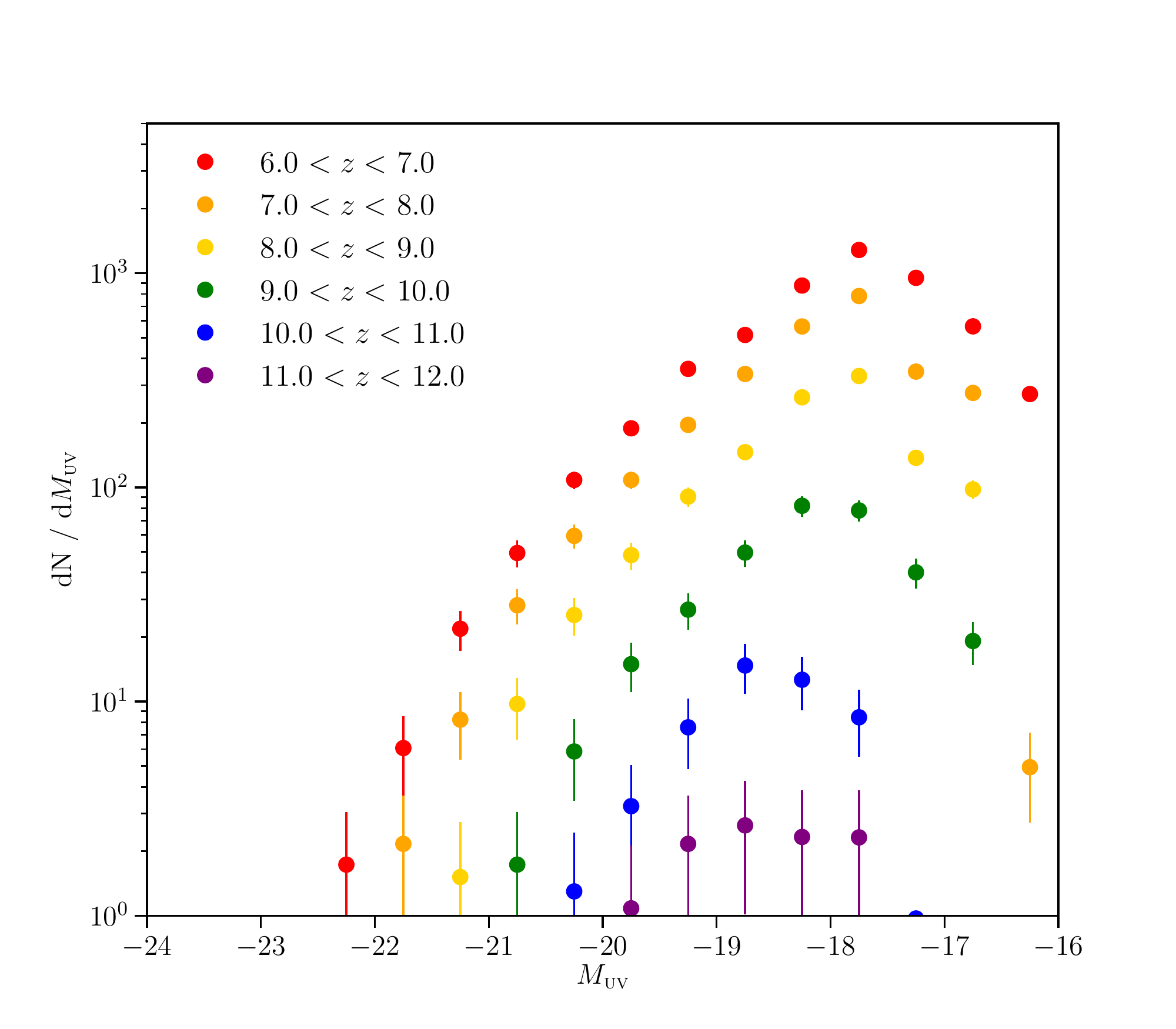}
\caption{Predicted total number of objects detected as a function of \Muv\ in bins of redshift for 
JADES. Objects are selected as ``detected'' if they are brighter than the $5\sigma$ limits in two photometric bands corresponding to the rest-frame UV, similar to common LBG selection techniques.  }
\label{gtonumbercounts}
\end{figure}

\begin{figure}
\includegraphics[width=0.5\textwidth]{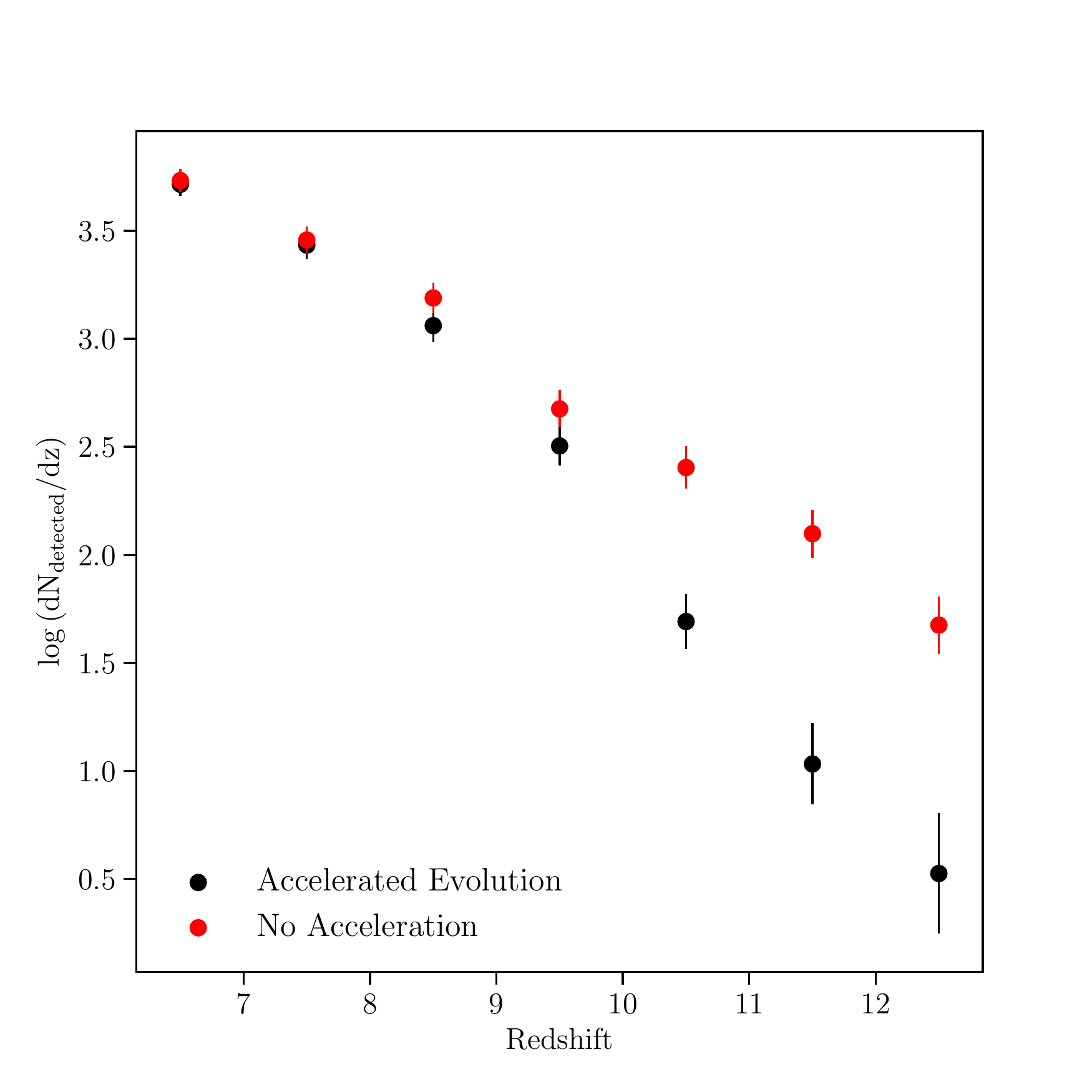}
\caption{Predicted total number of galaxies as a function of redshift in the full 
JADES area detected at $\gtrsim 5\sigma$. We plot results from the mock with the accelerated evolution above $z = 8$ with black points against those with the evolution predicted by \citet{Bouwens2015LF} with red points. The uncertainties include both Poisson error, and cosmic variance estimated from the \citet{Trenti2008} ``cosmic variance calculator''. Data from the JWST GTO program will help to discern between the two evolutionary scenarios at high redshift.  }
\label{gtodetectedgalaxies}
\end{figure}

In Figures~\ref{gtonumbercounts} and \ref{gtodetectedgalaxies}, we show predictions for the high-redshift detections with NIRCam imaging in the JADES GTO survey.
In these figures, we plot only those objects that are detected with at least $5\,\sigma$ (assuming point source detection limits) in two rest-frame UV photometric bands: that closest to 1500 \AA, and the nearest band at a longer wavelength.  
The detection bands correspond to F115W and F150W at $6<z<7$; F150W and F200W at $7<z<9.6$; and F200W and F277W at $9.6<z<13$. 

Figure~\ref{gtonumbercounts} shows the expected number of objects detected as a function of \Muv\ and redshift in the JADES survey based on our phenomenological model. 
We find that JADES should detect several thousands of galaxies at $z>6$, and several tens out to $z>10$. These predictions are based on the galaxy number counts evolution of our phenomenological model, which follows observations supporting a more rapid (``accelerated'') evolution of the UV luminosity function at $z\gtrsim8$ \citep[e.g.][]{Bouwens2016LF, Oesch2017}, implying a factor of $\sim 10$ decrease of galaxy counts between $z\sim 8$ and $z\sim 10$. This is substantially faster than the factor of $\sim 2$ decrease of galaxy counts per unit redshift seen at $3<z<8$ \citep[e.g.][]{Bouwens2015LF,Finkelstein2016} which are similar to the decrease measured by \citet{McLeod2015, McLeod2016} at $z>8$. 
However, due to large uncertainties from small sample sizes and survey volumes at $z\gtrsim8$ in current observations, there remain discrepancies in the literature of how fast the luminosity function evolves \citep{Oesch2012, Oesch2014, Zheng2012, Coe2013, Ellis2013, McLeod2015, McLeod2016, Bouwens2015LF, Bouwens2016LF, Calvi2016, Stefanon2017, Oesch2017,Ishigaki2017}. Accurately measuring the evolution at $z>8$ is an important goal for {\it JWST} because it will directly constrain the relationship between star-formation efficiency and the evolution of the halo mass function at early times \citep{Trenti2010, Tacchella2013, Mason2015, Liu2016, Mashian2016,SunFurlanetto2016,Ceverino2017,Cowley2018}, and is critical for our understanding of reionization. 

Therefore, we have used our mock catalog tool to predict if 
JADES will distinguish between the constant vs. ``accelerated" evolutionary model at $z\gtrsim8$. To do this, we have used the tool to produce a comparison mock realization based on 
the constant rate of evolution that was parametrized in \citet{Bouwens2015LF} at $4<z<8$, which we extrapolate to $z\gtrsim8$. This parametrization is in good agreement with the observations at $\range{z}{9}{10}$  by \citet{McLeod2015,McLeod2016}.
For this secondary set of realizations, we follow the same procedures for assigning physical properties and SEDs to mock galaxies as outlined previously in this work; the only difference is the surface density of counts of a given \Muv\ predicted by the UV luminosity functions. In Figure~\ref{gtodetectedgalaxies}, we additionally plot the total predicted number of galaxies that will be detected per redshift bin, expected for the constant rate of evolution from \citet{Bouwens2015LF} from the JADES survey (red points) in addition to the accelerated model (black points). The error bars in this figure represent both Poisson uncertainties as well as estimates for the uncertainty due to cosmic variance \citep{Trenti2008}, assuming 
the JADES survey area 
and a Press-Schechter formalism. From this figure it is clear that the future JADES survey will discriminate between the two evolutionary scenarios.

\begin{figure*}
\includegraphics[width=0.5\textwidth]{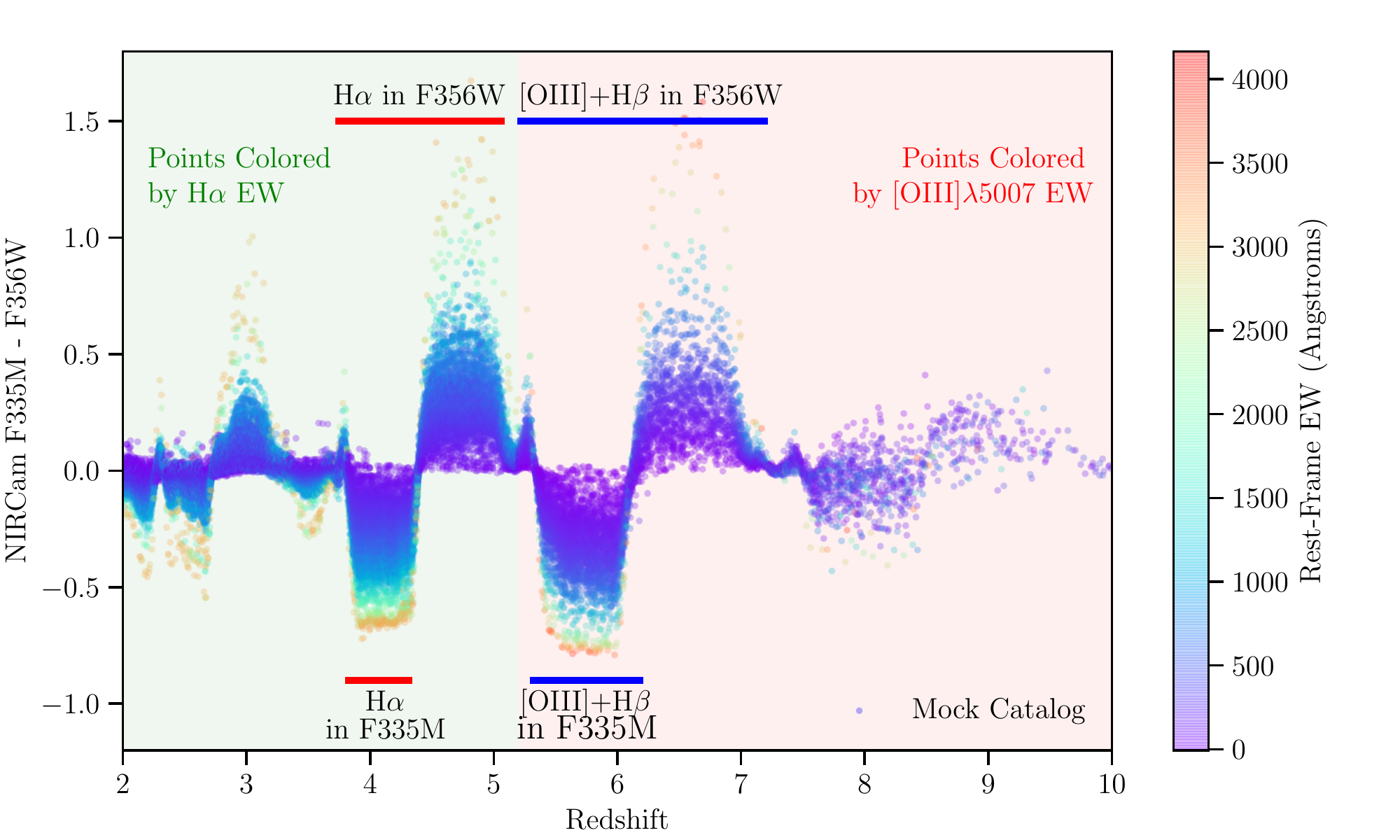}
\includegraphics[width=0.5\textwidth]{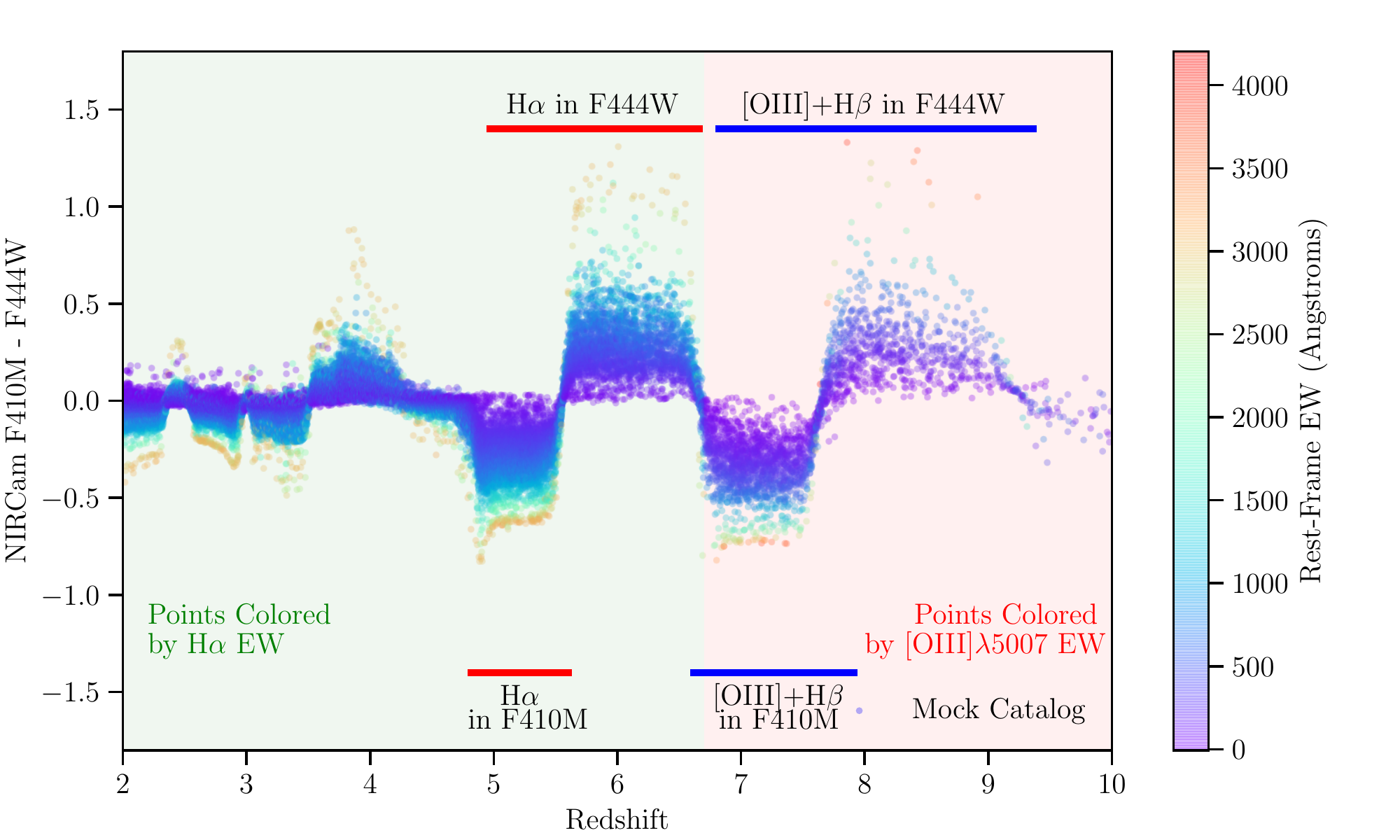}
\caption{Predicted {\it JWST}/NIRCam colors for mock galaxies in the realization of our model as a function of redshift. In both panels, we only plot those objects with fluxes above the 5$\sigma$ detection limit for that waveband in the Deep  
region of JADES as described in the text and Table~\ref{tab:nc}. Left panel: F335M-F356W colors. Right panel: F410M-F444W colors. The points are colored by the rest-frame EW of the H$\alpha$ emission line (light green region), and [OIII]$\lambda$5007 emission line (light red region). The redshift ranges where H$\alpha$, [OIII] and H$\beta$ enter the medium (wide) band filters are shown as the horizontal bars on the bottom (top) of the figure. In both panels, the mock predicts that we will be able to identify objects with powerful emission lines based on simple NIRCam color cuts. In the right panel, the features at $z = 2 - 3.5$ result from higher-order Paschen lines in the near-IR moving into the F410M and F444W filters.}
\label{fig:med_fluxexcess}
\end{figure*}

\subsection{Emission line predictions for NIRCam}\label{sec:nircammedflux}

The inclusion of nebular emission lines in the SEDs of mock galaxies enables analyses of the photometric contamination that is likely to be observed in future NIRCam imaging. Identifying the exact level of contamination in current surveys remains difficult, because there are only two deep {\it Spitzer} bands that provide rest-frame optical coverage of SEDs above $z\sim4.5$.  These bands are broad and there are only limited, narrow redshift ranges where one or other band is free from emission line contamination, and therefore able to provide constraints on the stellar continuum.  However, the extensive filter set of NIRCam, with three broad-band filters red-wards of current ground-based K-band imaging as well as a large array of medium- and narrow-band filters
will offer the opportunity to better characterize both the typical and extreme emission line EWs, which in turn will provide much better constraints on stellar masses at $z>4$. As summarized in Table~\ref{tab:nc}, the JADES survey will provide imaging in two medium bands at $\lambda > 3\,\micron$ (F335M and F410M).  These filters provide continuum anchor points when the emission line is outside the medium-band filter, or direct emission line contamination estimates of the broad-band fluxes when the line sits within the filter.

In Figure~\ref{fig:med_fluxexcess}, we plot the predicted redshift evolution of the NIRCam colors F335M-F356W and F410M-F444W of mock galaxies that would be detected at $>5\,\sigma$ in the deep region of the JADES survey. The galaxies are color-coded by the rest-frame EW of the emission lines that are contributing to the observed photometric flux at that redshift: H$\alpha$ at $z<5.2$ and $[\txn{O}\textsc{iii}]\lambda 5007$ at $z>5.2$ ($[\txn{O}\textsc{iii}]\lambda 5007$ dominates the EW of the $[\txn{O}\textsc{iii}]\lambda\lambda 4959,5007$ doublet). 
This figure demonstrates how strong emission lines can impact the observed infrared colors of galaxies in specific redshift ranges.
The multiple overlapping NIRCam bands means that not all bands are contaminated at a given time, and stellar mass estimates are likely to be better constrained than currently possible at $z\sim6$, where both IRAC bands are affected by emission line contamination. Additionally, Figure~\ref{fig:med_fluxexcess} 
demonstrates how NIRCam colors can be used to select high EW emission line objects at specific redshifts. 
Of those galaxies in the mock catalog with 5$\sigma$ Deep JADES detections, a color cut of F335M - F356W $> 0.8$ and $< -0.5$ will be successful at selecting a sample that is 99\% comprised of objects with rest-frame EW$_{\mathrm{H}\alpha} \gtrsim 1000$\AA\ at $z < 5.2$ (the number density of this sample is 3.85 objects per square arcminute). Similarly, a color cut of F335M - F356W $> 0.8$ and $< -0.6$ will be successful at selecting a sample that is 98\% comprised of objects with rest-frame EW$_{[\mathrm{OIII}]\lambda 5007} \gtrsim 1000$\AA\ at $z > 5.2$ (the number density of this sample is 1.62 objects per square arcminute). Objects selected with such extreme equivalent width emission lines would be obvious targets for follow-up spectroscopy with NIRSpec, and provide an independent probe of the H$\alpha$-derived star-formation rates 
free from the effects of slit losses produced by the MSA on NIRSpec.  They would additionally provide a consistent probe of the evolution of the number densities of extreme emission line galaxies from $z\sim4.8-8.3$.

\section{Summary}\label{summary}

We have developed a novel phenomenological model for the evolution of galaxies and their properties, 
based on empirical constraints from current surveys between $0.2 < z < 10$.
Our model follows 
observed stellar mass functions, UV luminosity functions, integrated distributions including \MStarMuv, \MuvBeta, and size-mass and size-\Muv\ distributions. Importantly, mock realizations of our model include galaxy SEDs that include strong nebular emission lines thanks to our
self-consistent modeling of the stellar and nebular emission with \beagle. These allow us to realistically model emission-line contamination to broad and medium-band filters but also provide emission line properties and low-resolution spectra for each galaxy in our mock catalog.
We have demonstrated that our phenomenological model is successful at matching the CSFRD, observations of the evolution of both the UV luminosity function and sSFR of galaxies over cosmic time as well as the evolution of the mass-metallicity relation.

We have created a mock catalog (our fiducial mock) using our phenomenological model and used this to make 
predictions for deep extragalactic surveys with NIRCam, in particular the joint NIRCam/NIRSpec GTO survey, JADES. We find that this survey will detect 1000s of galaxies at $z>6$ and 10s of galaxies at $z>10$, and will put firm constraints on the evolution of galaxy counts at $z>8$, resolving uncertainties on the rate of evolution that is currently debated in the literature. Additionally, we demonstrate how NIRCam colors can be used to select for high-equivalent width line emitters at high-redshift using the emission line information that is provided for the mock galaxies.
We make JAGUAR available for use, including both ready to use mock catalogs, and software to produce additional mock catalogs\footnote{available for download at http://fenrir.as.arizona.edu/jaguar/}.

\acknowledgments
Acknowledgments. 
We gratefully thank Darren Croton for thoughtful and constructive comments. Authors acknowledge helpful discussions with George Rieke, Michaela Hirschmann, and Jacob Magnusson, and gratefully thank Karl Misselt for computing and website assistance, and Pascal Oesch for providing his $z\sim10$ luminosity function measurements. CCW acknowledges enlightening conversations with Ivo Labb{\'e}. ECL, JC and SCh acknowledge support from the European Research Council (ERC) via an Advanced Grant under grant agreement no. 321323-NEOGAL. CCW acknowledges support from the National Science Foundation Astronomy and Astrophysics Fellowship grant AST-1701546. SCh acknowledges financial support from the Science and Technology Facilities Council (STFC).  All members of NIRCam (CCW, KNH, BER, RE, DPS, CNAW, SAl, SB, SCr, EE, DJE, MR) acknowledge funding from  JWST/NIRCam contract to the University of Arizona, NAS5-02015. BER acknowledges partial support through NASA contract NNG16PJ25C, grants 17-ATP17-0034 and HST-GO-14747. SAr is funded by MINECO under grant ESP2015-68964-P. RM and RA acknowledge ERC Advanced Grant 695671 "QUENCH'' and support by the Science and Technology Facilities Council (STFC). RS acknowledges a NWO Rubicon grant, project number 680-50-1518. This  work is  based on  observations taken  by  the CANDELS  Multi-cycle  Treasury Program with  the NASA/ESA HST. This research made use of Astropy, a community-developed core Python package for Astronomy \citep{Astropy}.

\appendix

\section{Re-fitting observed UV-luminosity and mass functions}

\label{app:MLmodelling}

\subsection{Fitting mass function parameters to $z\gtrsim4$ observed luminosity functions}
\label{app:MF_fit_to_LFs}

As described in Section~\ref{evolvemassfn}, we re-fit the individual \cite{Bouwens2015LF} UV-luminosity functions with \textit{mass} function Schechter parameters.  To do this we convolve a given mass function with our model of the evolving \MStarMuv\ relation (described in Section~\ref{MstarMuv}) to produce the corresponding UV luminosity function (general procedure described in Section~\ref{galaxy_counts_general}). We then use Markov-Chain Monte-Carlo (MCMC) sampling, employing the adaptive metropolis algorithm of \cite{haario2001}, to sample from the posterior probability distributions of the Schechter function parameters.  At each iteration, Schechter function parameters are proposed and corresponding CDF function generated.  The CDF function is used to randomly assign masses to a large population of  objects, which can then be used to assign each object an \Muv\ value following our adopted redshift-evolving \MStarMuv\ model.  The resulting  UV luminosity function is measured within the \Muv\ bins of the published measurements.  The likelihood of those given Schechter function parameters is evaluated using
the published $\Phi(\Muv,z)$ values and associated errors given in \cite{Bouwens2015LF}, Table~5, and the modeled luminosity function values evaluated in the same \Muv\ bins.

The $z\gtrsim4$ luminosity functions are well-described by a single Schechter function; thus fitting these data with a double Schechter function renders the parameters of $\Phi_1(\Mstar)$ unconstrained at $z\gtrsim4$.  We therefore constrain the values of each $\Phi_1(\Mstar)$ parameter while performing the luminosity function fits.  The values of \alphaOne\ and \phiStarOne\ are fixed in each redshift bin using the extrapolated best-fit linear and quadratic relations to the published T14 maximum likelihood measurements, and we also tie the value of \MFMstarOne\ to that of \MFMstarTwo with $\MFMstarOne=\MFMstarTwo=\MFMstar$.   The exact form of the redshift evolution of the parameters of $\Phi_1(\Mstar)$
do not affect the results of the luminosity function fits as long as the number density has fallen significantly by $z\sim4$.

The measurements are displayed in Figure~\ref{fig:MF_param_evolution}.

\subsection{Multi-level modeling to fit to Tomczak et al. (2014) star-forming galaxy stellar mass functions}
\label{app:MLmodelling}

To constrain the parameters $a_1$, $b_1$, $b_2$, $b_3$, $c_1$, $c_2$, $e_1$ and $f_1$ of our model of the redshift evolution of the mass function (equations \ref{eq:MFredshiftEvolutionStart} - \ref{eq:MFredshiftEvolutionEnd}), we fit to the T14 star-forming mass functions using a Bayesian multi-level modeling approach. 
Whereas the Bayesian fitting to each individual luminosity function entails sampling from the posterior distribution of each Schechter function parameter, the multi-level modeling involves sampling over the posterior distribution of the hyper-parameters (as we shall now refer to the parameters $a_1$, $b_1$, $b_2$, $b_3$, $c_1$, $c_2$, $e_1$ and $f_1$)  \textit{and}  the conditional probability distributions for each Schechter function parameter in each redshift bin.  The posterior distribution of the model free parameters can be expressed as:
\begin{equation}\label{eq:JointPosterior}
\begin{aligned}
\conditional{\mathbf{A},\mathbf{\Phi}}{\mathbf{X}} \propto
\prod_{i=1}^8 \txn{P}(A_i) \, 
\prod_{z=1}^{8} \conditional{\mathbf{x}_{z}}{\mathbf{\phi}_z} \,
\prod_{z=1}^{8} \conditional{\mathbf{\phi}_z}{\mathbf{A}} \, ,
\end{aligned}
\end{equation}
where $\mathbf{A} = [a_1, b_1, b_2, b_3, c_1, c_2, e_1,f_1]$ represents the free parameters describing the redshift evolution of the individual Schechter function parameters, $\txn{P}(A_i)$ is the prior on the $i$th free parameter,  $\mathbf{\Phi} = [\mathbf{\phi_1},\mathbf{\phi_2},...,\mathbf{\phi_n}]$ the set of $n$ mass function parameters, $\mathbf{X} = [\mathbf{x_0},\mathbf{x_1},...,\mathbf{x_8}]$ the measurements of the mass function in each redshift bin, and $\mathbf{\phi}_z = [\M^*_{1,\scM,z},\phi^*_{1,\scM,z},\alpha_{1,\scM,z},\M^*_{2,\scM,z},\phi^*_{2,\scM,z},\alpha_{2,\scM,z}]$ the Schechter function parameters in each redshift bin.   We fit with weakly informative priors (broad Gaussian distributions) on each hyper-parameter, with the added constraint that $b_3$ be negative to ensure that the normalization of $\Phi_1(\M)$ decreases with redshift, and $e_1$ and  $f_1$ be positive.  All priors are reported in Table~
\ref{table:MFhyperParams}.  We assume that the measurements represent true measurements of the mass function at the  mid-point of each redshift bin ($\mathbf{z}=[0.35, 0.625, 0.875, 1.125, 1.375, 1.75,$ $2.25, 2.75]$), and so do not include any modeling of the redshift distribution of the underlying sources that make up the mass bin volume density measurements.

The mass function estimates reported in T14 are supplied as $\log[\Phi(\Mstar)]$ (see their Table~1), with no information of the number of galaxies entering each bin.  In fact, the associated errors adopted from T14 account for Poisson noise, cosmic variance, the uncertainties arising from classifying galaxies as star-forming or quiescent and from the determination of stellar masses, and so are not simply Poissonian errors.  Thus, in the absence of the information required to construct the correct function for our count distribution, we resort to a Gaussian assumption, giving: 

\begin{equation}\label{eq:Likelihoods}
\begin{aligned}
\conditional{\mathbf{x}_z}{\phi_z} = \prod_i\,\frac{1}{\sqrt[]{2\pi}\sigma_i}\,\txn{exp}\Big(-\frac{(x_i-\hat{x_i}\,(\mathbf{\phi}_z))^2}{2\sigma_i^2}\Big)
\end{aligned}
\end{equation}
\noindent
where $x_i$ and $\sigma_i$ are the estimate and error of the mass function in bin $i$ respectively and $\hat{x_i}\,(\mathbf{\phi}_z)$ is the model prediction.
 
We sample the posterior distribution of model parameters with a Metropolis-within-Gibbs sampler (see, e.g., \citealt{Sharma2017}).
Gibbs sampling involves drawing directly from the conditional distribution for each parameter in turn. However, when the conditional probability is tricky to derive, it is possible to use the Metropolis update step, where the next step in the chain is sampled from a proposal distribution, often a Gaussian, and accepted or rejected based on comparison of the posterior probability between the current and last steps of the chain.  The advantage of using a Gibbs sampler for this problem is that the conditional distribution of the model parameters in each redshift bin are independent of each other, meaning that  $\conditional{\mathbf{\phi}_{i}}{...\mathbf{\phi}_{i-1},\mathbf{\phi}_{i+1},...,\mathbf{A},\mathbf{X}} =\conditional{\mathbf{x}_i}{\mathbf{\phi}_i}\,\conditional{\mathbf{\phi}_i}{\mathbf{A}}$ and so the mass function parameters for each redshift bin can be sampled in turn, before updating the values of the hyper-parameters.

During each iteration of the chain our multi-level modeling algorithm performs the following steps:
\begin{enumerate}
\item Values for the double Schechter function parameters are proposed for each individual redshift bin in turn, sampling from $\conditional{\mathbf{\phi}_{i}}{...\mathbf{\phi}_{i-1},\mathbf{\phi}_{i+1},...,\mathbf{A},\mathbf{X}}$, using a Metropolis update step.  
\item Once the mass function parameters of each redshift bin has been updated, new values for the hyper parameters are proposed.  They are updated simultaneously, sampling from $\conditional{\mathbf{A}}{\mathbf{\Phi},\mathbf{X}}$, again using a Metropolis update step.
\end{enumerate}
\noindent
After an initial burn-in stage of 10000 iterations, the width of the Gaussian proposal distribution is set using the covariance matrix of the past history of the chains, following the adaptive Metropolis algorithm of \cite{haario2001}.  We run two different chains with independent starting points for 40000 iterations each, after which we test for convergence using the scale reduction factor, $\hat{R}$, which compares the within-chain and between-chain variance.  We require $\hat{R} < 1.05$ in each free parameter.

After rejecting the initial 10000 iterations, we estimate each parameter value as the median of the values in the chain and uncertainties as their standard deviation.

\section{Uncertainties in star-forming/quiescent selection criteria and impact on creating parent catalog for high-mass quiescent galaxies from 3D-HST data}
\label{app:robust_Q_determination}

When drawing multiple realizations from SED fits to individual quiescent galaxies we often
find that the uncertainties on the $U$, $V$ and $J$-band absolute magnitudes allow for solutions that would place the object in the star-forming region of the $UVJ$ color space.  
This is because galaxies are selected from a combined $J_{125}$, $J_{140}$ and $H_{160}$ band image in the 3D-HST catalog  (Section~\ref{section:3DHST_SEDfitting}). Since quiescent galaxies have red SEDs, their fainter fluxes at shorter wavelengths can therefore suffer from large observational uncertainties. Although the uncertainties on the star-forming and quiescent galaxy
classifications are included in the T14 stellar mass function estimates,  we must ensure that galaxies designated as quiescent in our mock catalog are assigned quiescent SEDs.  In the case of assigning SEDs to our star-forming galaxy mock, the prior on \MStarMuv\ used to weight the draws from the \beagle\ fitting prevents the assignment of quiescent SEDs.  However, for the quiescent parent catalog we only require that the realizations are drawn from within the 68\% contour on stellar mass and this requirement does not explicitly require the SED to also be quiescent.

\begin{figure}
\includegraphics[width=0.53\textwidth]{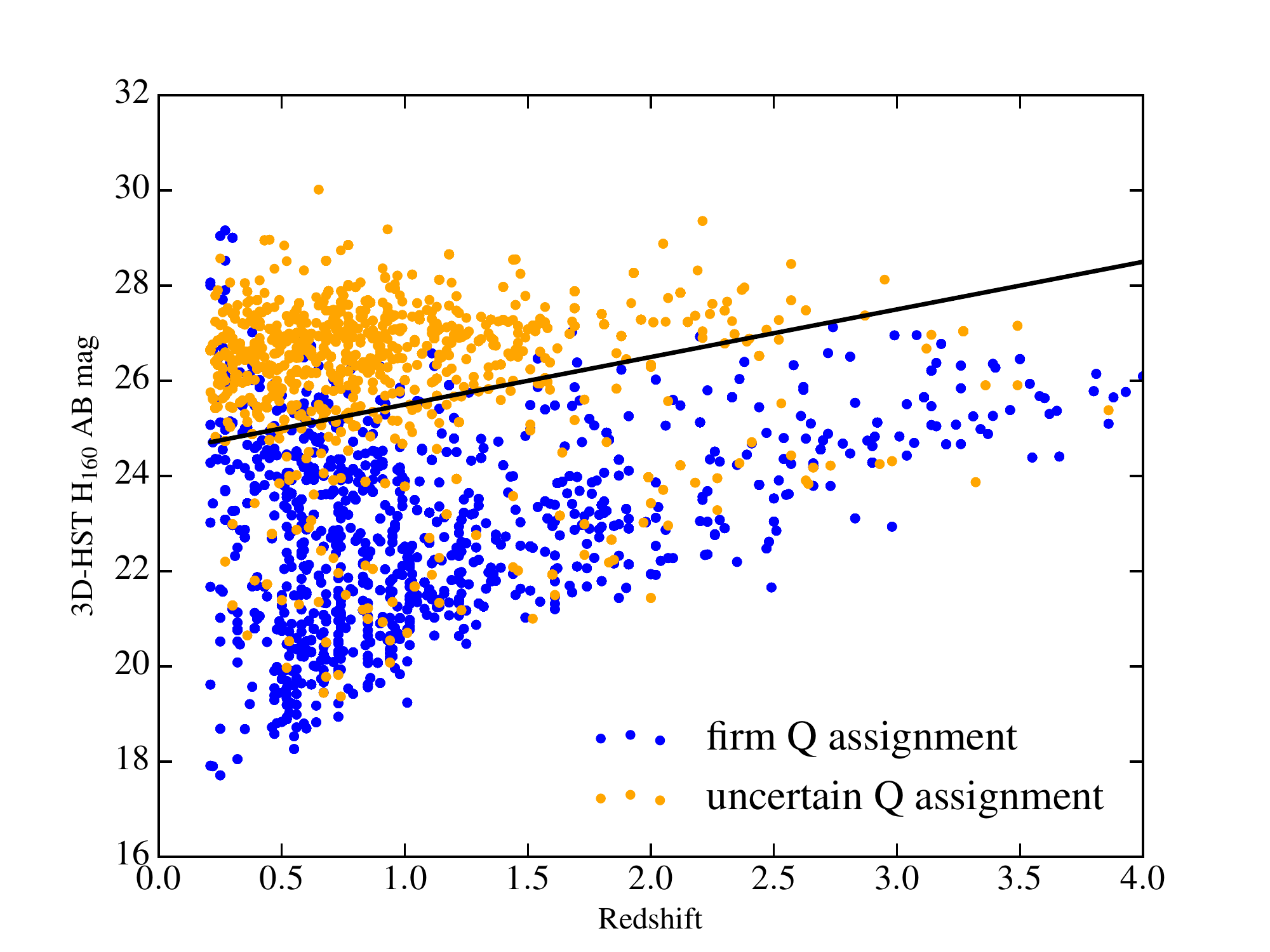}
\includegraphics[width=0.53\textwidth]{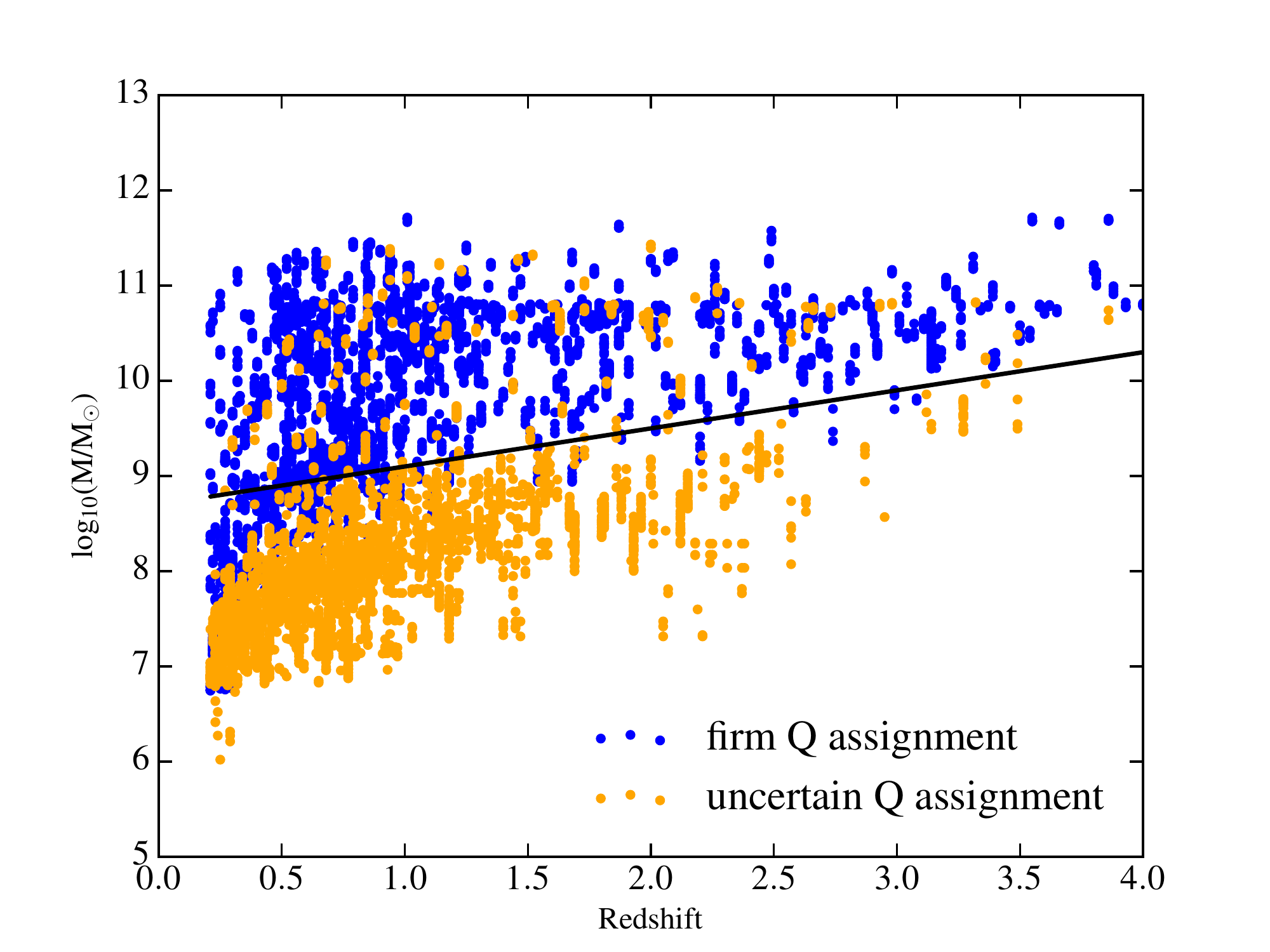}
\caption {Left panel: 3D-HST H$_{160}$ magnitude plotted against redshift for a subset of the galaxies classified as quiescent (Q) from their rest-frame $U-V$ and $V-J$ colors.  The blue points are objects that have at least 45 realizations that are assigned as quiescent from a random sampling of 50 realizations within the 68\% confidence contour on stellar mass.  The orange points have between 20-30 realizations assigned as quiescent.  The black line is $H_{160} = 24.5+z$ shows the approximate division between the two categories and is used to determine objects which are robustly assigned as quiescent. Right panel: model \Mstar vs redshift for each  }
\label{fig:quiescent_vs_H160}
\end{figure}

To characterize the redshift-dependent limiting magnitude at which the quiescent assignment based on $UVJ$ colors is robust,  we randomly sample 50 solutions from within the 68\% uncertainty in mass for each quiescent object in the 3D-HST catalog and re-measure the $UVJ$ colors, providing a fraction of draws per object that reside in the quiescent region of $UVJ$ color space.  We compare in Figure~\ref{fig:quiescent_vs_H160} the distribution of $H_{160}$ magnitudes of objects for which at least 45 of these 50 realizations fall in the quiescent region of the $UVJ$ color space (in blue) to that of objects for which between 20 and 30 realizations fall outside of the quiescent region (in orange).  Based on these results,
we limit the objects that enter the 3D-HST-derived parent catalog based on their
$UVJ$ colors to $H_{160} < 24.5 + z$ (black line in Figure~\ref{fig:quiescent_vs_H160}).  Given that the $H_{160}$ band samples the SED red-wards of the 4000\AA\ break for $z\lesssim2.5$, we expect this limit in $H_{160}$ to approximately correspond to a limit in stellar mass.  We therefore plot the stellar mass as a function of redshift for those objects with firm and uncertain quiescent assignments in Figure~\ref{fig:quiescent_vs_H160}, right panel.  Using this figure we set a mass limit of $\LMstarMsun > 8.7+0.4z$  for firm quiescent galaxy assignment.  We note that this limit is likely to be conservative at $z>2.5$, where the $H_{160}$-band magnitude no longer correlates strongly with stellar mass. In practice, only objects from our mock catalog with masses above this limit will be matched to the observationally-derived parent galaxy catalog.  Objects below this mass will be matched to the parent galaxy catalog produced by \beagle.

\bibliographystyle{yahapj}
\bibliography{MockJWST}

\end{document}